\documentclass[letterpaper,11pt]{article}
\pdfoutput=1

\usepackage{jheppub}
\usepackage{hyperref}
\usepackage{amsmath}
\usepackage{amsfonts}
\usepackage{graphicx}
\usepackage{caption}
\usepackage{subcaption}
\usepackage{placeins}
\usepackage{tikz}
\usepackage[english]{babel}

\usepackage{array}
\usepackage{mathtools}
\def\be{\begin{equation}}
\def\ee{\end{equation}}
\def\ba{\begin{eqnarray}}
\def\ea{\end{eqnarray}}

\newcommand{\MCG}{\text{MCG}}

\title{
All Loop Scattering as a  Counting Problem
} 

\author[a]{N.~Arkani-Hamed,}\author[b]{H.~Frost,}\author[c]{G.~Salvatori,}\author[d]{P-G.~Plamondon}\author[e]{H.~Thomas}

\affiliation[a]{School of Natural Sciences, Institute for Advanced Study, Princeton, NJ, 08540, USA}

\affiliation[b]{Mathematical Institute, Andrew Wiles Building, Woodstock Rd, Oxford, UK}

\affiliation[c]{Max-Plank-Instit\"ut fur Physik, Werner-Heisenberg-Institut, D-80805 M\"unchen, Germany}

\affiliation[d]{Laboratoire de Math\'ematiques de Versailles, UVSQ, CNRS, Universit\'e Paris-Saclay, IUF, France}

\affiliation[e]{LaCIM, D\'epartement de Math\'ematiques, Universit\'e du Qu\'ebec \`a Montr\'eal, Montr\'eal, QC, Canada}

\emailAdd{arkani@ias.edu}
\emailAdd{frost@maths.ox.ac.uk}
\emailAdd{giulios@mpp.mpg.de}
\emailAdd{pierre-guy.plamondon@uvsq.fr}
\emailAdd{thomas.hugh\_r@uqam.ca}

\date{\today}

\abstract{
This is the first in a series of papers presenting a new understanding of scattering amplitudes based on fundamentally combinatorial ideas in the kinematic space of the scattering data.  We study the simplest theory of colored scalar particles with cubic interactions, at all loop orders and to all orders in the topological 't Hooft expansion. We find a novel formula for loop-integrated amplitudes, with no trace of the conventional sum over Feynman diagrams, but instead determined by a beautifully simple counting problem attached to any order of the topological expansion. These results represent a significant step forward in the decade-long quest to formulate the fundamental physics of the real world in a radically new language, where the rules of spacetime and quantum mechanics, as reflected in the principles of locality and unitarity, are seen to emerge from deeper mathematical structures.}

\begin{document}
  \maketitle
  
\section{Introduction and Summary}
Scattering amplitudes are perhaps the most basic and important observables in fundamental physics. The data of a scattering process---the on-shell momenta and spins of the particles---are specified at asymptotic infinity in Minkowski space. The conventional textbook formalism for computing amplitudes ``integrates in" auxiliary structures that are not present in the final amplitude, including the bulk spacetime in which particle trajectories are imagined to live, and the Hilbert space in which the continuous bulk time evolution of the wavefunction takes place. These auxiliary structures are reflected in the usual formalism for computing amplitudes, using Feynman diagrams, which manifests the rules of spacetime (locality) and quantum mechanics (unitarity). As has been increasingly appreciated over the past three decades, this comes at a heavy cost---the introduction of huge redundancies in the description of physics, from field redefinitions to gauge and diffeomorphism redundancies, leading to enormous complexities in the computations, that conceal a stunning hidden simplicity and seemingly miraculous mathematical structures revealed only in the final result \cite{Parke:1986gb,Bern_1994,Bern_1995,Witten_2004,Cachazo_2004,Britto_2005,Britto_2005W}. 

This suggests that we should find a radically different formulation for the physics of scattering amplitudes. The amplitudes should be the answer to entirely new mathematical questions that make no reference to bulk spacetimes and Hilbert space, but derive locality and unitarity from something more fundamental. A number of concrete examples of this have already been found in special cases. The discovery of deep and simple new structures in combinatorics and geometry has led to new definitions of certain scattering amplitudes, without reference to spacetime or quantum mechanics. Notably, the amplituhedron determines the scattering amplitudes in planar {\cal N} =4 SYM, and associahedra and cluster polytopes determine colored scalar amplitudes at tree-level and one-loop \cite{Arkani_Hamed_2014,2017ABHY,Salvatori_2019,2019AHHST}.

Up to now, these results have been limited in how much of the perturbative expansion they describe---at all loop orders for maximally supersymmetric theories, but only in the planar limit, and only through to one loop for non-supersymmetric theories. Furthermore, the connection between combinatorial geometry and scattering amplitudes at loop level has only been made through the integrand (pre-loop integration) of the amplitudes, and not the amplitudes themselves. Both of these limitations must be transcended to understand all aspects of particle scattering in the real world.

This article is the first in a series reporting on what we believe is major new progress towards this goal. These ideas set the foundation for a number of other interrelated threads and results that will appear in various groups of papers. So we take this opportunity to give a birds-eye view of the nature of these developments and the new concepts that are driving this progress.

Our departure point is a new formulation of a simple theory,---colored scalar particles with cubic interactions,---at all loop orders and to all orders in the topological 't Hooft expansion, in the form of what we call a \emph{curve integral}. This approach has no hint of a sum over Feynman diagrams anywhere in sight and is instead associated with a simple counting problem defined at any order in the topological expansion. This counting problem defines a remarkable set of variables, $u_C$, associated with every curve, $C$, on a surface. The $u$-variables non-trivially define \emph{binary geometries} \cite{Arkani-Hamed:2019mrd} by dint of satisfying the remarkable non-linear equations \cite{us_us}
\begin{equation}
u_C + \prod_D u_D^{n(C,D)} = 1,
\end{equation}
where $n(C,D)$ is the intersection number of the curves $C,D$. In the \emph{positive region}, where all the $u_C$ are non-negative, the $u$-equations force all the $u_C$ to lie between 0 and 1: $0\leq u_C \leq 1$. Of mathematical interest, this positive region is a natural and invariant compactification of \emph{Teichm\"uller space}. This algebraic presentation of Teichm\"uller space is a counterpart to the famous synthetic compactification of Teichm\"uller spaces and surface-type cluster varieties given by Fock-Goncharov \cite{2003math.....11149F,penner2012decorated}. The new compactifications defined by the $u_C$ variables are immediately relevant for physics, and lead to the new \emph{curve integral} formulation of all-loop amplitudes presented in this article.

The curve integral does more than reformulate the perturbative series in a new way. It also exposes basic new structures in field theory. For instance, a striking consequence of our formulation is that amplitudes for large $n$ particles at $L$-loops effectively factorise into a tree and a loop computation. The full large $n$ amplitudes can be reconstructed from computations of $n$-point tree amplitudes and low-point $L$-loop amplitudes. Moreover, our curve integral formulas make manifest that amplitudes satisfy a natural family of differential equations in kinematic space. The solutions of these equations give novel and efficient recursion relations for all-loop amplitudes. 

This article focuses on colored scalar amplitudes. However, the results here have extensions to other theories. New curve integral formulations have been discovered for theories of colored scalar particles with arbitrary local interactions, as well as for the amplitudes of pions and non-supersymmetric Yang-Mills theories. These formulas reveal striking inter-relations between these theories, together with surprising hidden properties of their amplitudes that are made manifest by the curve integral formalism. 

Our results also have implications for the understanding of strings and UV completion. The counting problem at the heart of this paper not only defines QFT amplitudes, it also defines amplitudes for bosonic strings, via the $u$-variables, $u_C$, mentioned above. This gives a combinatorial formulation of string amplitudes that makes no reference to worldsheet CFTs and vertex operators. This new approach to string amplitudes differs from the conventional theory in a very fundamental way. The $u$-variables, which are derived from a simple counting problem, have a beautiful and direct connection to the geometry of two-dimensional surfaces. But this connection is via the \emph{hyperbolic geometry} of Teichm\"uller space, and {\it not} via the conventional picture of Riemann surfaces with a complex structure. The new string formulas are not just an exercise in passing between the complex and the hyperbolic pictures for Teichm\"uller space. We find that we can reproduce bosonic strings at loop level, but other choices are just as consistent, at least insofar as the field theory limit is concerned.  This allows us to deform string amplitudes into a larger, but still highly constrained, space of interesting objects. This runs counter to the lore that string theory is an inviolable structure that cannot be modified without completely breaking it. Our larger class of string amplitudes transcends the usual strictures on spacetime dimension, as well as the famous instabilities of non-supersymmetric strings. Moreover, our new combinatorial-geometric point of view also makes it easier to recover particle amplitudes from strings in the $\alpha^\prime \to 0$ limit. By contrast, recovering field theory from conventional string theory involves vastly (technically, infinitely!) more baggage than is needed \cite{us_strings}. 

There are several other related developments, including the discovery of a remarkable class of polytopes, \emph{surfacehedra}, whose facet structure captures, mathematically, the intricate boundary structure of Teichm\"uller space, and, physically, the intricate combinatorics of amplitude singularities at all loop orders, and whose \emph{canonical form} determines (an appropriate notion of the) loop integrand at all orders in the topological expansion. 

The results of all these parallel threads of investigation will be presented in various groups of papers. We end this preview of coming attractions by explaining a quite different sort of motivation for our works that will be taken up in near-future work. The counting problem that lies at the heart of this paper has an entirely elementary definition. But the central importance of this counting problem will doubtless seem mysterious at first sight. It finds its most fundamental origin in remarkably simple but deep ideas from the ``quiver representation theory" \cite{2017DW,haupt2012} of (triangulated) surfaces. Arrows between the nodes of a quiver can be associated with maps between vector spaces attached to the nodes. Choosing compatible linear maps between the nodes defines a \emph{quiver representation}. In this context, our counting problem is equivalent to counting the \emph{sub-representations} of these quiver representations. This perspective illuminates the mathematical structure underlying all of our formulas. But these ideas also hint at a fascinating prospect. The amplitudes we study are associated with the class of surface-type quivers, which are dual to triangulated surfaces. Nothing in our formulas forces this restriction on us: we are free to consider a much wider array of quivers. {\it All} of these quivers can be associated with amplitude-like functions. This vast new class of functions enjoys an intricate (amplitude-like) structure of ``factorisations" onto simpler functions. This amounts to a dramatic generalisation of the notion of an ``amplitude", and in a precise sense also generalises the rules of spacetime and quantum mechanics to a deeper, more elementary, but more abstract setting.

Having outlined this road map, we return to the central business of this first paper. We will study the simplest theory of $N^2$ colored particles with any mass $m$, grouped into an $N \times N$ matrix $\Phi^I_J$ with $I,J = 1, \cdots, N$. The Lagrangian, with minimal cubic coupling, is 
\begin{equation} 
{\cal L} = {\rm Tr} (\partial \Phi)^2 + m^2 {\rm Tr} (\Phi^2) + g {\rm Tr} (\Phi^3),
\end{equation}
in any number $D$ of spacetime dimensions. This theory is a simpler cousin of all theories of colored particles, including Yang-Mills theories, since the singularities of these amplitudes are the same for all such theories, only the \emph{numerators} differ from theory to theory. The singularities of amplitudes are associated with some of the most fundamental aspects of their conventional interpretation in terms of spacetime processes respecting unitarity. So understanding the amplitudes for this simple theory is an important step towards attacking much more general theories.

We will show that {\it all} amplitudes in this theory, for any number $n$ of external particles, and to all orders in the genus (or $1/N$) expansion \cite{1973H}, are naturally associated with a strikingly simple counting problem. This counting problem is what allows us to give \emph{curve integral} formulas for the amplitudes at all orders. The curve integral makes it easy to perform the loop integrations and presents the amplitude as a single object. 

As an example, consider the single-trace amplitude for $n$-point scattering at 1-loop. Let the particles have momenta $p_i^\mu$, $i=1,...,n$. The curve integral for this amplitude (pre-loop integration) is
\begin{equation}\label{eq:intro:1loop}
   \mathcal{A}^{1-{\rm loop}}_n = \int d^D l \int\limits_{\sum_i t_i \geq 0} d^n t \,{\rm exp} \left[-\sum_{i=1}^n \alpha_{i} (l + p_1 + \cdots + p_i)^2 - \sum_{i,j} \alpha_{i,j} (p_i + \cdots + p_{j-1})^2 \right]
\end{equation}
where 
\begin{align}
    \alpha_{i,j} &= f_{i,j} + f_{i+1,j+1} - f_{i,j+1} - f_{i+1,j},\\ \alpha_i &= \alpha_{i,i+n}, \nonumber \\ 
    f_{i,j}& = {\rm  max} (0,t_j,t_{j} + t_{j-1}, \cdots, t_j + t_{j-1} + \cdots t_{i+2}).
\end{align}
The propagators that arise in the 1-loop Feynman diagrams are either loop propagators, with momenta $(l + p_1 + \cdots + p_i)$, or tree-like propagators, with momenta $(p_i + p_{i+1} + \cdots + p_{j-1})$. The exponential in \eqref{eq:intro:1loop} looks like a conventional Schwinger parametrisation integral, except that \emph{all} the propagators that arise at 1-loop are included in the exponent. Instead of Schwinger parameters, we have \emph{headlight functions}: $\alpha_i$ (for the loop propagators) and $\alpha_{i,j}$ (for the tree propagators). The headlight functions are piecewise linear functions of the $t_i$ variables. The magic is that \eqref{eq:intro:1loop} is a {\it single} integral over an $n$-dimensional vector space. Unlike conventional Schwinger parametrisation, which is done one Feynman diagram at a time, our formulas make no reference to Feynman diagrams.  Amazingly, the exponent in \eqref{eq:intro:1loop} breaks $t$-space into different cones where the exponent is linear. Each of these cones can be identified with a particular Feynman diagram, and the integral in that cone reproduces a Schwinger parameterisation for that diagram. This miracle is a consequence of the properties of the headlight functions $\alpha_i(t)$ and $\alpha_{i,j}(t)$. These special functions arise from a simple counting problem associated with the corresponding propagator.

As in conventional Schwinger parametrisation, the dependence on the loop momentum variable, $l^\mu$, in the curve integral, \eqref{eq:intro:1loop}, is Gaussian. We can perform the loop integration to find the a second curve integral for the amplitude (post loop integration),
\begin{equation}\label{intro:1loop2}
\mathcal{A}^{1-{\rm loop}}_n  =  \int\limits_{\sum_i t_i \geq 0} d^n t \left( \frac{2\pi}{\mathcal{U}} \right)^{\frac{D}{2}} e^{-\frac{\mathcal{F}}{\mathcal{U}}}.
\end{equation}
In this formula, the polynomials ${\cal U}$ and ${\cal F}$ are given by 
\begin{equation}
    \mathcal{U} = \sum_i \alpha_i, \qquad \mathcal{F} = \sum_{i,j} \alpha_i \alpha_j (p_i + \cdots p_{j-1})^2 -  \left(m^2 \sum_i \alpha_i + 2 \sum_{i,j} \alpha_{i,j} \, p_i.p_j \right) \mathcal{U}.
\end{equation}
These polynomials are analogs of the familiar Symanzik polynomials, but whereas the Symanzik polynomials appear in individual Feynman integrals, this one curve integral above computes the whole amplitude.

These 1-loop curve integrals generalise to all orders in perturbation theory, at any loop order and genus. In the rest of this introductory section we give a birds-eye view of the key formulas and results.

\subsection{Kinematic space}
\begin{figure}
    \centering
    \includegraphics[width=\textwidth,trim={0cm 0cm 0cm 0cm},clip]{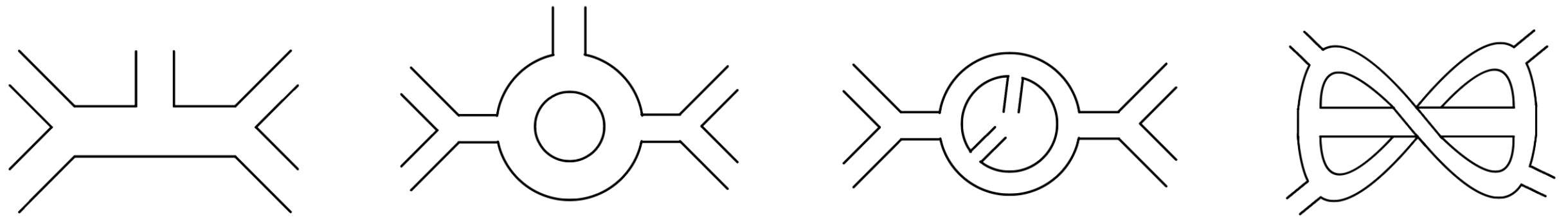}
    \caption{Fat graphs at tree-level, 1-loop single trace, 1-loop double trace, and 2-loop single trace, respectively.}
      \label{intro:fatgraphs}
\end{figure}

To begin with, we have to define the \emph{kinematic space} where all the action will take place. In our theory, each Feynman diagram is what is called a `double-line notation diagram', `ribbon graph' or `fatgraph' in the literature; we will call them fatgraphs in what follows. Examples of fatgraphs are shown in Figure \ref{intro:fatgraphs}. Order by order, in the 't Hooft expansion, these Feynman diagrams get organised into partial amplitudes, labeled by their shared \emph{color structure}. Conventionally, when we do a 't Hooft expansion, we think of these fat graphs as `living on' or `being drawn on' a surface with some genus and number of boundary components. We will think of them in a different way: a {\it single} fat graph itself {\it defines} a surface. In fact, we will use a single fat graph to define all the data we need to compute an amplitude!

\begin{figure}
    \centering
    \includegraphics[width=0.5\textwidth,trim={0cm 0cm 0cm 0cm},clip]{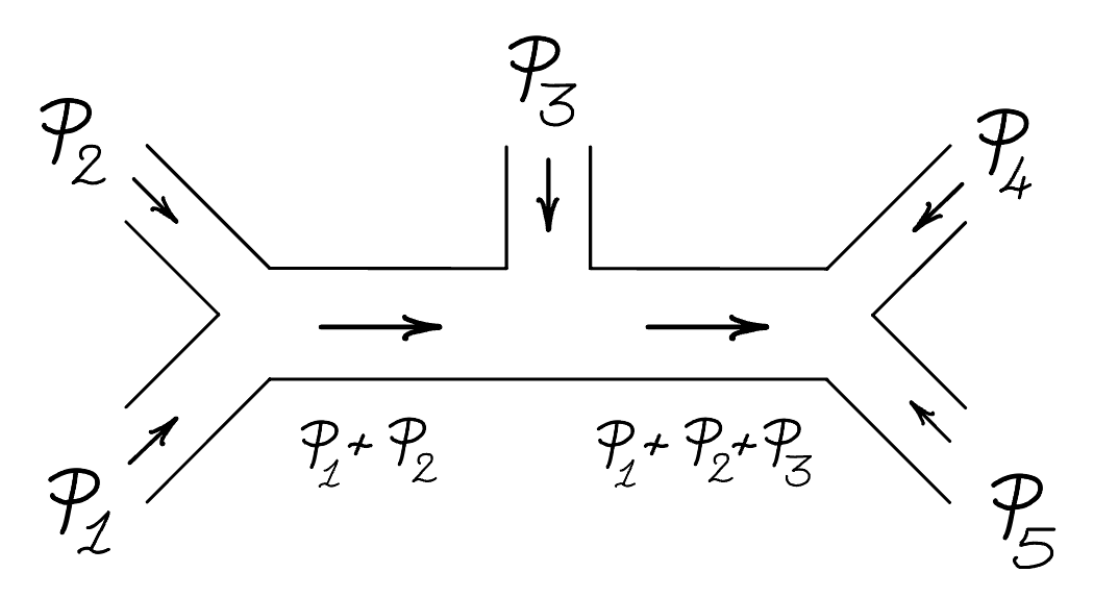}
    \caption{A tree fat graph with momenta assigned to all edges.}
      \label{intro:diagmom}
\end{figure}

Take some fatgraph, $\Gamma$, at any order in the 't Hooft expansion. Suppose that it has $n$ external lines and $E$ internal edges. Then this fat graph has loop order, $L$, with
\begin{equation}
E = n + 3(L-1).
\end{equation}
Let the external lines have momenta $p_1,\ldots, p_n$, and introduce $L$ loop variables, $\ell_1,\ldots,\ell_L$. Then, by imposing momentum conservation at each vertex of $\Gamma$, we can find a consistent assignment of momenta to all edges of the fat graph in the usual way: if each edge, $e$, gets a momentum $p_e^\mu$, then whenever three edges, $e_1,e_2,e_3$, meet at a vertex, we have
\begin{equation}
p_{e_1}^\mu + p_{e_2}^\mu+p_{e_3}^\mu = 0.
\end{equation}
For example, Figure \ref{intro:diagmom} is an assignment of momenta to the edges of a tree graph.

\begin{figure}
    \centering
    \includegraphics[trim={0cm 0cm 0cm 0cm}, clip, width=.4\textwidth]{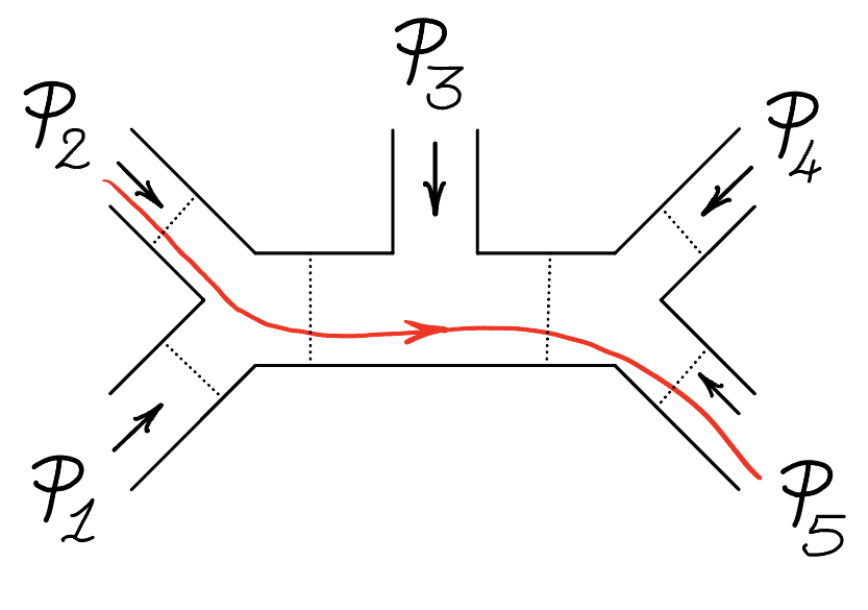}
    \caption{Every curve, $C$, drawn on the fat graph inherits a momentum, $P_C^\mu$, from the momenta assigned to the fat graph itself.}
    \label{intro:curveex}
\end{figure}

The amplitude itself depends on momenta only through Lorentz invariant combinations. So we want to define a collection of Lorentz invariant kinematic variables. Consider a curve, $C$, drawn on the fatgraph $\Gamma$ that starts at an external line, passes through the graph and exits at another external line. For example, the curve in Figure \ref{intro:curveex} starts at $p_2$, and exits at $p_5$. Every such curve can be assigned a unique momentum. It is given by the momentum of the first edge plus the sum of all momenta on the graph entering the curve `from the left'. For example, in Figure \ref{intro:curveex}, the curve starts with momentum $p_2$, and then takes two right turns. At the first right turn, momentum $p_3$ enters from the left. At the second right turn, momentum $p_4$ enters from the left. The total momentum of the curve is then given by
\begin{equation}
p_C^\mu = p_2^\mu + p_3^\mu + p_4^\mu.
\end{equation}
Notice that if we had gone in the opposite direction (starting at $p_5$), we would have got
\begin{equation}
- p_C^\mu = p_5^\mu + p_1^\mu.
\end{equation}
But by total momentum conservation ($p_1+\ldots+p_n = 0$), it does not matter which direction we take.

For a general curve, $C$, on any fatgraph, this rule can be written as:
\begin{equation}\label{intro:momrule}
P^\mu_C = p^\mu_{{\rm start}} + \sum_{{\rm right \, turns}} p^\mu_{{\rm from \, left}}.
\end{equation}
This rule assigns to every curve $C$ on our fatgraph $\Gamma$ some momentum, $P_C^\mu$. Each $P_C^\mu$ is a linear combination of external momenta, $p_i$, and loop variables, $\ell_a$. Each curve, $C$, then also defines a Lorentz invariant kinematic variable
\begin{equation}
X_C = P_C^2 + m^2.
\end{equation}
The collection of variables $X_C$, for \emph{all} curves $C$ on the fatgraph, defines a complete set of kinematic variables in our kinematic space. Modulo a small detail about how to deal with internal color loops, this completes the description of our kinematic space.

It is significant in our story that we can use the momenta of a \emph{single} fat graph (or Feynman diagram) to define a complete set of kinematic variables $X_C$. As we will see in more detail in Section \ref{sec:integrands}, this basic idea ends up solving the long-standing problem of defining a good notion of loop integrand beyond the planar limit!

\subsection{The First Miracle: Discovering Feynman diagrams}
We now look for a question whose answer produces scattering amplitudes. We just saw how we can define all our kinematics using a single fatgraph. So with this starting point, what would make us consider \emph{all} possible Feynman diagrams (i.e. all spacetime processes)? And why should these be added together with equal weights (as demanded by quantum mechanics)? Amazingly, the answer to both of these fundamental questions is found right under our noses, once we think about how to systematically describe all the curves on our fatgraph.

\begin{figure}
    \centering
    \includegraphics[trim={0cm 0cm 0cm 0cm}, clip, width=.75\textwidth]{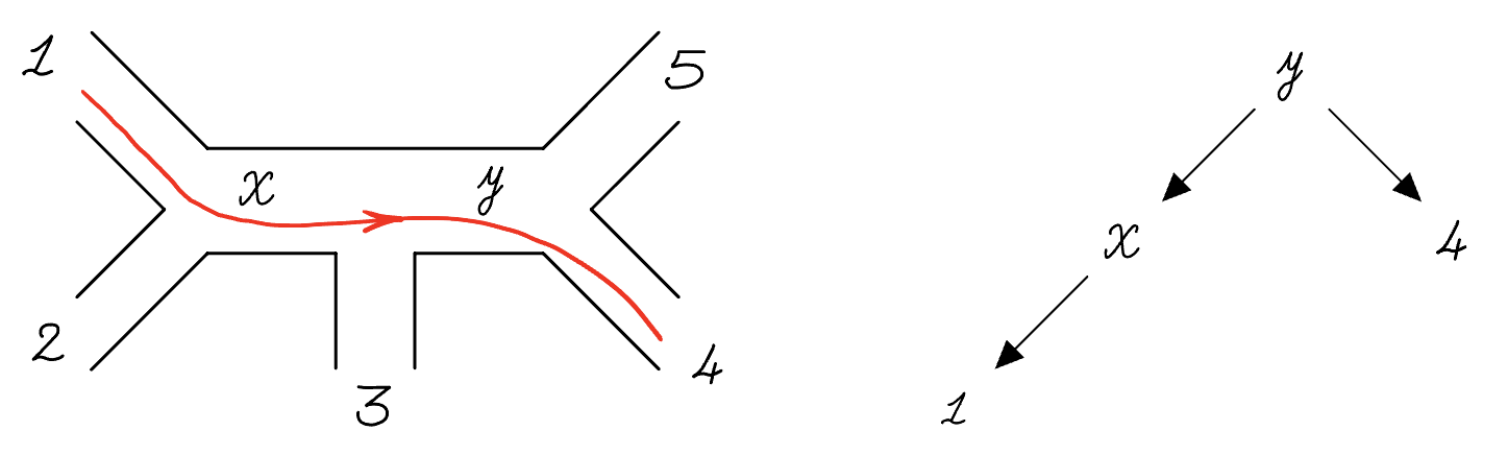}
    \caption{Describing a curve on a fatgraph (left) using a mountainscape diagram (right).}
    \label{intro:mex}
\end{figure}

How can we describe a curve on our fat graph without drawing it? We can do this by labeling all the edges, or ``roads", on the fatgraph. Any curve passes through a series of these roads. Moreover, at each vertex, we demand that the curve must turn either left or right: we do not allow our curves to do a `U turn'. It follows that a curve is fully described by the order of the roads and turns it takes as it passes through the graph. For example, the curve in Figure \ref{intro:mex} enters through edge `$1$', takes a left turn, goes down `$x$', takes a left turn, goes down `$y$', takes a right turn, and then exits via `$4$'. 

We can represent this information graphically as a \emph{mountainscape}, where left turns are represented by upward slopes, and right turns are represented by downward slopes. The mountainscape for the curve in Figure \ref{intro:mex} is shown in the Figure.
 
Once again, let our fatgraph have $E$ internal edges. To every curve $C$, we will associate a vector ${\bf g}_C$ in \emph{curve space}. As a basis for this vector space, take $E$ vectors ${\bf e}_i$, associated to each internal edge. Then ${\bf g}_C$ can be read off from the mountainscape for $C$ using the following rule:
\begin{equation}
    {\bf g}_X = \sum_{{\rm peaks \, p}} {\bf e}_{{\rm p}} - \sum_{{\rm valleys\, v}} {\bf e}_{{\rm v}}. 
\end{equation}
For example, the curve in Figure \ref{intro:mex} has a peak at `$y$' and no valleys. So the $g$-vector for this curve is
\begin{equation}
{\bf g}_C =  {\bf e}_{y}.
\end{equation}
 
 \begin{figure}
    \centering
    \includegraphics[trim={0cm 0cm 0cm 0cm}, clip, width=\textwidth]{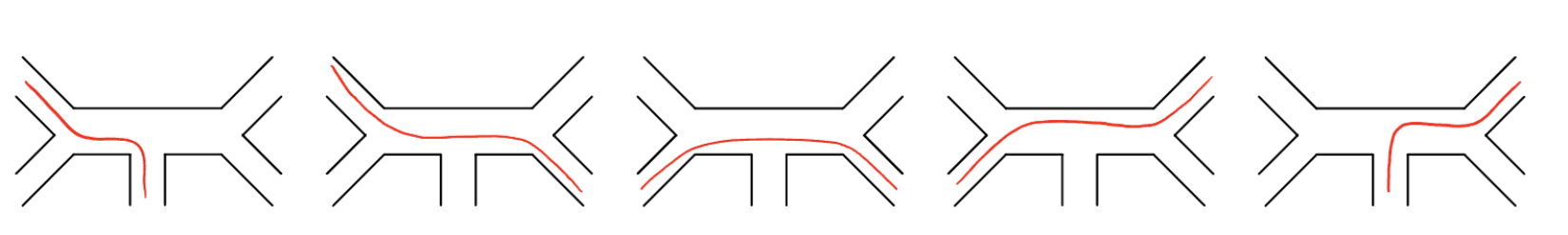}
    \caption{The five (non boundary) curves that we can draw on a 5-point tree fatgraph.}
    \label{intro:gex}
\end{figure}

\begin{figure}
    \centering
    \includegraphics[trim={0cm 0cm 0cm 0cm}, clip, width=.55\textwidth]{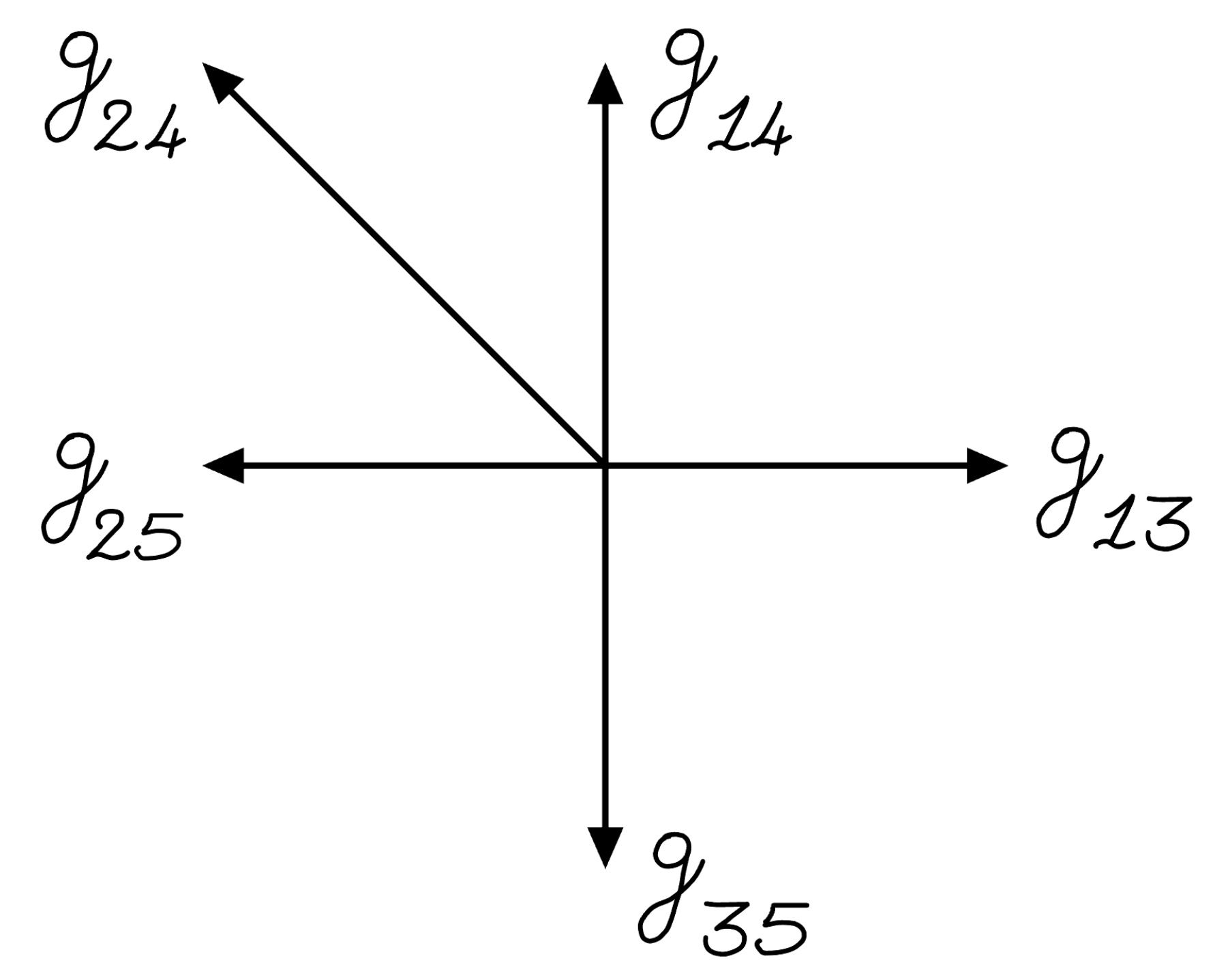}
    \caption{The collection of ${\bf g}$-vectors for the fat graph in Figure \ref{intro:mex} cuts the 2-dimensional vector space into five regions.}
    \label{intro:fanex}
\end{figure}

Now consider \emph{every} curve that we can draw on the fatgraph in Figure \ref{intro:mex}. There are 10 possible curves. 5 of these are `boundaries', and their g-vectors end up vanishing (because their mountainscapes have no peaks or valleys). The remaining 5 curves are drawn in Figure \ref{intro:gex}. If we label the external lines, each curve can be given a name $C_{ij}$ ($i,j=1,2,3,4,5$), where $C_{ij}$ is the curve connecting $i$ and $j$. Their g-vectors are 
\begin{equation} 
{\bf g}_{13} = {\bf e}_{x}, ~~ {\bf g}_{14} = {\bf e}_{y}, ~~ {\bf g}_{24} = -{\bf e}_{x} + {\bf e}_{y}, ~~ {\bf g}_{25} = - {\bf e}_{x}, ~~ {\bf g}_{35} = - {\bf e}_{y}.
\end{equation}
If we draw these five g-vectors, we get Figure \ref{intro:fanex}. This has revealed a wonderful surprise! Our g-vectors have divided curve space into five regions or \emph{cones}. These cones are spanned by the g-vectors for the following pairs of curves:
\begin{equation}
(C_{13},C_{14}), ~(C_{14},C_{24}),~(C_{24},C_{25}),~(C_{25},C_{35}),~\text{and}~(C_{35},C_{13}).
\end{equation}
These pairs of curves precisely correspond to {\it all} the five Feynman diagrams of the 5-point tree amplitude!

This is a general phenomenon. The collection of g-vectors for all the curves $C$ on a fatgraph is called \emph{the g-vector fan}\cite{2005math.....10312F,fomin2006cluster,Fomin_2018}, or \emph{the Feynman fan}, associated to that fatgraph. Each top-dimensional cone of the fan is spanned by an $E-$tuple of curves, $C_{a_1}, \cdots, C_{a_E}$, and these $E-$tuples of curves are precisely the propagators of Feynman diagrams. Moreover, the cones are non-overlapping, and together they densely cover the entire vector space! The g-vector fan is telling us that all the Feynman diagrams for the amplitude are combined in curve space.

Even better, each of the cones in the g-vector fan have the same size. It is natural to measure the size of a cone, bounded by some g-vectors ${\bf g}_1, \cdots, {\bf g}_E$, using the determinant of these vectors: $\langle {\bf g}_1 \cdots {\bf g}_E \rangle$. Remarkably, the cones of the g-vector fan all satisfy: $\langle {\bf g}_1 \cdots {\bf g}_E \rangle = \pm 1$.

To summarise, starting with a \emph{single} fatgraph at any order in perturbation theory, simply recording the data of the curves on the fatgraph, via their g-vectors, brings {\it all} the Feynman diagrams to life. Furthermore, we see why they are all naturally combined together into one object, since they collectively cover the entire curve space! This represents a very vivid and direct sense in which the most basic aspects of spacetime processes and the sum-over-histories of quantum mechanics arise as the answer to an incredibly simple combinatorial question.

\subsection{An infinity of diagrams and the spectre of Gravity}
\begin{figure}
    \centering
    \includegraphics[trim={0cm 0cm 0cm 0cm}, width=.65\textwidth]{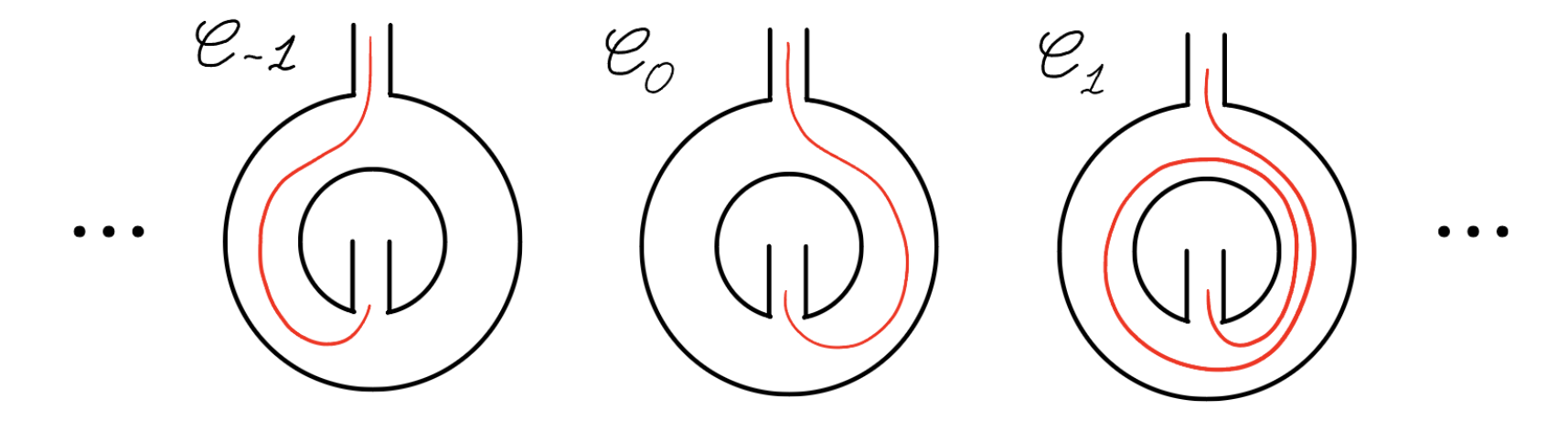}
    \caption{A double-trace 1-loop fat graph, which has infinitely many possible curves.}
    \label{intro:A11}
\end{figure}

An important novelty appears with the first non-planar amplitudes. Consider the double-trace one-loop amplitude at 2-points. A fatgraph for this amplitude is given in Figure \ref{intro:A11}. There are now infinitely many curves that we can draw on this fat graph: they differ from one another only in how many times they wind around the graph.

\begin{figure}
    \centering
    \includegraphics[trim={0cm 0cm 0cm 0cm}, clip, width=.5\textwidth]{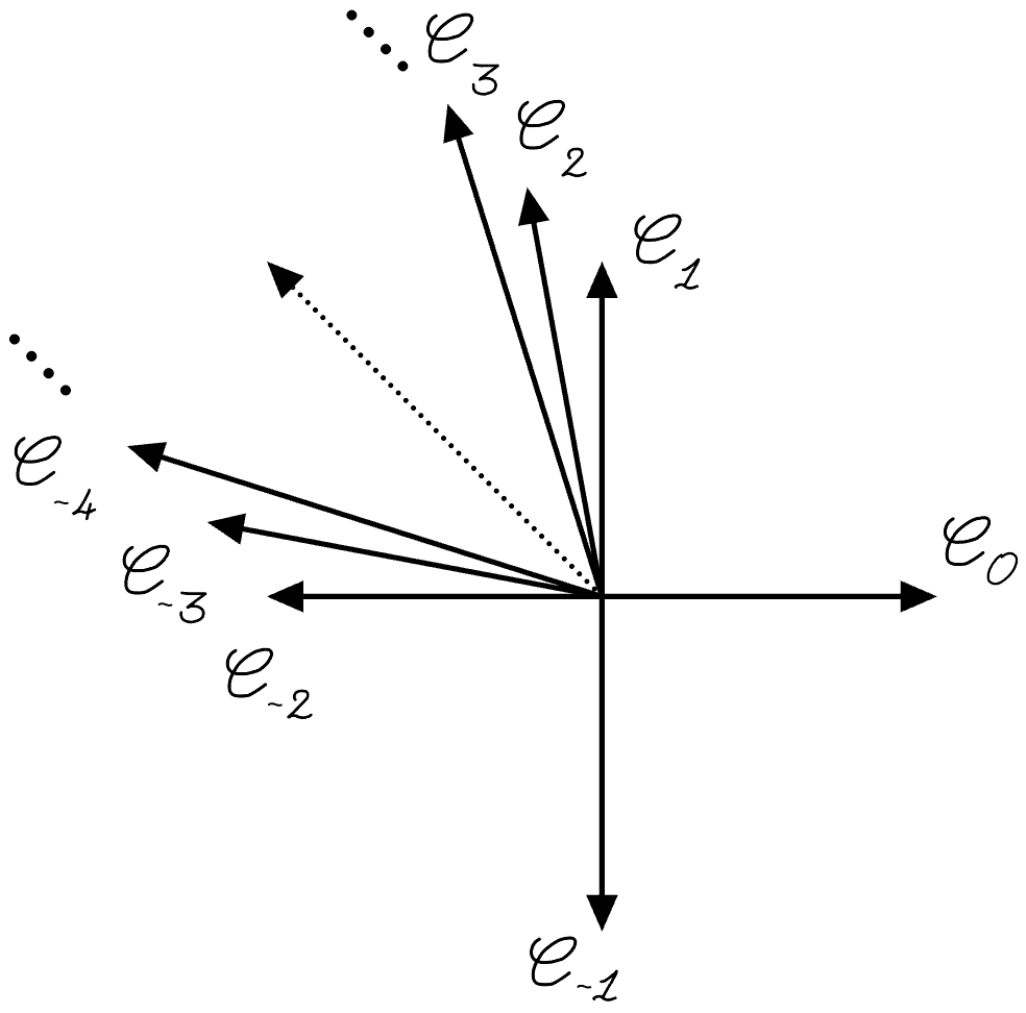}
    \caption{The g-vector fan for the 2-point double-trace 1-loop fat graph, which has infinitely many regions.}
    \label{intro:A11fan}
\end{figure}

The g-vector fan for this infinity of curves is shown in Figure \ref{intro:A11fan}. These g-vectors break curve space up into infinitely many cones. Each of these cones is bounded by a pair of g-vectors, $g_{C_m}$ and $g_{C_{m+1}}$, where $C_m$ and $C_{m+1}$ are two curves that differ by exactly one \emph{winding}. If we use our rule for the momenta of curves, \eqref{intro:momrule}, the momenta of these curves are
\begin{equation}
P_{C_m}^\mu = mk^\mu + \ell^\mu, ~\text{and}~P_{C_{m+1}}^\mu = (m+1)k^\mu + \ell^\mu.
\end{equation}
So the momenta associated to each cone are related to each other by a translation in the loop variable, $\ell^\mu \mapsto \ell^\mu + k^\mu$. It follows that \emph{every} cone in Figure \ref{intro:A11fan} corresponds to a copy of the \emph{same} Feynman diagram.

What has gone wrong? The g-vector fan is telling us to include infinitely many copies of one Feynman diagram. This is a consequence of the \emph{mapping class group} of the fat graph in Figure \ref{intro:A11}. The mapping class group of this fatgraph acts by increasing the winding of curves drawn on the fatgraph. In fact, this infinity of windings is the heart of the well-known difficulty in defining a loop integrand for non-planar amplitudes. Fortunately, as we will see, it is easy to \emph{mod out} by the action of the mapping class group, which we will do using what we call the \emph{Mirzakhani trick}\cite{mirzakhani2007}. Getting rid of these infinities using the Mirzakhani trick is the final ingredient we need in order to define amplitudes directly from the combinatorics of a single fatgraph.

As an aside, note that the infinite collection of cones in Figure \ref{intro:A11fan} does not quite cover the entire vector space! The g-vectors asymptotically approach the direction $(-1,1)$, but never reach it. This is the beginning of fascinating story: it turns out that the vector $(-1,1)$ is the g-vector for the {\it closed} curve that loops once around the fat graph. Nothing in our story asks us to consider these closed curves, but the g-vector fan forces them on us. Physically, these new closed curves are associated with the appearance of a new {\it uncoloured} particle, $\sigma$. These missing parts of the fan are then seen to have a life of their own: they tell us about a theory with uncoloured self-interactions, $\sigma^3$, that is minimally coupled to our colored particle by an interaction $\sigma$ Tr $(\Phi)$.
The appearance of $\sigma$ is a scalar avatar of how the graviton is forced on us in string theory even if we begin only with open strings. From our perspective, however, this has absolutely nothing to do with the worldsheet of string theory; it emerges directly from the combinatorics defined by a fatgraph.

    \subsection{The Amplitudes}
The g-vector fan gives a beautiful unified picture of all Feynman diagrams living in an $E$-dimensional vector space, \emph{curve space}. This result suggests a natural formula for the full amplitude in the form of an integral over curve space. To find this formula, we need one extra ingredient. For every curve, $C$, we will define a piecewise-linear \emph{headlight function}, $\alpha_C({\bf t})$. We will define the headlight function $\alpha_C$ so that it ``lights up" curve space in the direction ${\bf g}_C$, and vanishes in all other g-vector directions: 
\begin{equation}
    \alpha_C({\bf g}_D) = \delta_{C,D}
\end{equation}
This definition means that $\alpha_C$ vanishes everywhere, except in those cones that involve ${\bf g}_C$. Moreover, $\alpha_C$ is \emph{linear} inside any given cone of the Feynman fan.

Using linear algebra, we can give an explicit expression for $\alpha_C$ in any cone where it is non-vanishing. Suppose that the g-vectors of such a cone are $({\bf g}_C, {\bf g}_{D_1}, \cdots, {\bf g}_{D_{E-1}})$. The unique linear function of ${\bf t}$ which evaluates to $1$ on ${\bf g}_C$ and 0 on all the other g-vectors is
\begin{equation}
\alpha_C = \frac{\langle {\bf t}\, {\bf g}_{D_1} \cdots {\bf g}_{D_{E-1}} \rangle}{\langle {\bf g}_C {\bf g}_{D_1} \cdots {\bf g}_{D_{E-1}} \rangle}.
\end{equation}
In what follows, imagine that we already know these functions, $\alpha_C({\bf t})$.

We now define an \emph{action}, $S$, given by a sum over all curves on a fatgraph: 
\begin{equation}
S({\bf t}) = \sum_C \alpha_C({\bf t}) X_C,\qquad \text{with}~ X_C = P_C^2 + m^2.
\end{equation}
Recall that $P_C^\mu$ is the momentum we associate to a curve $C$. If we restrict $S({\bf t})$ to a single cone, bounded by some g-vectors, ${\bf g}_{C_1},\ldots,{\bf g}_{C_E}$, then the only $\alpha$'s that are non-zero in this cone are precisely $\alpha_{C_1}, \ldots, \alpha_{C_E}$. Moreover, $S({\bf t})$ is linear in this cone. It is natural to parametrise the region inside this cone by ${\bf t} = \rho_1 {\bf g}_{C_1} + \cdots \rho_{E} {\bf g}_{C_E}$, with $\rho_i \geq 0$ positive. Then we can integrate $\exp(-S)$ in this cone. The result is identical to the result of a standard Schwinger parametrisation for a single Feynman diagram:
\begin{equation}
\int\limits_{\text{cone}} d^E t \, e^{-S} = \int\limits_0^\infty d^E\rho \, |\langle g_{C_1} \cdots g_{C_E} \rangle| \prod_{i=1}^E e^{-\rho_i X_{C_i}} = \prod_{i=1}^E \frac{1}{P_{C_i}^2 + m^2}.
\end{equation}
The factor $|\langle g_{X_1} \cdots g_{X_E}\rangle|$ is the Jacobian of the change of variables from $(t_1,\cdots,t_E)$ to $(\rho_1, \cdots, \rho_E)$. As we have remarked, the cones are \emph{unimodular} and these Jacobian factors are all equal to 1!

In order to get the full amplitude, all we have to do now is integrate $\exp(S)$ over the whole vector space, instead of restricting it to just a single cone. However, to account for the infinity resulting from the \emph{mapping class group}, we also need to factor out this $\MCG$ action in our integral, which we denote by writing the measure as
\begin{equation}
\frac{d^E t}{{\rm MCG}}.
\end{equation}
Before doing the loop integrations, the full amplitude is then given by a \emph{curve integral}:
\begin{equation}
    {\cal A} = \int d^D \ell_1 \cdots d^D \ell_L \int \frac{d^E t}{{\rm MCG}} \, {\rm exp}\left(-\sum_X \alpha_X({\bf t}) (p_X^2 + m^2) \right).
\end{equation}
The dependence on loop momenta in this formula is Gaussian. When we integrate the loop momenta, we find the final amplitude is given by a curve integral
\begin{equation}\label{intro:globsym}
    {\cal A} = \int \frac{d^E t}{{\rm MCG}} \, \left( \frac{\pi^L}{{\cal U}(\alpha)}\right)^{\frac{D}{2}} {\rm exp}\left(\frac{{\cal F}(\alpha)}{{\cal U}(\alpha)}\right).
\end{equation}
${\cal U}(\alpha)$ and ${\cal F}(\alpha)$ are homogeneous polynomials in the headlight functions. They are analogous to Symanzik polynomials, but are not associated with any particular Feynman diagram. We give simple formulas for ${\cal A}$ and ${\cal F}$ in Section \ref{sec:amplitudes}. 

The key to using these curve integral formulas lies in how we mod out by the MCG. One way of doing this would be to find a \emph{fundamental domain} in ${\bf t}$-space that would single out one copy of each Feynman diagram. However, in practice this is no easier than enumerating Feynman diagrams. Instead, we will use an elegant way of modding out that we call \emph{the Mirzakhani trick}, which is analogous to the Fadeev-Popov trick familiar from field theory. As we will see, any MCG invariant function, $f$, can be integrated as,
\begin{equation}
    \int \frac{d^E t}{{\rm MCG}} f = \int d^E t \, {\cal K}(\alpha) f,
\end{equation}
where the \emph{Mirzakhani kernel} ${\cal K}(\alpha)$ is a simple rational function of the $\alpha_C$'s.\footnote{The restriction on the integration region $\sum_i t_i \geq 0$ in equation \eqref{intro:1loop2} for 1-loop amplitudes can be thought of as the smallest example of a Mirzakhani kernel. In this formula, we are modding out by a discrete $Z_2$ symmetry, described more in Section \ref{sec:delta}.} We will describe several formulas for these kernels. In all cases, ${\cal K}$ has support on a finite region of the fan, so that only a small number of the $\alpha_C$'s is ever needed to compute the amplitude. We will also show how some of our methods for producing ${\cal K}$ give new systematic recursive methods for computing amplitudes.

\subsection{The Second Miracle: The Counting Problem}
We have given a formula, \eqref{intro:globsym}, for partial amplitudes at any order in the `t Hooft expansion of our theory. However, the meat of this formula is in the headlight functions, $\alpha_C$. The problem is that headlight functions are, naively, hard to compute!

The issue can already been seen at tree level. For $n$-points at tree level, the number of possible curves, $C$, is $\sim n^2$, whereas the number of Feynman diagrams (or cones) grows exponentially as $\sim 4^n$. Each $\alpha_C$ restricts to a different linear function on each of the $\sim 4^n$ cones. So we would expect that it takes an exponentially-growing amount to work to compute all of the $\alpha_C$,---about as much work as it would take us to just enumerate all the Feynman diagrams to begin with! So, is there an easier way to compute $\alpha_C$? 

This is where a second miracle occurs. It turns out that headlight functions can be computed efficiently by matrix multiplication. In fact, the calculation is completely \emph{local to the curve}, in the sense that we only need to know the path taken by $C$, and nothing else about the fatgraph it lives in. There are always many fewer curves than there are Feynman diagrams. This means that the amount of work to compute the $\alpha_C$'s should grow slower than the amount of work it takes to enumerate all Feynman diagrams.

This way of computing $\alpha_C$ is based on a simple combinatorial problem. For a curve, $C$, draw its \emph{mountainscape}. We are going to record all the ways in which we can pick a subset of letters of $C$, subject to a special rule: if we pick a letter $y$, we also have to pick any letters \emph{downhill} of $y$. We will then define an \emph{F polynomial} for the curve, $F(C)$, which records the valid subsets.

\begin{figure}
\centering
\begin{subfigure}{.3\textwidth}
  \centering
  \includegraphics[width=.7\linewidth]{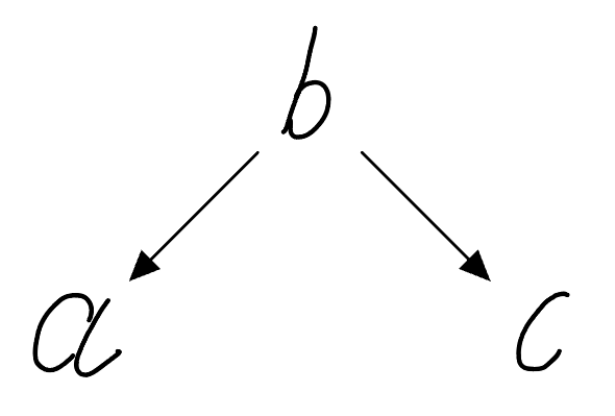}
  \caption{~}
\end{subfigure}%
\begin{subfigure}{.3\textwidth}
  \centering
  \includegraphics[width=.7\linewidth]{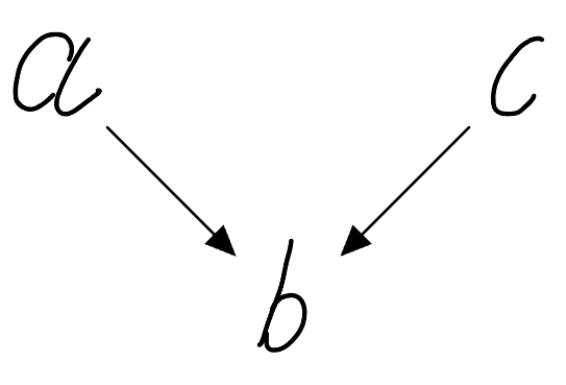}
  \caption{~}
\end{subfigure}
\begin{subfigure}{.3\textwidth}
  \centering
  \includegraphics[width=.7\linewidth]{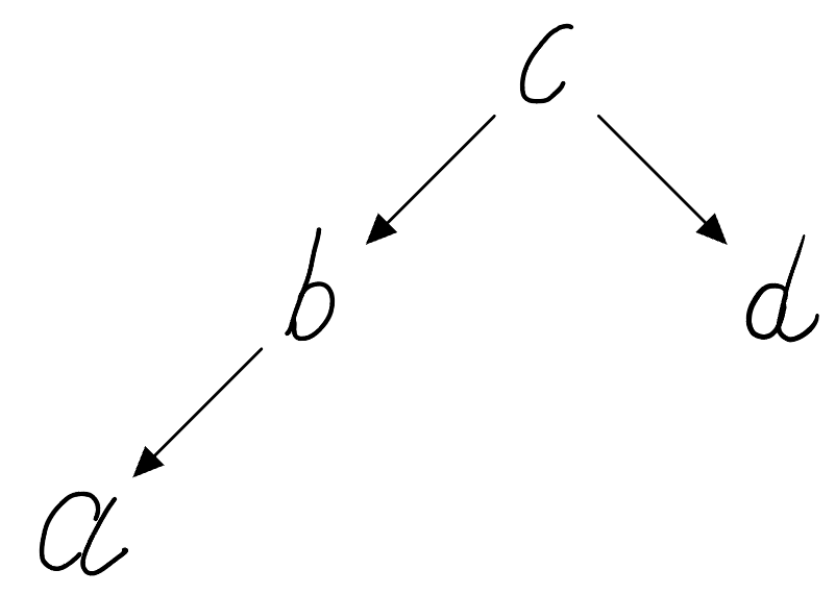}
  \caption{~}
\end{subfigure}
  \caption{Three mountainscapes.}
\label{intro:Fex}
\end{figure}

For example, for the mountainscape in Figure \ref{intro:Fex}(a), we get
\begin{equation}
F = 1 + a + c + a c + a b c.
\end{equation}
This is because we can choose the following subsets: no-one (``1"); just $a$; just $c$; $a$ and $c$ together; or finally we can pick $b$, but if we do, we must also pick $a$ and $c$, which are both downhill of $b$. In Figure \ref{intro:Fex}(b), we get
\begin{equation}
F = 1 + b + a b + b c + a b c,
\end{equation}
because in this example we can choose: no-one; just $b$; we can pick $a$, but if we do we must also pick $b$; we can pick $c$, but we must then also pick $b$; and finally we can can pick both $a$ and $c$, but then we must also pick $b$. Finally, we leave Figure \ref{intro:Fex}(c) as an exercise. The result is
\begin{equation}
F = 1 + a + d + a d + ab + ab d + abcd.
\end{equation}

In general, there is a fast method for computing $F(C)$ by reading the mountainscape for $C$ from left to right. Say the leftmost letter is $Y$, and call the next letter $y$. Then write $F(C) = F_{{\rm no}} + F_{{\rm yes}}$, where we group the terms in $F(C)$ according to whether they include $Y$ ($F_\text{yes}$) or not ($F_\text{no}$). Similarly write $f_{{\rm no}}, f_{{\rm yes}}$ for what we would get starting instead from $y$. Suppose that in our mountainscape we move ``up" from $Y$ to $y$. Then if we do not pick $Y$, then we cannot pick $y$ either, since if we choose $y$ we must choose $Y$. On the other hand if we do choose $Y$, we can either pick or not pick $y$. Thus, in this case, we have
\begin{equation}
F_{{\rm no}} = f_{{\rm no}},\qquad F_{{\rm yes}} = Y (f_{{\rm no}} + f_{{\rm yes}}).
\end{equation}
Similarly if, in our mountainscape, we move down from $Y$ to $y$, we find that
\begin{equation}
F_{{\rm no}} = f_{{\rm no}} + f_{{\rm yes}},\qquad F_{{\rm yes}} = Y f_{{\rm yes}}.
\end{equation}
In matrix form, we find that
\begin{equation}\label{intro:FFff}
\left(\begin{array}{c} F_{{\rm no}} \\ F_{{\rm yes}} \end{array}\right) = M_{L,R}(Y) \left(\begin{array}{c} f_{{\rm no}} \\ f_{{\rm yes}} \end{array}\right),
\end{equation}
 where $M_{L}$ and $M_R$ are the matrices
\begin{equation}\label{intro:MLMR}
    M_L(Y) = \left(\begin{array}{cc} 1 & 0 \\ Y & Y\end{array}\right), \qquad M_R(Y)=\left(\begin{array}{cc} 1 & 1 \\ 0 & Y \end{array} \right).
\end{equation}

Now suppose that the curve $C$ is given explicitly by the following series of edges and turns:
\begin{equation}
(y_1,{\rm turn}_1,y_2,{\rm turn}_2, \cdots, y_{m-1}, {\rm turn}_{m-1}, y_m),
\end{equation}
where ${\rm turn}_i$ is either a left or right turn, immediately following $y_i$. Given \eqref{intro:FFff}, we find
\begin{equation}
\left(\begin{array}{c} F_{{\rm no}} \\ F_{{\rm yes}} \end{array}\right) = M \left(\begin{array}{c} 1 \\ y_{{m}}  \end{array} \right),
\end{equation}
where
\begin{equation}
M(C) = M_{{\rm turn}_1}(y_1) M_{{\rm turn}_2}(y_2) \cdots M_{{\rm turn}_{m-1}}(y_{m-1}).
\end{equation}
So our counting problem is easily solved simply by multiplying a series of $2 \times 2$ matrices (equation \ref{intro:MLMR}) associated with the left and right turns taken by the curve $C$.

Suppose that the initial edge of $C$, $y_1$, and the final edge, $y_m$, are external lines of the fatgraph. It is natural to write $F(C)$ as a sum over four terms:
\begin{equation}
F(C) = F_{{\rm no}, \, {\rm no}} + F_{{\rm no}, \, {\rm yes}} + F_{{\rm yes}, \, {\rm no}} + F_{{\rm yes}, \, {\rm yes}},
\end{equation}
where we group terms in $F(C)$ according to whether they do or do not include the first and last edges: $y_1$ and/or $y_m$. Indeed, these terms are also the entries of the matrix $M(C)$,
\begin{equation}
M(C)=\left(\begin{array}{cc} F_{{\rm no},{\rm no}} & F_{{\rm no},{\rm yes}} \\ F_{{\rm yes},{\rm no}} & F_{{\rm yes},{\rm yes}} \end{array} \right),
\end{equation}
if we now set $y_m=1$. In fact, we will also set $y=1$ for every external line of the fatgraph, and will reserve $y$-variables for internal edges of the fatgraph.

Notice that $\det M_L(y) = \det M_R(y) = y$, so that
\begin{equation}
\det M(C) = \prod_{i=2}^{m-1} y_i.
\end{equation}
In other words, we have the identity
\begin{equation}\label{intro:FFFFy}
F_{{\rm no}, {\rm no}} F_{{\rm yes}, {\rm yes}} = F_{{\rm no}, {\rm yes}} F_{{\rm yes}, {\rm no}} + \prod_i y_i.
\end{equation}
Motivated in part by this identity, we will define $u$-variables for every curve,
\begin{equation}
u_C = \frac{F(C)_{{\rm no},{\rm yes}} \,F(C)_{{\rm yes},{\rm no}}}{F(C)_{{\rm no}, {\rm no}} \,F(C)_{{\rm yes},{\rm yes}}} = 
\frac{M(C)_{12} M(C)_{21}}{M(C)_{11} M(C)_{22}}.
\end{equation}
These $u_C$ variables are most interesting to us in the region $y_i \geq 0$. Equation \eqref{intro:FFFFy} implies that $0 \leq u_C \leq 1$ in this region. They vastly generalise the $u$-variables defined and studied in \cite{brown2009,Arkani_Hamed_2021}.

We now define the headlight functions. We define them to capture the asymptotic behaviour  of the $u$-variables when thought of as functions of the ${\bf y}$ variables. We define
\begin{equation}
\alpha_C = - {\rm Trop\ } u_C.
\end{equation}
where ${\rm Trop\ }u_C$ is the so-called \emph{tropicalization} of $u_C$.

The idea of tropicalization is to look at how functions behave asymptotically in ${\bf y}$-space. To see how this works, parameterise the $y_i\geq 0$ region by writing $y_i = \exp t_i$, where the $t_i$ are real variables. Then, as the $t_i$ become large, ${\rm Trop\ }u_C$ is defined such that
\begin{equation}
u_C({\bf t}) \rightarrow \exp \left( {\rm Trop\ } u_C \right).
\end{equation}
For example, consider a simple polynomial, $P(y_1,y_2) = 1 + y_2 + y_1 y_2 = 1 + e^{t_2} + e^{t_1 + t_2}$. As we go to infinity in ${\bf t} = (t_1, t_2)$ in different directions, different monomials in $P$ will dominate. In fact, we can write, as we go to infinity in ${\bf t}$,
\begin{equation}
P \rightarrow \exp \max(0,t_2,t_1+t_2),
\end{equation}
and so ${\rm Trop\ }(P) = \max(0,t_2,t_1+t_2)$. If we have a product of polynomials, $F = \prod_a P_a^{c_a}$, then as we go to infinity in ${\bf t}$ we have $F \to e^{{\rm Trop (F)}}$, where ${\rm Trop\ }F = \sum c_a{\rm Trop\ }(P_a)$.

Returning to headlight functions, our definition can also be written as
\begin{equation} 
\alpha_C = {\rm Trop\ }(M(C)_{11}) + {\rm Trop\ }(M(C)_{22}) - {\rm Trop\ }(M(C)_{12}) - {\rm Trop\ }(M(C)_{21}).
\end{equation}

For example, consider again the $n=5$ tree amplitude. Take the curve $C$ from Figure \ref{intro:mex} (left). This curve has path $(1, L, x, R, y, R,4)$. So it has a matrix (with $y_{23},y_{15}\equiv 1$)
\begin{equation} 
M(C) = M_L(1) M_R(x) M_R(y)=\left(\begin{array}{cc} 1 & 1+y \\ 1 & 1+y+xy \end{array} \right).
\end{equation}
Using this matrix, we find that its $u$-variable is
\begin{equation}
u_C = \frac{1 + y}{1+y+xy},
\end{equation}
and so its headlight function is
\begin{equation}
\alpha_C = {\rm max}(0,t_{y},t_{x} + t_{y}) - {\rm max}(0,t_{y}).
\end{equation}
Amazingly, this function satisfies the key property of the headlight functions: $\alpha_C$ vanishes on every g-vector, except for its own g-vector, ${\bf g}_C = (1,0)$.

\subsection{Back to the Amplitude!}
We have now formulated how to compute all-order amplitudes in Tr$\Phi^3$ theory as a counting problem. The final expression for the integrated amplitude at any order of the topological expansion associated with a surface ${\cal S}$ is given as 
\begin{eqnarray}
{\cal A} = \int d^E t \, {\cal K}(\alpha) \left( \frac{\pi^L}{{\cal U}(\alpha)}\right)^{\frac{D}{2}} {\rm exp}\left(\frac{{\cal F}(\alpha)}{{\cal U}(\alpha)}\right),
\end{eqnarray}
where ${\cal F}(\alpha), {\cal U}(\alpha)$ are homogeneous polynomials in the $\alpha_C$'s, ${\cal K}(\alpha)$ is the Mirzakhani kernel that mods out by the mapping-class-group, and crucially, each $\alpha_C$ is determined entirely by the path of its curve, using a simple counting problem on the curve. The presence of ${\cal K}$ ensures that only a finite number of $\alpha_C$'s ever appear in our computations, which makes the formula easy to apply. There is no trace of the idea of `summing over all spacetime processes' in this formula. Instead, small combinatorial problems attached to the curves on a fatgraph, treated completely independently of each other, magically combine to produce local and unitary physics, pulled out of the platonic thin air of combinatorial geometry.

Our goal in the rest of this paper is to describe these ideas systematically. Our focus in here will exclusively be on simply presenting the formulas for the amplitudes.  
This presentation will be fully self-contained, so that the interested reader will be fully equipped to find the curve integrals for the $\text{Tr}\phi^3$ theory at any order in the topological expansion. The methods can be applied at any order in the topological expansion, but there are a number of novelties that need to be digested. We illustrate these subtleties one at a time, as we progress from tree level examples through to one and two loops, after which no new phenomena occur. We begin at tree-level to illustrate the basic ideas. At one-loop single-trace, we show how to deal with \emph{spiralling} curves. Then, as we have seen above, double-trace amplitudes at 1-loop expose the first example of the infinities associated with the mapping class group. Finally, we study the leading $1/N$ correction to single-trace at 2-loops---the genus one amplitude---to show how to deal with a non-abelian mapping class group. This non-abelian example illustrates the generality and usefulness of the Mirzakhani trick. 

In all cases discussed in this paper we will use use the smallest example amplitudes possible to illustrate the new conceptual points as they arise. The next paper in this series will give a similarly detailed set of formulae for amplitudes for any number of particles, $n$. In this sense this first pair of papers can be thought of as a ``user guide" for the formalism. A systematic accounting of the conceptual framework underlying these formulae, together with an exposition of the panoply of related developments, will be given in the later papers of this series.

\section{The partial amplitude expansion}\label{sec:partial}
Consider a single massive scalar field with two indices in the fundamental and anti-fundamental representations of $\mathrm{SU}(N)$, $\phi = \phi^I_J \, t_I t^J$, and with a single cubic interaction,
\begin{align}\label{eq:feyn}
  \mathcal{L}_{int} = g \mathrm{Tr}\left[\phi^3\right] = g\,\phi_I^J\phi_J^K\phi_K^I.
\end{align}
The trace of the identity is $\mathrm{Tr}(1) = \delta_I^I = N$. The propagator for the field $\phi$ can be drawn as a double line and the Feynman diagrams are \emph{fatgraphs} with cubic vertices. The Feynman rules follow from \eqref{eq:feyn}. To compute the $n$ point amplitude, $\mathcal{A}_n$, fix $n$ external particles with momenta $k_i^\mu$ and colour polarisations $t_i^{IJ}$. A fatgraph $\Gamma$ with $V$ cubic vertices contributes to the amplitude as
\begin{align}
(ig)^V \,C_\Gamma \,\mathrm{Val}(\Gamma),
\end{align}
where $C_\Gamma$ is the tensorial contraction of the polarisations $t_i^{IJ}$ according to $\Gamma$. The kinematical part is given by an integral of the form
\begin{align}
\mathrm{Val}(\Gamma) = \int \prod_{i=1}^L d^D \ell_{i} \prod_{\mathrm{edges}~e} \frac{1}{P_{e}^2 + m^2},
\end{align}
for some assignment of loop momenta to the graph. Each momentum $P_e^\mu$ is linear in the external momenta $k_i^\mu$ and in the loop momentum variables $\ell_i^\mu$. To find $P_e^\mu$, the edges of $\Gamma$ need to be oriented, so that momentum conservation can be imposed at each cubic vertex.

\begin{figure}
\begin{center}
\includegraphics[width=0.75\textwidth]{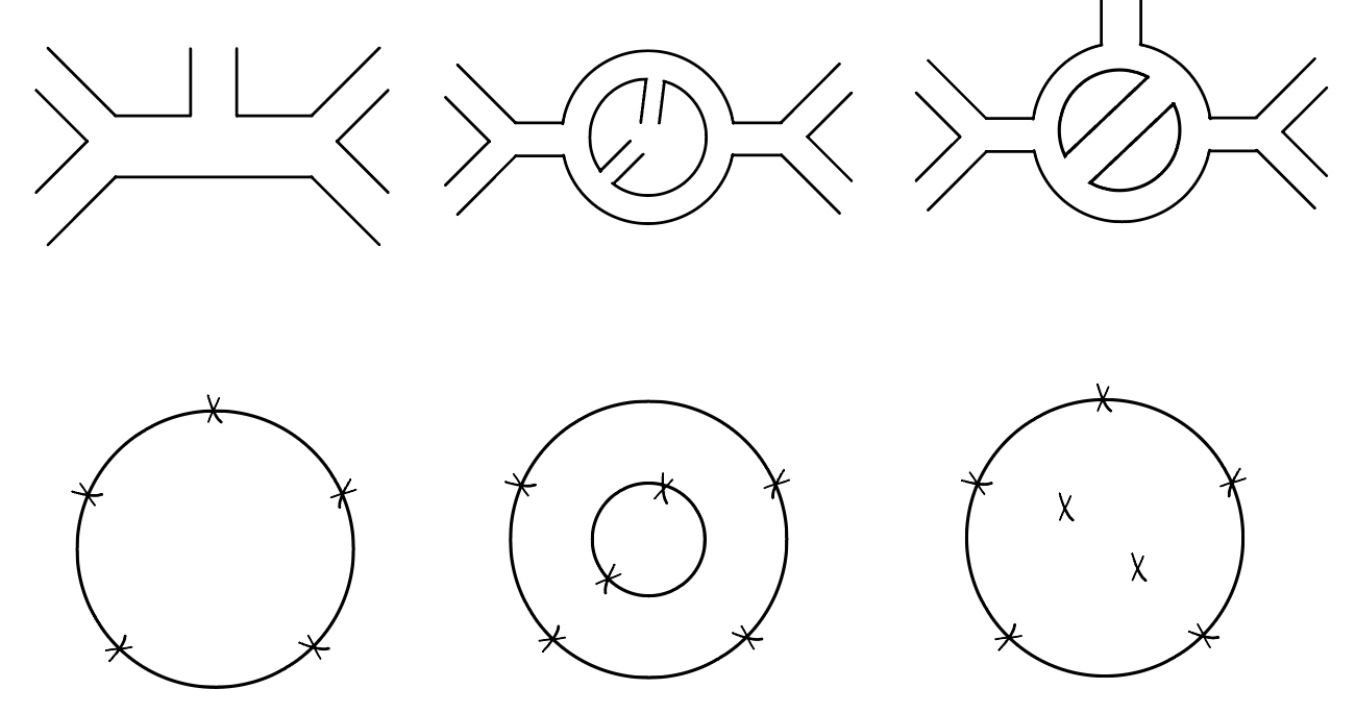}
\caption{Feynman graphs $\Gamma$ and the surfaces $S(\Gamma)$ that label their colour factors.}
\label{fig:surfex}
\end{center}
\end{figure}

The colour factors $C_\Gamma$ organise the amplitude $\mathcal{A}_n$ into partial amplitudes. This is because $C_\Gamma$ depends only on the topology of $\Gamma$ regarded as a surface, and forgets about the graph. Write $S(\Gamma)$ for the surface obtained from the fatgraph $\Gamma$ by `forgetting' the graph. Two fatgraphs $\Gamma_1, \Gamma_2$ share the same colour factor, $C_\Sigma$, if they correspond to the same marked surface, $\Sigma = S(\Gamma_1)=S(\Gamma_2)$. The amplitude can therefore be expressed as
\begin{align}\label{eq:ampn}
\mathcal{A}_n = \sum_{L=0}^\infty (ig)^{n-2+2L} \sum_{\substack{\Sigma~\mathrm{s.t.}\\h+2g=L+1}} C_\Sigma {\cal A}_\Sigma,
\end{align}
where we sum over marked bordered surfaces $\Sigma$ having $n$ marked points on the boundary. At loop order $L$, this second sum is over all surfaces $\Sigma$ with $h$ boundary components and genus $g$, subject to the Euler characteristic constraint: $h+2g=L+1$. The partial amplitudes appearing in \eqref{eq:ampn} are
\begin{align}\label{eq:para}
{\cal A}_\Sigma = \sum_{\substack{\Gamma \\ S(\Gamma)=\Sigma}} \mathrm{Val}(\Gamma).
\end{align}
Examples of some ribbon graphs $\Gamma$ and their corresponding surfaces are shown in Figure \ref{fig:surfex}.

Our aim is to evaluate ${\cal A}_\Sigma$. It is conventional to compute $\mathrm{Val}(\Gamma)$ using \emph{Schwinger parameters}. Schwinger parameters are introduced via the identity
\begin{align}
\frac{1}{P^2+m^2} = \int_0^\infty d \alpha \, e^{-\alpha (P^2+m^2)}.
\end{align}
The integration in $\ell_i^\mu$ loop variables then becomes a Gaussian integral, and the result can be written as
\begin{align}\label{eq:syman}
\mathrm{Val}(\Gamma) = \int\limits_{\alpha_i \geq 0} d^E\alpha \, \left(\frac{2\pi}{{\cal U}_\Gamma}\right)^{\frac{D}{2}} \exp \left( \frac{\mathcal{F}_\Gamma}{\mathcal{U}_\Gamma} - m^2 \sum_{i}\alpha_i \right),
\end{align}
where $\mathcal{U}_\Gamma$ and $\mathcal{F}_\Gamma$ are the Symanzik polynomials of $\Gamma$. The Symanzik polynomials depend on $\Gamma$ regarded as a graph (i.e. forgetting that it is a surface). The first Symanzik polynomial is given by
\begin{align}
\mathcal{U}_\Gamma = \sum_{T}\prod_{e\not\in T} \alpha_e,
\end{align}
where the sum is over all spanning trees, $T$, of $\Gamma$. The second Symanzik polynomial is given by a sum over all spanning 2-forests, $(T_1,T_2)$, which cut $\Gamma$ into two tree graphs:
\begin{align}
\mathcal{F}_\Gamma = - \sum_{(T_1,T_2)} \left(\prod_{e\not\in T_1 \cup T_2} \alpha_e \right) \left( \sum_{e\not\in T_1 \cup T_2} P_e \right)^2,
\end{align}
where $P_e^\mu$ is the momentum of the edge $e$. It can be shown that $\mathcal{F}_\Gamma$ depends only on the external momenta, and not on the loop momentum variables.

The partial amplitudes ${\cal A}_\Sigma$ are given by sums over integrals of this form, as in \eqref{eq:para}. But it is the purpose of this paper to show how ${\cal A}_\Sigma$ can be written more compactly as a \emph{single} Symanzik-like integral. It does not work to naively sum the integrands of $\mathrm{Val}(\Gamma)$ for different Feynman diagrams $\Gamma$. One problem is that there is no conventional way to relate the loop momentum variables for different Feynman graphs. We will see how this is solved by basic facts from surface geometry. Moreover, a simple counting problem associated to surfaces will allow us to define tropical functions we call \emph{headlight functions}. These simple functions allow us to evaluate the full partial amplitude without enumerating the Feynman diagrams.

\section{Momenta and curves}\label{sec:momcur}
Curves on fatgraphs are the key building block for our formulation of amplitudes. In this section we show how a fatgraph can be used to assign momenta to its curves. This momentum assignment solves the problem of finding a consistent choice of momentum variables for all Feynman diagrams contributing to an amplitude. This generalizes the \emph{dual momentum variables} that can be used for planar amplitudes.

\subsection{Mountainscapes}
A \emph{curve} is a path on the fatgraph that enters from an external line, passes through the fatgraph without self-intersections, and exits on an external line. It is sometimes useful to separately consider \emph{closed curves}, which are paths on the fatgraph that form a closed loop.

Curves are important because they define \emph{triangulations} of fatgraphs. A triangulation is a maximal collection of pairwise non-intersecting curves. The key point is that each triangulation of $\Gamma$ corresponds, by graph duality, to some fatgraph $\Gamma'$. These fatgraphs $\Gamma'$ all have the same colour factor and so contribute, as Feynman diagrams, to the same amplitude.\footnote{There is also a duality between triangulations of a fatgraph $\Gamma$, and triangulations of the surface $S(\Gamma)$. Defining this requires some care and is not needed for the results here.} The methods in this paper can be used to automatically find all the triangulations of $\Gamma$ without having to list them, using only the data of the curves on $\Gamma$.

\begin{figure}
\begin{center}
  \includegraphics[width=.4\linewidth]{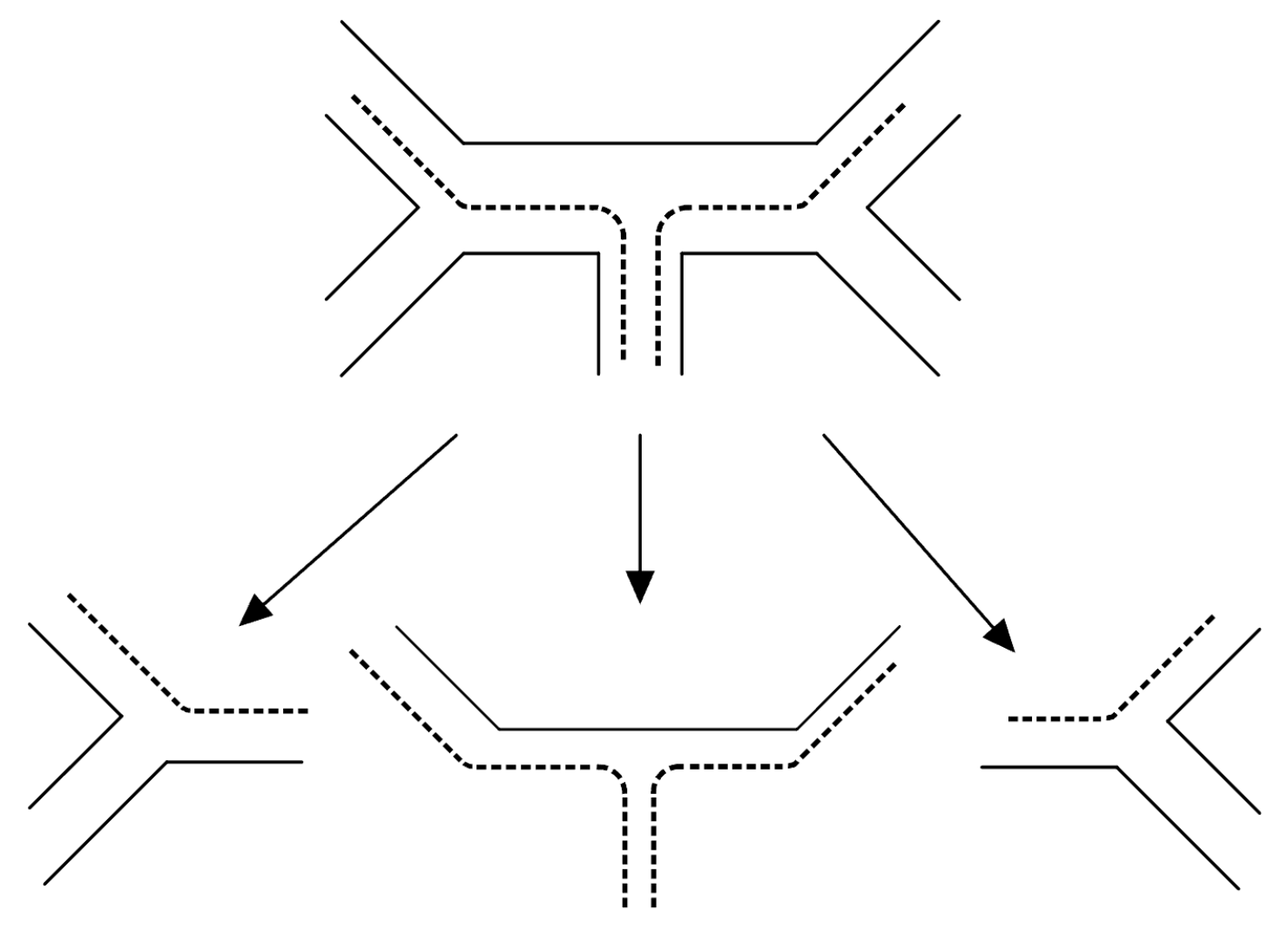}
\caption{A triangulation of a fatgraph is a maximal set of curves that cuts the fatgraph into cubic vertices.}
\label{fig:adual}
\end{center}
\end{figure}

A curve $C$ on $\Gamma$ is completely specified by reading off the order in which $C$ passes through the edges of $\Gamma$. It is also helpful to record the \emph{left and right turns} made by the curve. We present this information using \emph{mountainscape diagrams}. The vertices of a mountainscape are labelled by the edges of $\Gamma$. Each left turn made by $C$ is recorded with a left arrow (and a step up), while each right turn is written with a right arrow (and a step down):
\begin{center}
\begin{tabular}{c c c}
 \begin{tikzpicture} \draw[<-] (-1,-1) node[below left]{$i$} -- (-0.5,-0.5) node[above right]{$j$}; \end{tikzpicture} \qquad &\qquad&\qquad  \begin{tikzpicture} \draw[->] (-1,-1) node[above left]{$i$} -- (-0.5,-1.5) node[below right]{$k$}; \end{tikzpicture}\\
\emph{Turn left from $i$ to $j$.} &   & \emph{Turn right from $i$ to $k$.}
\end{tabular}
\end{center}

\begin{figure}
\begin{center}
\includegraphics[width=0.75\textwidth]{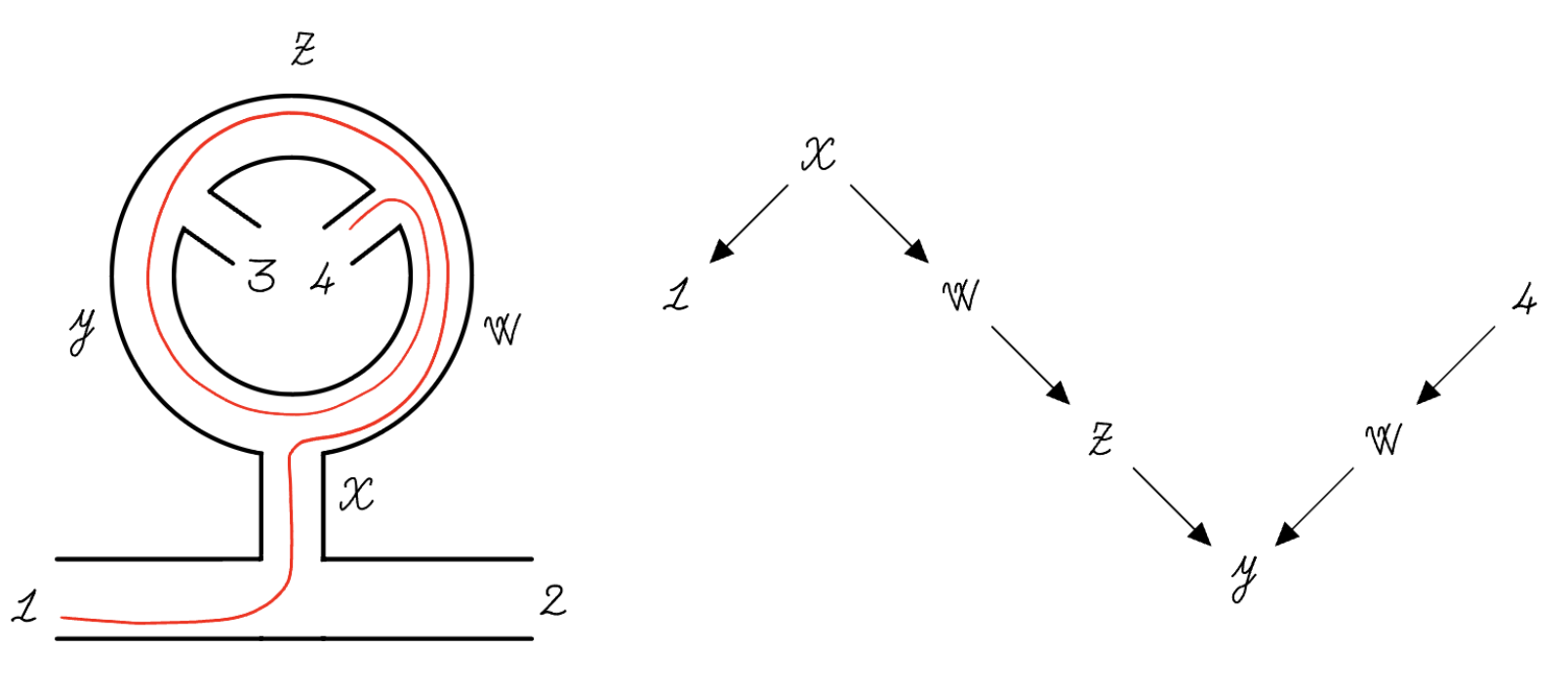}
\caption{A curve on a fatgraph (left) and its mountainscape diagram (right).}
\label{fig:apathex}
\end{center}
\end{figure}

For example, the curve in Figure \ref{fig:apathex}(a) passes through the edges $1,x,w,z,y,w,4$. Its mountainscape is shown in Figure \ref{fig:apathex}(b). If we traverse $C$ in the opposite direction we obtain the left-right reflection of this mountainscape. We regard these as being the \emph{same} mountainscape. For brevity, it is convenient to write mountainscapes as a \emph{word}, writing `L' for a left turn, and `R' for a right turn. The mountainscape in Figure \ref{fig:apathex}(b) is given by the word
\begin{equation}
    C = 1 L x R w R z R y L w L 4.
\end{equation}

\subsection{Intersections}\label{sec:intersections}
Mountainscape diagrams encode the intersections of curves. In fact, it is not necessary to know the whole fatgraph in order to determine if two curves intersect: their mountainscapes alone have all the data needed.

\begin{figure}
\begin{center}
\includegraphics[width=0.75\textwidth]{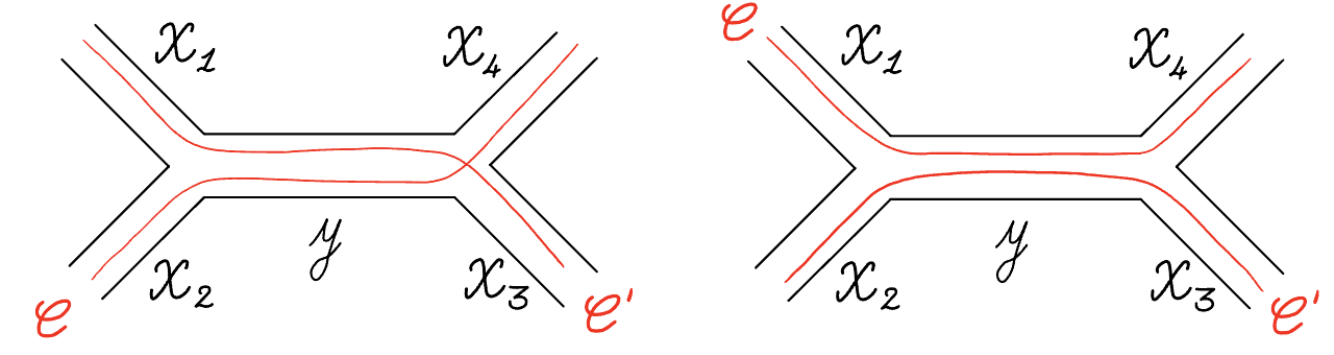}
\caption{A pair of intersecting curves (left), and a pair of non-intersecting curves (right).}
\label{fig:int}
\end{center}
\end{figure}

For example, consider Figure \ref{fig:int}. The two curves in Figure \ref{fig:int}(a) are
\begin{align}
C = x_2 R y L x_4 \qquad \text{and}\qquad C'= x_1 L y R x_3.
\end{align}
These two mountainscapes \emph{overlap} on the edge $y$, which they share in common. For $C$, $y$ is a \emph{peak}, whereas for $C'$, $y$ is a \emph{valley}. This is equivalent to the information that $C$ and $C'$ \emph{intersect} at $y$. By contrast, the two curves in Figure \ref{fig:int}(b) are
\begin{align}
C= x_1 L y L x_4 \qquad \text{and}\qquad C' = x_2 R y R x_3.
\end{align}
These curves also overlap on the edge $y$. But $y$ does not appear in these curves as a peak or valley. This is equivalent to the information that $C$ and $C'$ do not intersect.

In general, if two curves $C$ and $C'$ intersect, their paths must overlap near the intersection. So suppose that $C$ and $C'$ share some sub-path, $W$, in common. Then $C$ and $C'$ \emph{intersect along $W$} only if $W$ is a peak for one and a valley for the other. In other words, $C$ and $C'$ intersect at $W$ if they have the form
\begin{align}\label{eqn:int}
C = W_1 R W L W_2\qquad\text{and}\qquad
C' = W_3 L W R W_4,
\end{align}
or
\begin{align}
C = W_1 L W R W_2 \qquad\text{and}\qquad
C' = W_3 R W L W_4,
\end{align}
for some sub-paths $W_1,W_2,W_3,W_4$. The left/right turns are very important. If the two curves have the form, say,
\begin{align}
C = W_1 R W R W_2\qquad\text{and}\qquad
C' = W_3 L W L W_4,
\end{align}
then they \emph{do not intersect} at $W$.

Using this general rule, we can find triangulations of fatgraphs using only the data of the curves.

\begin{figure}
\begin{center}
\includegraphics[width=0.75\textwidth]{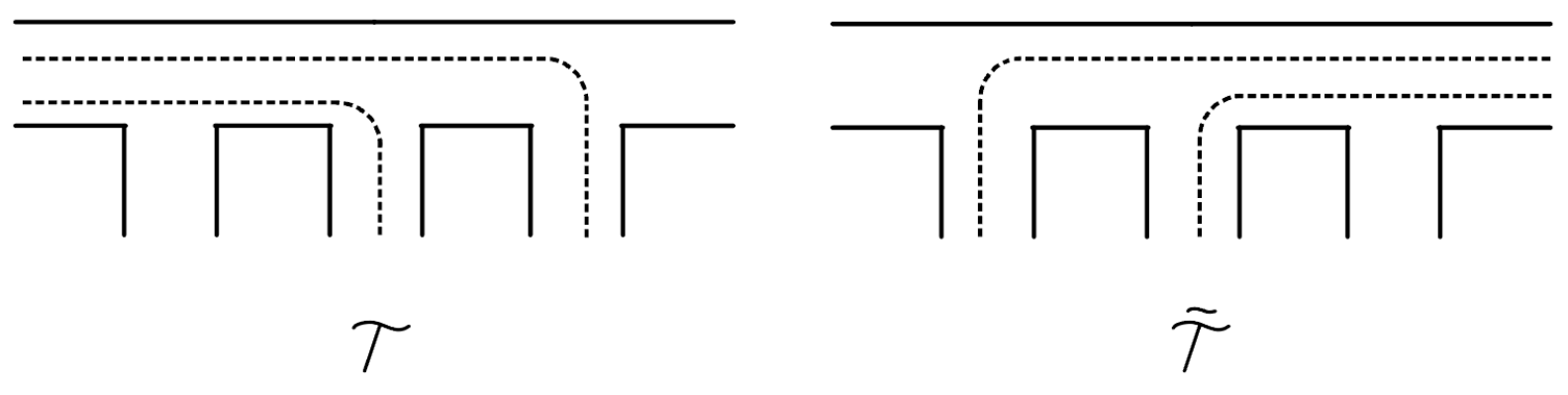}
\caption{The two special triangulations of a fatgraph, $T$ and $\tilde{T}$, are defined by curves with exactly one peak (left) and curves with exactly one valley (right).}
\label{fig:tttilde}
\end{center}
\end{figure}

For every fatgraph $\Gamma$, there are two special triangulations. Suppose that $\Gamma$ has edges $e_i$, $i=1,\ldots,E$. Let $C_i$ be the curve that, starting from $e_i$, turns right in both directions away from $e_i$. Then
\begin{align}\label{eqn:TCi}
C_i = \cdots L e L e' L e_i R e'' R e''' R \cdots.
\end{align}
$C_i$ has exactly one peak, which is at $e_i$. The intersection rule, \eqref{eqn:int}, shows that no pair of such curves $C_i,C_j$ ($i\neq j$) intersect. So the $C_i$ give $E$ nonintersecting curves, and these form a triangulation, $T$. We can also consider the curves
\begin{align}\label{eqn:TCii}
\tilde C_i = \cdots R e R e' R e_i L e'' L e''' L \cdots,
\end{align}
that turn left going in both directions away from $e_i$. These $\tilde C_i$ each have exactly one valley, at $e_i$, and so they are mutually nonintersecting. Together, they give another triangulation of the fatgraph, $\tilde T$. An example of these special triangulations is given in Figure \ref{fig:tttilde}.

\subsection{Momentum Assignments}\label{sec:mom}
The edges of a fatgraph $\Gamma$ are naturally decorated with momenta, induced by the \emph{external momenta} of the graph. Let $\Gamma$ have $n$ external momenta $p_1^\mu,\ldots,p_n^\mu$, directed \emph{into} the graph (say). By imposing momentum conservation at each cubic vertex, we obtain a momentum $p_e^\mu$ for every edge. If $\Gamma$ has loops (i.e. $E> n-3$), then there is a freedom in the definition of the $p_e^\mu$ that we parametrise by some $L$ \emph{loop momentum variables}, $\ell_1^\mu,\ldots,\ell_L^\mu$. This is the standard rule for assigning momenta to a fatgraph, $\Gamma$.

\begin{figure}
\centering
\begin{subfigure}{.5\textwidth}
  \centering
  \includegraphics[width=.9\linewidth]{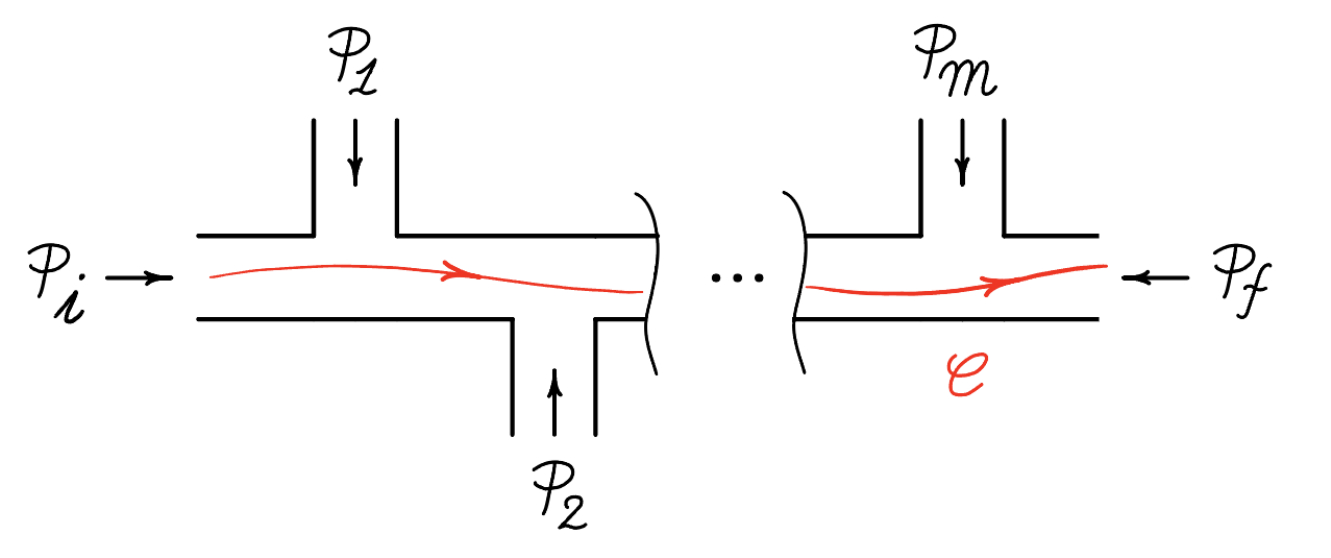}
  \caption{~}
  \label{fig:mom1}
\end{subfigure}%
\begin{subfigure}{.5\textwidth}
  \centering
  \includegraphics[width=.9\linewidth]{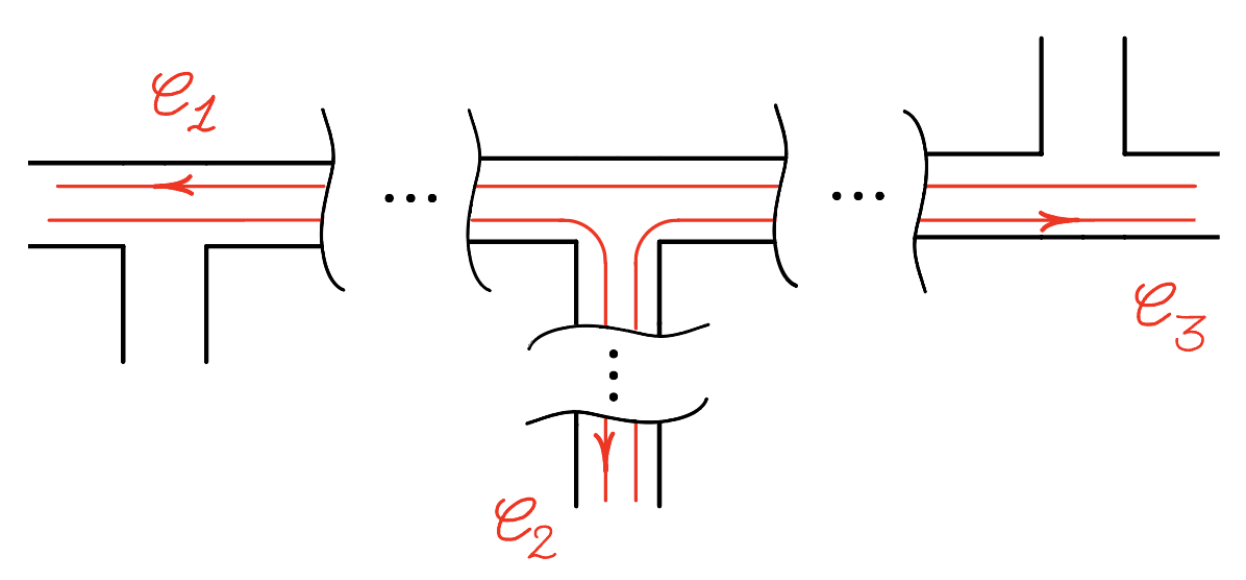}
  \caption{~}
  \label{fig:mom2}
\end{subfigure}
\caption{(a) Reversing a curve reverses its momentum assignment. (b) The momenta of three curves that cut out a cubic vertex satisfy momentum conservation.}
\label{fig:mom}
\end{figure}

To go further, we now introduce a way to also assign a momentum to every \emph{curve} on $\Gamma$. For a curve with an orientation, $\overrightarrow{C}$, will assign a momentum $P_{\overrightarrow{C}}^\mu$. This momentum assignment should satisfy two basic rules. If $\overleftarrow{C}$ is the curve with reversed orientation (Figure \ref{fig:mom1}), then
\begin{align}\label{eqn:mom1}
P_{\overleftarrow{C}}^\mu= - P_{\overrightarrow{C}}^\mu.
\end{align}
And if three curves, $\overrightarrow{C}_1,\overrightarrow{C}_2,\overrightarrow{C}_3$, cut out a cubic vertex (Figure \ref{fig:mom1}), then we impose momentum conservation at that vertex:
\begin{align}\label{eqn:mom2}
P_{\overrightarrow{C}_1}^\mu+P_{\overrightarrow{C}_2}^\mu+P_{\overrightarrow{C}_3}^\mu = 0.
\end{align}

The solution to satisfying both \eqref{eqn:mom1} and \eqref{eqn:mom2} is very simple, if we start with the momenta $p_e^\mu$ assigned to the edges of $\Gamma$. Suppose $\overrightarrow{C}$ enters $\Gamma$ via the external line $i$. Then assign this curve
\begin{equation}\label{eq:momwalk}
    P_{\overrightarrow{C}}^\mu = p_i^\mu + \sum_\text{right turns} p_\text{left}^\mu,
\end{equation}
where $p_\text{left}^\mu$ is the momentum of the edge incident on $C$ from the left, at the vertex where $\overrightarrow{C}$ makes a right turn. The momentum assignment, \eqref{eq:momwalk}, can easily be checked to satisfy \eqref{eqn:mom1} and \eqref{eqn:mom2}.

\begin{figure}
\begin{center}
\includegraphics[width=0.90\textwidth]{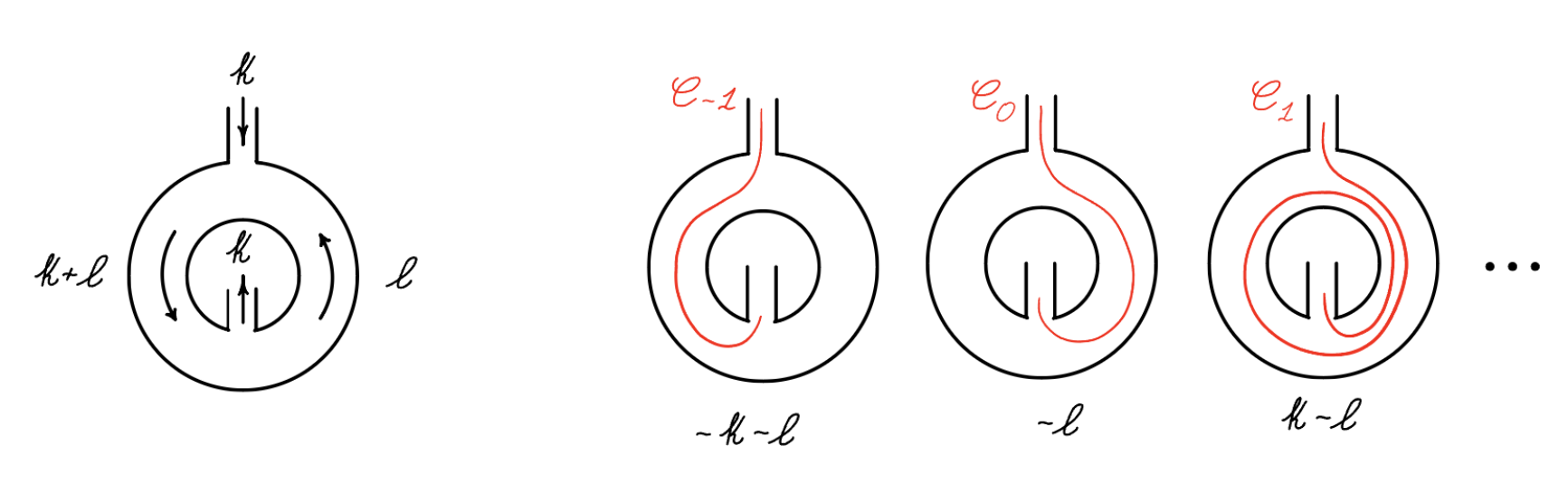}
\caption{An assignment of momenta to the edges of a fatgraph (left) induces as assignment of momenta to curves on the fatgraph (right).}
\label{fig:momex2}
\end{center}
\end{figure}

For example, take the fatgraph in Figure \ref{fig:momex2}. The assignment of momenta to the edges of the graph is shown in the Figure. The curve $C_0$ in Figure \ref{fig:momex2} enters the graph with momentum $p^\mu$. Then it turns left, traverses an edge, and then turns right. At the right turn, the momentum incident on the curve from the left is $-p - \ell^\mu$. So the momentum assignment of this curve is
\begin{equation}
    P_{\overrightarrow{C}_0}^\mu = - \ell^\mu.
\end{equation}
The curve $C_1$ in Figure \ref{fig:momex2} has two right turns. At its first right turn, it gains momentum $p^\mu$. At its second right turn, it gains momentum $-p^\mu-\ell^\mu$. So the momentum assignment of this curve is
\begin{equation}
    P_{\overrightarrow{C}_1}^\mu = p^\mu - \ell^\mu.
\end{equation} 

For \emph{any} triangulation, $T$, the above rules assign a momentum to every curve in the triangulation. By construction, these momenta satisfy momentum conservation at each of the cubic vertices cut out by $T$. The upshot of this is that we can \emph{re-use} the same loop momentum variables, $\ell_1,...,\ell_L$, when assigning momenta to \emph{any} triangulation of $\Gamma$. This simple idea makes it possible to do the loop integrations for all diagrams at once, instead of one Feynman diagram at a time, which is a key step towards our formulas for amplitudes. This idea also makes it possible to compute well-defined \emph{loop integrands}, even beyond the planar limit.

\subsubsection{Aside on Homology}\label{sec:homology}
There is a more formal way to understand the assignment of momenta to curves: these momentum assignments are an avatar of the homology of the fatgraph. Let $H_1(\Gamma,\Gamma_\infty)$ be the homology of $\Gamma$ (regarded as a surface) relative to the \emph{ends} of the external edges of the fatgraph, $\Gamma_\infty$. An oriented curve $\overrightarrow{C}$ represents a class $[\overrightarrow{C}]\in H_1(\Gamma,\Gamma_0)$, and
\begin{align}
[\overrightarrow{C}] + [\overleftarrow{C}] = 0
\end{align}
in homology. Moreover, if three curves cut out a cubic vertex, their classes satisfy
\begin{align}
[\overrightarrow{C}_1]+[\overrightarrow{C}_2]+[\overrightarrow{C}_3] = 0
\end{align}
in homology. This means that a momentum assignment to curves satisfying \eqref{eqn:mom1} and \eqref{eqn:mom2} defines a linear map
\begin{align}
P: H_1(\Gamma,\Gamma_\infty) \rightarrow \mathbb{R}^{1,D-1},
\end{align}
from $H_1(\Gamma,\Gamma_\infty)$ to Minkowski space.

\subsection{Spirals}\label{sec:spiral}
The colour factor $C_\Gamma$ is a product of trace factors $\text{tr}(t_1...t_k)$ formed from the colour polarisations ${t_i}_I^J$. If $\Gamma$ has a closed colour loop, this boundary contributes $\text{tr}(1) = N$ to the colour factor. For such a fatgraph, there are curves that infinitely spiral around this closed loop. These spiral curves can be treated just the same as all the other curves. In fact, the momentum assignment for spiral curves follows again from the same rule above, \eqref{eq:momwalk}.

\begin{figure}
\begin{center}
\includegraphics[width=0.35\textwidth]{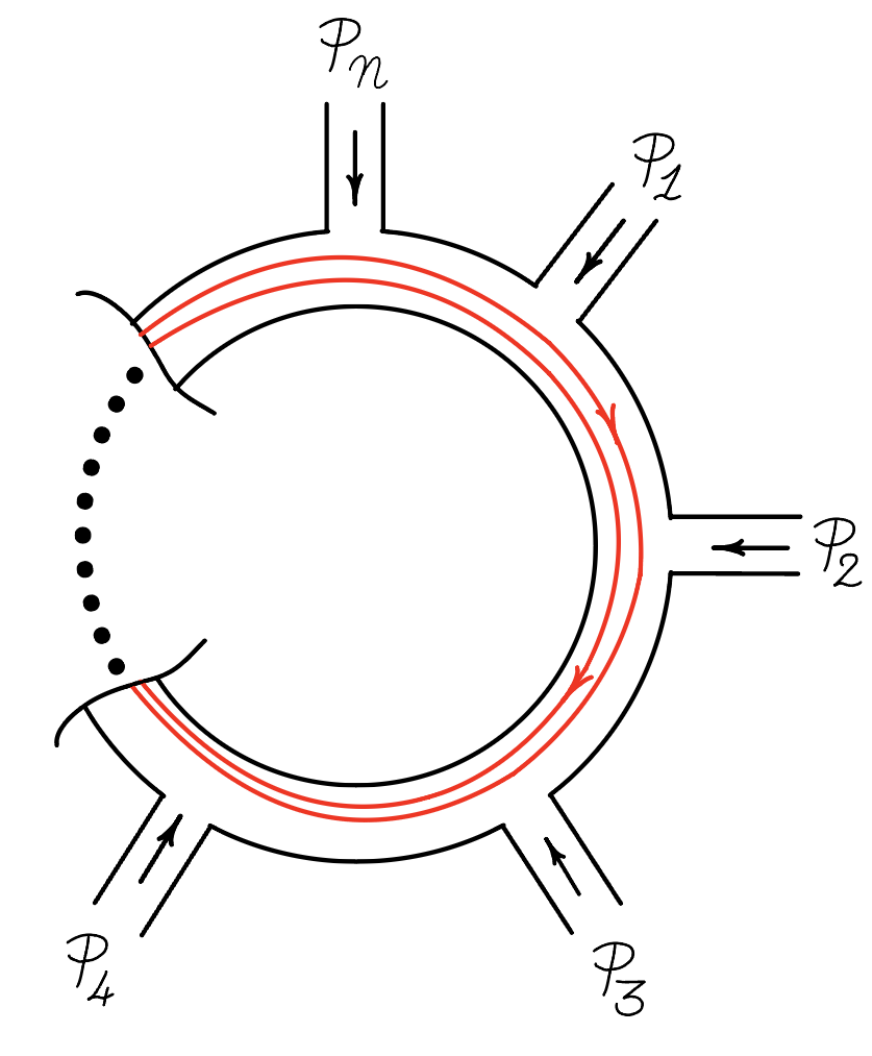}
\caption{The momenta incident on a closed loop in a fatgraph sum to zero. This ensures that the assignment of momentum to a spiral curve is well defined.}
\label{fig:spiralsum}
\end{center}
\end{figure}

Suppose that $\Gamma$ has a closed colour loop, $\beta$. Suppose that there are some $m\geq 1$ edges incident on the loop, as in Figure \ref{fig:spiralsum}. By momentum conservation, the momenta of these edges, $p_1,\ldots,p_m$, must sum up to zero: $\sum_{i=1}^m p_i = 0$. This ensures that \eqref{eq:momwalk} assigns a well-defined momentum to a curve that spirals around this boundary, because the contributions from the $p_i^\mu$ vanish after every complete revolution. 

\section{The Feynman Fan}\label{sec:fan}
For a fatgraph $\Gamma$ with $E$ edges ($e_1,\ldots,e_E$), consider the $E$-dimensional vector space, $V$, generated by some vectors, ${\bf e}_1,\ldots,{\bf e}_E$. To every curve $C$ on the fatgraph, we can assign a \emph{$g$-vector}, ${\bf g}_C \in V$. These simple integer vectors contain all the key information about the curves on $\Gamma$. Moreover, the $g$-vectors define a \emph{fan} in $V$ that we can use to rediscover the Feynman diagram expansion for the amplitude.

To define the $g$-vector of a curve, $C$, consider the \emph{peaks} and \emph{valleys} of its mountainscape. $C$ has a \emph{peak at $e_i$} if it contains
\begin{align}
\cdots L e_i R \cdots.
\end{align}
$C$ has a \emph{valley at $i$} if it contains
\begin{align}
\cdots R e_i L \cdots.
\end{align}
Let $a^i_C$ be the number of times that $C$ has a peak at $e_i$, and let $b^i_C$ be the number of times that $C$ has a valley at $e_i$. This information about the peaks and valleys is recorded by the \emph{$g$-vector of $C$},
\begin{align}
{\bf g}_C \equiv \sum_{i=1}^E g_C^i \,{\bf e}_i,\qquad\text{where}~g_C^i = a^i_C - b^i_C.
\end{align}
Each curve has a distinct $g$-vector. The converse is even more surprising: a curve is completely specified by its $g$-vector.

For example, consider the curve, $C_i$, in the triangulation $T_\Gamma$, which has only one peak, at $e_i$. The $g$-vector for $C_i$ is then
\begin{align}
{\bf g}_{C_i} = {\bf e}_i.
\end{align}
So the $g$-vectors of this triangulation $T_\Gamma$ span the positive orthant of $V$.

\subsection{Example: tree level at 5-points}\label{ex:A2:fan}
\begin{figure}
\centering
  \includegraphics[trim={0cm 0cm 0cm 0cm}, clip, width=\textwidth]{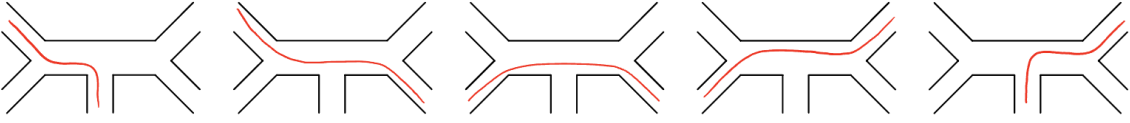}
  \caption{The five curves on the $n=5$ tree fatgraph.}
  \label{fig:exA2}
\end{figure}

\begin{figure}
  \centering
  \includegraphics[trim={0cm 0cm 0cm 0cm}, clip, width=.45\textwidth]{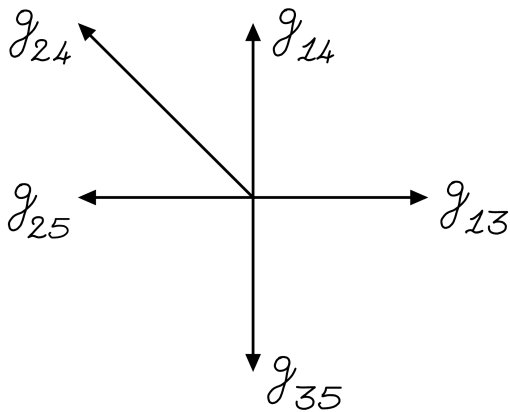}
\caption{The Feynman fan for $n=5$ tree level.}
  \label{fig:fan1}
\end{figure}

Take the comb graph $\Gamma$, with edges labelled by variables $x$ and $y$, as in Figure \ref{fig:exA2}. The five curves on $\Gamma$ are
\begin{align}
C_{13} = 1LxR3,\qquad
C_{14} = 1LxLyR4,\qquad
C_{24} = 2RxLyR4,
\end{align}
\begin{align}
C_{25} = 2RxLyL5,\qquad
C_{35} = 3RyL5.
\end{align}
Counting the peaks and valleys of these mountainscapes gives
\begin{align}
{\bf g}_{13} = \begin{bmatrix} 1 \\ 0 \end{bmatrix},~{\bf g}_{14} = \begin{bmatrix} 0 \\ 1 \end{bmatrix},~{\bf g}_{24} = \begin{bmatrix} -1 \\ 1 \end{bmatrix},~{\bf g}_{25} = \begin{bmatrix} -1 \\ 0 \end{bmatrix},~{\bf g}_{35} = \begin{bmatrix} 0 \\ -1 \end{bmatrix}.
\end{align}
These $g$-vectors are shown in Figure \ref{fig:fan1}. They define a \emph{fan} in the 2-dimensional vector space. The top-dimensional cones of this fan are spanned by pairs of $g$-vectors, such as ${\bf g}_{14}$ and ${\bf g}_{24}$, whose corresponding curves define triangulations.

\subsection{The Fan}\label{sec:fansyms}
The $g$-vectors of all the curves on $\Gamma$ together define an integer fan $\mathfrak{F} \subset V$. To define a fan, we must specify all of its \emph{cones}. We adopt the rule that two or more $g$-vectors span a cone in $\mathfrak{F}$ if and only if their curves do not intersect. The main properties of $\mathfrak{F}$ are:
\begin{enumerate}
\item It is a polyhedral fan that is dense $V$.\footnote{A fan is \emph{polyhedral} if the intersection of any two cones is itself, if nonempty, a cone in the fan, and the faces of each cone are cones in the fan. A fan is \emph{dense} if any integer vector is contained in some cone of the fan. In general, irrational vectors are not always contained in our fans, but this will not play any role in this paper.}
\item Its top dimensional cones are in 1:1 correspondence with triangulations.
\item The $g$-vectors of each top-dimensional cone span a parallelepiped of unit volume.
\end{enumerate}
Since the top-dimensional cones of $\mathfrak{F}$ correspond to triangulations, and hence to Feynman diagrams, we call $\mathfrak{F}$ the \emph{Feynman fan}, or sometimes, the \emph{$g$-vector fan}.

The property that $\mathfrak{F}$ is \emph{polyhedral and dense} means that every rational vector ${\bf g} \in V$ is contained in \emph{some} cone in the fan. This implies that every such ${\bf g}$ can be \emph{uniquely} written as a positive linear combination of $g$-vectors. In Section \ref{sec:counting}, we solve the problem of how to do this expansion explicitly.

\subsection{The Mapping Class Group}

\begin{figure}
  \centering
  \includegraphics[trim={0cm 0cm 0cm 0cm}, clip, width=.65\textwidth]{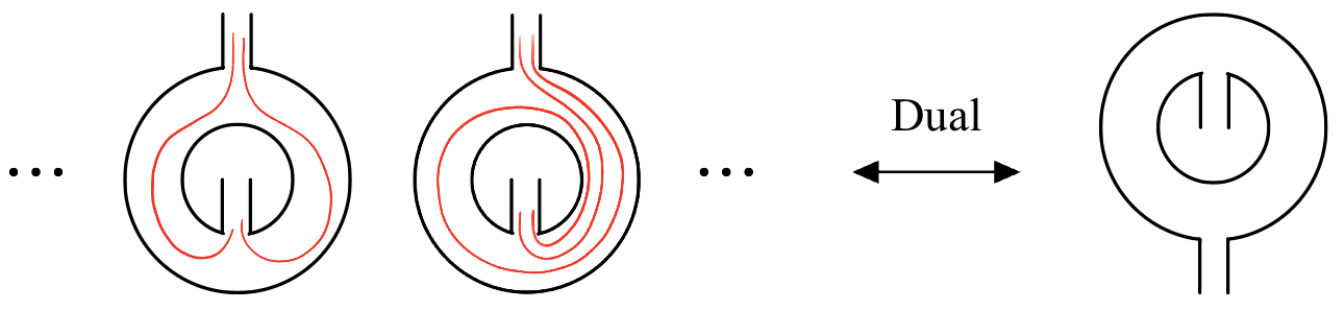}
\caption{Triangulations (left) that are related to each other by the action of of the $\MCG$. These triangulations are all dual to the same Feynman diagram (right).}
  \label{fig:mcgdual}
\end{figure}

The Feynman fan of a fat graph $\Gamma$ inherits from $\Gamma$ an action of a discrete, finitely generated group called the \emph{mapping class group}, $\MCG$. The $\MCG$ of a fatgraph, $\Gamma$, is the group of homeomorphisms of $\Gamma$, up to isotopy, that restrict to the identity on its boundaries. The action of $\MCG$ on the fatgraph can be studied by considering its action on curves. Since we only ever consider curves up to homotopy, a group element $\gamma \in \MCG$ induces a map on curves
\begin{equation}
    \gamma: C\mapsto \gamma C.
\end{equation} 
Since $\MCG$ acts via homeomorphisms, it does not affect curve intersections and non-intersections. If $C$ and $C'$ are two non-intersecting curves, then $\gamma C$ and $\gamma C'$ are likewise non-intersecting. Similarly, if $C,C'$ intersect, so do $\gamma C$ and $\gamma C'$. This means that if some curves, $C_1,\ldots, C_E$, form a triangulation, so do their images under $\MCG$. Moreover, if the triangulation $\{C_1,\ldots, C_E\}$ is dual to a fatgraph $\Gamma'$, then each image $\{\gamma C_1,\ldots, \gamma C_E\}$ is \emph{also} dual to the same fatgraph, $\Gamma'$.

For example, take the 2-point non-planar fatgraph $\Gamma$ in Figure \ref{fig:mcgdual}. The $\MCG$ acts on $\Gamma$ by \emph{Dehn twists} that increase the number of times a curve winds around the fatgraph. All triangulations of $\Gamma$ are related to each other by the $\MCG$ and they are all dual to the same fatgraph (right in Figure \ref{fig:mcgdual}).

In general, if $\Gamma$ has loop number $L$, then $\MCG$ has a presentation with $L$ generators \cite{penner2012decorated}. These can be identified with Dehn twists around annuli in the fatgraph.

The $\MCG$ action on curves induces a piecewise linear action on the vector space, $V$,
\begin{align}
\gamma: {\bf g}_C \mapsto {\bf g}_{\gamma C}.
\end{align}
It follows from the above properties of the $\MCG$ action on curves that the action of $\MCG$ on $V$ leaves the fan $\mathfrak{F}$ invariant (if we forget the labels of the rays). Furthermore, two top-dimensional cones of the fan correspond to the same Feynman diagram if and only if they are related by the $\MCG$ action. 

\subsubsection{Aside on automorphisms}
\begin{figure}
\begin{center}
\includegraphics[width=0.5\textwidth]{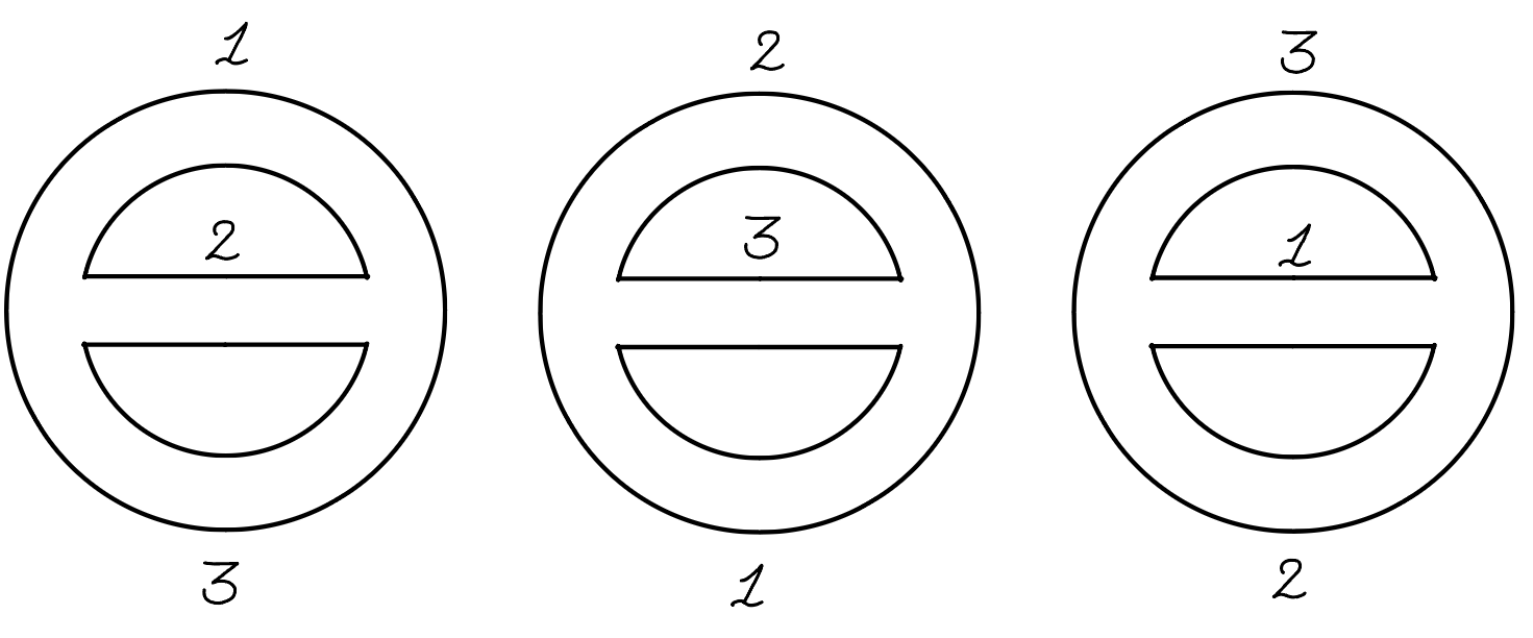}
\caption{A fatgraph with $|\text{Aut}(\Gamma)| = 3$. Cyclic permutations of the edges leave it unchanged.}
\label{fig:aut}
\end{center}
\end{figure}

There is another discrete group that acts on the Feynman fan: the group of graph automorphisms, $\text{Aut}(\Gamma)$. The elements of $\text{Aut}(\Gamma)$ are permutations of the labels of the edges of $\Gamma$. A permutation is an \emph{automorphism} if it leaves the list of fat vertices of $\Gamma$ unchanged (including the vertex orientations). Each fat vertex can be described by a triple of edge labels with a cyclic orientation, $(ijk)$. 

$\text{Aut}(\Gamma)$ has a linear action on $V$ given by permuting the basis vectors ${\bf e}_1,\ldots,{\bf e}_E$. The action of $\text{Aut}(\Gamma)$ leaves the fan invariant (again if we forget the labels of the rays).

An example of a fatgraph with nontrivial automorphisms is Figure \ref{fig:aut}. In this example, cyclic permutations of the 3 edges preserve the fatgraph. Most fatgraphs that we will consider have trivial automorphism groups, and so the action of $\text{Aut}(\Gamma)$ will not play a big role in this paper.

\subsection{Example: the non-planar 1-loop propagator}\label{ex:A11:fan}
\begin{figure}
\centering
  \includegraphics[trim={0cm 0cm 0cm 0cm}, width=.65\textwidth]{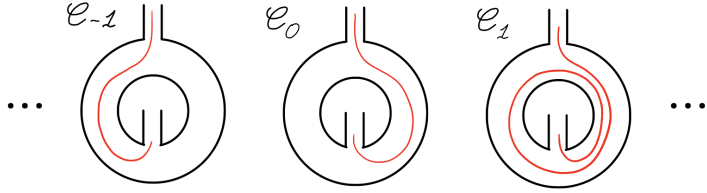}
  \caption{The infinite family of curves, $C_n$, for the non-planar one loop propagator.}
  \label{fig:exA11}
\end{figure}

Take the 1-loop fatgraph $\Gamma$ in Figure \ref{fig:exA11}, with edges labeled by variables $x$ and $y$. Some of the curves on $\Gamma$, $C_n$, are shown in the Figure. These curves are related to each other by the action of $\MCG$, which is generated by a Dehn twist, $\gamma$. With the labelling in Figure \ref{fig:exA11}, the action of $\gamma$ is
\begin{equation}
    \gamma: C_n \mapsto C_{n+1}.
\end{equation}
There are infinitely many such curves on the fatgraph.

The paths of the curves on $\Gamma$ are
\begin{align}
C_n &= 1 L (x L y R)^n x R 2 \qquad\text{for}~n\geq 0,\\
C_n &= 1 R y (R x L y)^{1+n} L 2 \qquad\text{for}~ n <0, \\
\Delta &= x L y R,
\end{align}
where $\Delta$ is the closed loop. Note that the curves $C_n$ differ from one another by multiples of the closed path $\Delta$. In this way, we can see the $\MCG$ directly in terms of the mountainscapes of the curves.

Counting peaks and valleys in the mountainscapes, the $g$-vectors for these curves are:
\begin{align}
{\bf g}_{n} & = \begin{bmatrix}-n+1\\n\end{bmatrix}\qquad \text{for}~n\geq 0,\\
{\bf g}_{n} & = \begin{bmatrix}n+1\\-n-2\end{bmatrix}\qquad  \text{for}~n< 0,\\
{\bf g}_\Delta & = \begin{bmatrix}-1\\ 1\end{bmatrix}. &~
\end{align}
These $g$-vectors define the fan in Figure \ref{fig:fan1}. There are infinitely many rays of this fan. The action of $\MCG$ on curves lifts to a piecewise linear action on the fan, generated by the action of the Dehn twist $\gamma$. $\gamma$ acts on the fan as
\begin{align}
{\bf g}_{{n+1}} &= {\bf g}_{n} + {\bf g}_\Delta,\qquad \text{for}~n \geq 0,\\
{\bf g}_{0} &= {\bf g}_{{-1}} + (1,1),\\
{\bf g}_{{n+1}} &= {\bf g}_{{n}} - {\bf g}_\Delta,\qquad \text{for}~n < -1.
\end{align}
This is (trivially) an isomorphism of the fan.

\begin{figure}
  \centering
  \includegraphics[trim={0cm 0cm 0cm 0cm}, clip, width=.45\textwidth]{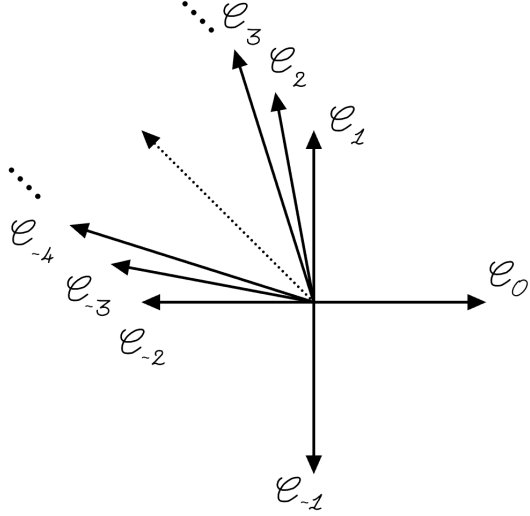}
  \caption{The Feynman fan for the non-planar 1-loop propagator.}
  \label{fig:fan2}
\end{figure}

\subsection{The Delta plane}\label{sec:delta}
A \emph{closed curve} is a curve $\Gamma$ that forms a loop. For a closed curve $\Delta$, consider the series of left and right turns that it makes. We can record this series of turns as a \emph{cyclic word}, like $W_\Delta= (RRLRL)$. Whenever $RL$ appears in $W_\Delta$ it corresponds to a \emph{valley} in the mountainscape, which happens where the curve switches from turning right to turning left. Likewise, $LR$ corresponds to a \emph{peak}. If the cyclic word $W_C$ has $n$ occurrences of `$RL$', it must also have exactly $n$ occurrences of `$LR$'. For example, the cyclic word
\begin{align}
(RRLRLLLRRLL),
\end{align}
switches from right-to-left 3 times, and from left-to-right 3 times.

In other words, the mountainscape for a closed curve has exactly as many peaks as valleys. It follows that the $g$-vector, ${\bf g}_\Delta$, for any closed curve $\Delta$ is normal to the vector ${\bf n} = (1,1,1,...,1)^T$. We call the plane normal to ${\bf n}$ the \emph{$\Delta$ plane}: $V_\Delta \subset V$.

For example, in the previous subsection, the closed curve $\Delta$ had $g$-vector ${\bf g}_\Delta = (-1,1)$, which is normal to the vector $(1,1)$.

Finally, note that a closed curve that makes \emph{only} right-turns (resp. left-turns) corresponds to a path around a loop boundary of $\Gamma$. These curves have no peaks and no valleys. So these loop boundaries are assigned zero $g$-vector. They are also assigned zero momentum (by the reasoning in Section \ref{sec:spiral}).

\subsection{Example: the planar 1-loop propagator}\label{ex:D2:fan}
\begin{figure}
\begin{center}
\includegraphics[width=0.7\textwidth]{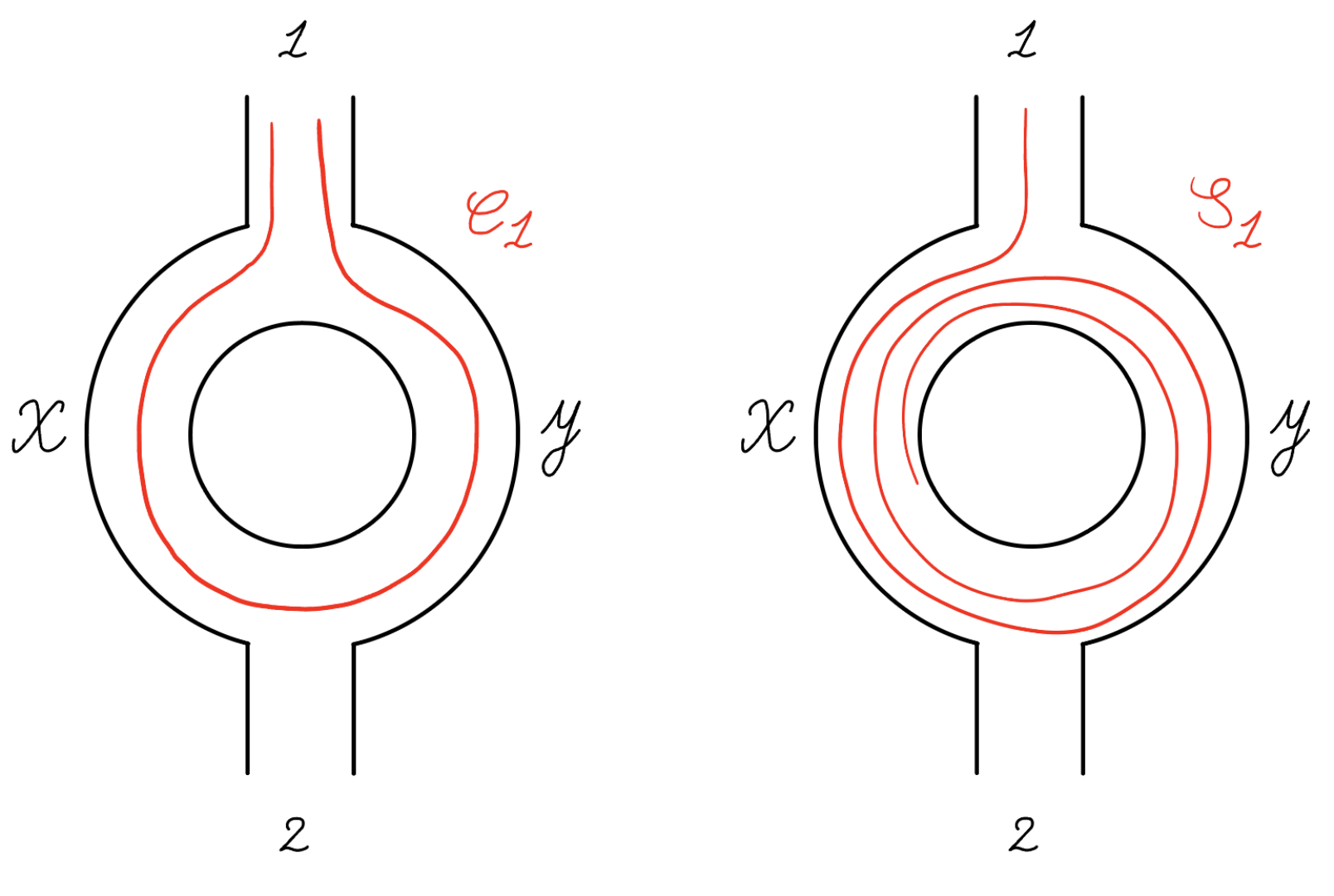}
\caption{Curves on the bubble fatgraph.}
\label{fig:D2curves}
\end{center}
\end{figure}

\begin{figure}
\begin{center}
\includegraphics[width=0.35\textwidth]{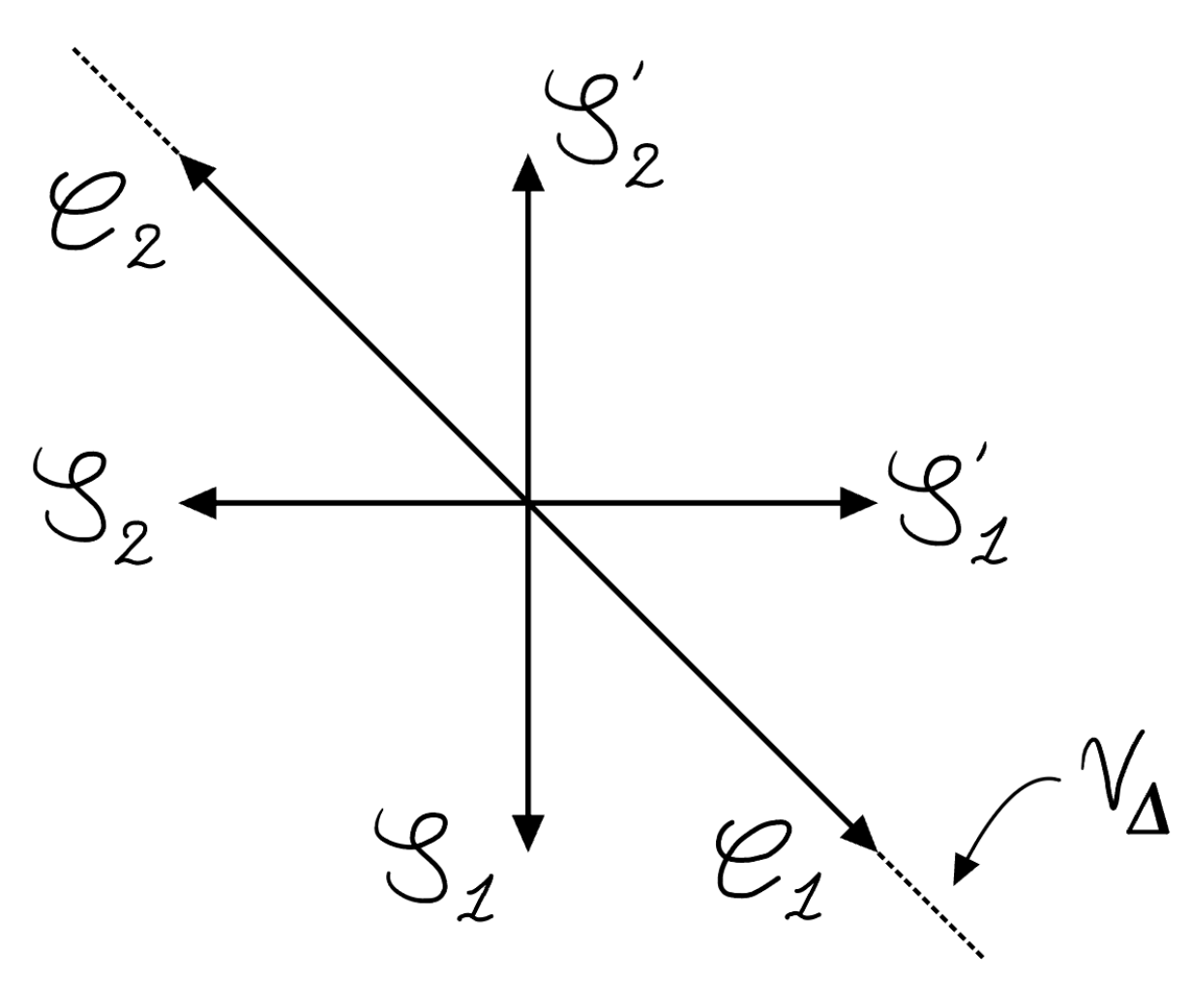}
\caption{The Feynman Fan for the 1-loop planar propagator.}
\label{fig:D2fan}
\end{center}
\end{figure}

Take the 1-loop bubble diagram, $\Gamma$, with edges $x$ and $y$, and external edges $1$ and $2$, as in Figure \ref{fig:D2curves}. Consider the four curves, $C_1,C_2,S_1,S_2$, shown in the Figure. These have paths
\begin{align}
C_1&= 1 R x L y R 1\\
C_2&= 2 R y L x R 2\\
S_1&= 1 R x L y L x L y L \cdots \\
S_2&= 2 R y L x L y L x L \cdots.
\end{align}
The curves $S_1,S_2$ end in anticlockwise spirals around the closed loop boundary. There are also two curves, $S_1'$ and $S_2'$, which spiral \emph{clockwise} into the puncture:
\begin{align}
S_1'&= 1 L y R x R y R\cdots \\
S_2'&= 2 L x R y R x R \cdots.
\end{align}
Counting peaks and valleys, the $g$-vectors of these curves are
\begin{align}
{\bf g}_{C_1} = \begin{bmatrix} -1\\ 1 \end{bmatrix},~{\bf g}_{S_1'} = \begin{bmatrix} 1\\ 0 \end{bmatrix},~{\bf g}_{S_2'} = \begin{bmatrix} 0\\ 1 \end{bmatrix},~{\bf g}_{C_2} = \begin{bmatrix} 1\\ -1 \end{bmatrix},~{\bf g}_{S_1} = \begin{bmatrix} 0\\ -1 \end{bmatrix},~{\bf g}_{S_2} = \begin{bmatrix} -1\\ 0 \end{bmatrix}.
\end{align}
These $g$-vectors give the fan in Figure \ref{fig:D2fan}. Notice that the $g$-vectors of the curves $C_1,C_2$ lie on the Delta plane: $x+y=0$.

Including the anticlockwise spirals would lead to us counting every Feynman diagram twice. This is because the triangulation with $C_1, S_1$ is dual to the same diagram as the triangulation by $C_1, S_1'$, and so on. To prevent overcounting, it makes sense to restrict to the part of the fan that involves only $C_1,S_1,S_2$, and $C_2$. This part of the fan is precisely the half space, $x+y\leq 0$, cut out by the Delta plane.

\section{A Counting Problem For Curves}\label{sec:counting}
There is a natural counting problem associated to mountainscapes, and this counting problem plays the central role in our amplitude computations.

For a mountainscape, $C$, the idea is to form subsets of $C$ by \emph{filling up} the mountainscape from the bottom. A subset is valid if it includes everything \emph{downhill} of itself in the mountainscape.

\begin{figure}
\begin{center}
\includegraphics[width=0.6\textwidth]{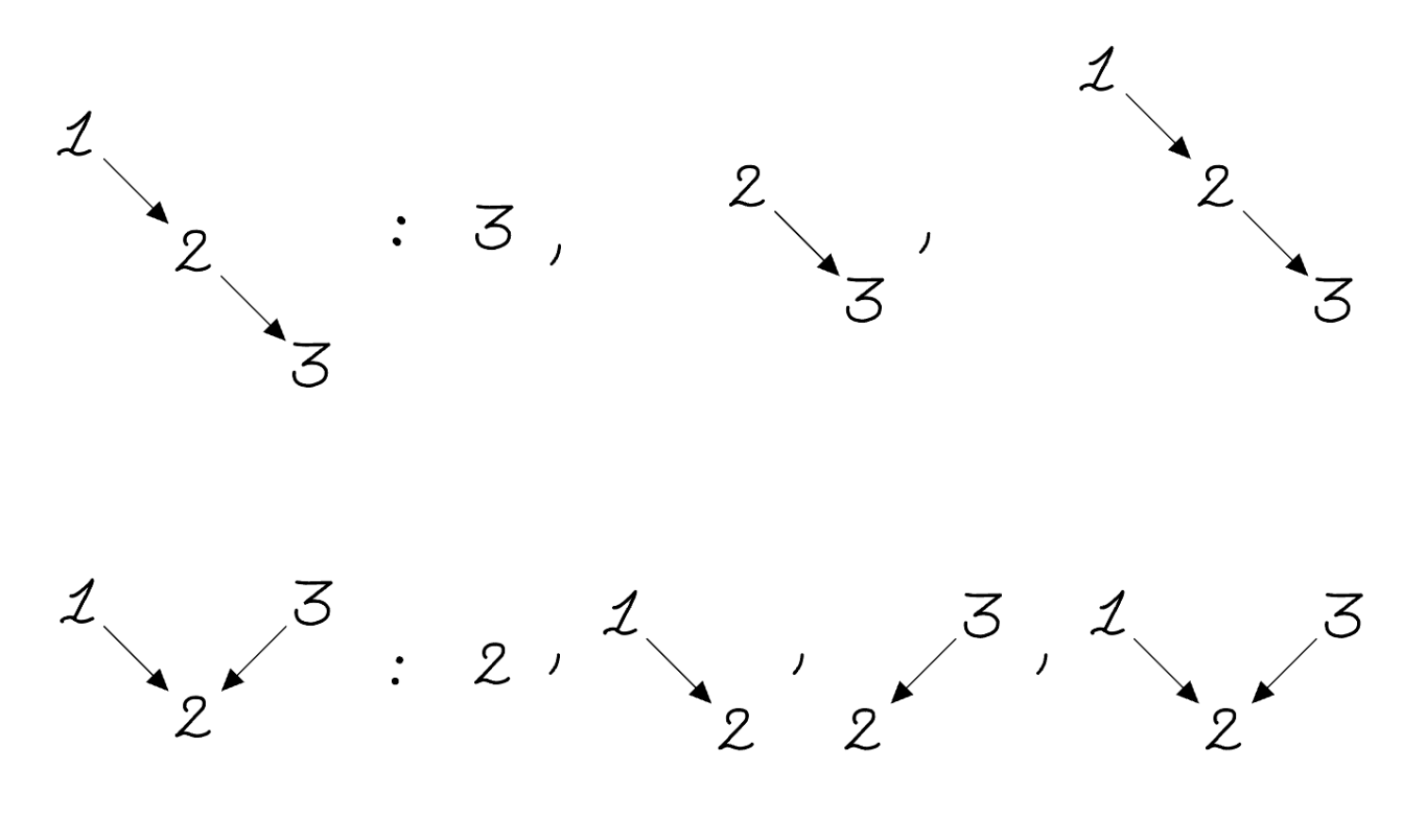}
\caption{Two examples of mountainscapes and their sub-mountainscapes.}
\label{fig:Fex}
\end{center}
\end{figure}

For example, consider the curve in Figure \ref{fig:Fex}, 
\begin{align}\label{eq:exF}
C = 1 R 2 L 3.
\end{align}
The valid subsets of $C$, shown in the Figure, are $2, 1R2, 2L3$, and $1R2L3$. In other words, if $3$ is in the subset, then $2$ must also be included, because it is downhill of (left of) $3$. Likewise, if $1$ is in the subset, then $2$ must also be included, because 2 is downhill of (right of) $3$. 

This information can be summarised using a generating function or \emph{$F$-polynomial}. Introduce variables $y_i$, $i=1,\ldots,E$, labelled by the edges of $\Gamma$. Then the $F$-polynomial of a curve $C$ is
\begin{align}
F_C = 1+\sum_{C' \subset C} \,\prod_{i \in C'} y_i,
\end{align}
where the sum is over all valid (non-empty) subsets of $C$, including $C$ itself.

In the example, \eqref{eq:exF}, we have four valid subsets, and the $F$-polynomial is
\begin{align}\label{eq:Fex}
F_C = 1 + y_2 + y_1y_2 + y_2y_3 + y_1y_2y_3.
\end{align}

\subsection{Curve Matrices}\label{sec:matrices}
Consider a curve $C$ that starts at any edge $e_i$ and ends at any edge $e_j$. It is natural to decompose its $F$-polynomial as a sum of four terms,
\begin{equation}\label{eq:Fdecom}
    F_C = F_{--}+F_{-+}+F_{+-}+F_{++},
\end{equation}
where: $F_{--}$ counts subsets that exclude the first and last edges; $F_{-+}$ counts subsets that exclude the first edge and include the last edge; and so on.

Now consider what happens if we \emph{extend} $C$ along one extra edge. Let $C'$ extend $C$ by adding a left turn before $i$:
\begin{equation}
    C' = e_kLC,
\end{equation}
for some edge $e_k$. The $F$-polynomial of $C'$ can be deduced using \eqref{eq:Fdecom}. Terms that involve $y_i$ \emph{must} contain $y_k$, since $e_k$ is \emph{downhill} of $e_i$ in the curve. So
\begin{equation}\label{eq:Fd1}
    F_{C'} = (1+y_k)F_{--}+(1+y_k)F_{-+}+y_kF_{+-}+y_kF_{++}.
\end{equation}
Similarly, if $C''$ is obtained from $C$ by adding a right turn before $e_i$, then $C'' = e_lRC$, for some edge $e_l$, and we find that the new $F$-polynomial is
\begin{equation}\label{eq:Fd2}
    F_{C''} = F_{--} + F_{-+} + (1+y_l)F_{+-} + (1+y_l)F_{++}.
\end{equation}
This equation follows because any term not containing $y_i$ \emph{cannot} contain $y_l$, since $e_i$ is \emph{downhill} of $e_l$ in the curve.

Equations \eqref{eq:Fd1} and \eqref{eq:Fd2} can be used to compute the $F$-polynomial for any curve. It simple to do implement this is by defining a \emph{curve matrix}, whose entries are given by the decomposition, \eqref{eq:Fdecom}:
\begin{equation}
    M_C = \begin{bmatrix} F_{--} & F_{-+}\\ F_{+-} & F_{++} \end{bmatrix}.
\end{equation}
The curve matrix $M_{C'}$ is obtained from the curve matrix $M_C$ via the matrix version of \eqref{eq:Fd1}:
\begin{equation}\label{eq:Md1}
    M_{C'} = \begin{bmatrix} 1 & 0 \\ y_k & y_k \end{bmatrix} M_C.
\end{equation}
The matrix multiplying $M_C$ in this equation represents what happens when $C$ is extended by adding a left turn at the start. Similarly, the matrix version of \eqref{eq:Fd2} is
\begin{equation}\label{eq:Md2}
    M_{C''} = \begin{bmatrix} 1 & 1 \\ 0 & y_l \end{bmatrix} M_C,
\end{equation}
which represents what happens when $C$ is adding a right turn at the start.

It can be convenient to decompose the new matrices appearing in \eqref{eq:Md1} and \eqref{eq:Md2} as a product,
\begin{equation}
    \begin{bmatrix} 1 & 0 \\ y_k & y_k \end{bmatrix} = \begin{bmatrix} 1 & 0 \\ 0 & y_k \end{bmatrix} \begin{bmatrix} 1 & 0 \\ 1 & 1\end{bmatrix}, \qquad \begin{bmatrix} 1 & 1 \\ 0 & y_l \end{bmatrix} = \begin{bmatrix} 1 & 0 \\ 0 & y_l \end{bmatrix} \begin{bmatrix} 1 & 1 \\ 0 & 1\end{bmatrix}.
\end{equation}
Then, for any curve, $C$, we can compute its curve matrix, $M_C$, directly from the word specifying the curve. To do this, we just replace each turn and edge with the associated matrix:
\begin{equation}
    L \rightarrow \begin{bmatrix} 1 & 0 \\ 1 & 1\end{bmatrix},\qquad R \rightarrow \begin{bmatrix} 1 & 1 \\ 0 & 1\end{bmatrix},\qquad e_i \rightarrow \begin{bmatrix} 1 & 0 \\ 0 & y_i \end{bmatrix}.
\end{equation}
Every curve matrix $M_C$ is then a product of these simple matrices.

For example, for the curve $C=1R2L3$ considered above, its matrix is
\begin{equation}\label{eq:goncharov}
    M_C = \begin{bmatrix} 1 & 0 \\ 0 & y_1 \end{bmatrix} \begin{bmatrix} 1 & 1 \\ 0 & 1 \end{bmatrix} \begin{bmatrix} 1 & 0 \\ 0 & y_2 \end{bmatrix} \begin{bmatrix} 1 & 0 \\ 1 & 1 \end{bmatrix} \begin{bmatrix} 1 & 0 \\ 0 & y_3 \end{bmatrix} = \begin{bmatrix} 1+y_2 & y_2y_3\\ y_1y_2 & y_1y_2y_3 \end{bmatrix}.
\end{equation}
The sum of the entries of this curve matrix recovers the curve's $F$-polynomial, \eqref{eq:Fex}.

Curve matrices neatly factorise. If several curves all begin with the same word, $W$, their words can be written as $C_i = W C_i'$. Their matrices are then $M_{C_i} = M_W M_{C_i'}$, so that we only have to compute $M_W$ once to determine all the $M_{C_i}$. Moreover, if we add extra legs to a fatgraph $\Gamma$, to form a larger fatgraph, $\Gamma'$, the matrices $M_C$ for the larger fatgraph can be obtained directly from the matrices for the smaller fatgraph. In practice, this is very useful, and allows us to exploit the methods in this paper to compute all-$n$ formulas for amplitudes. \cite{us_alln}

\begin{figure}
\begin{center}
\includegraphics[width=0.75\textwidth]{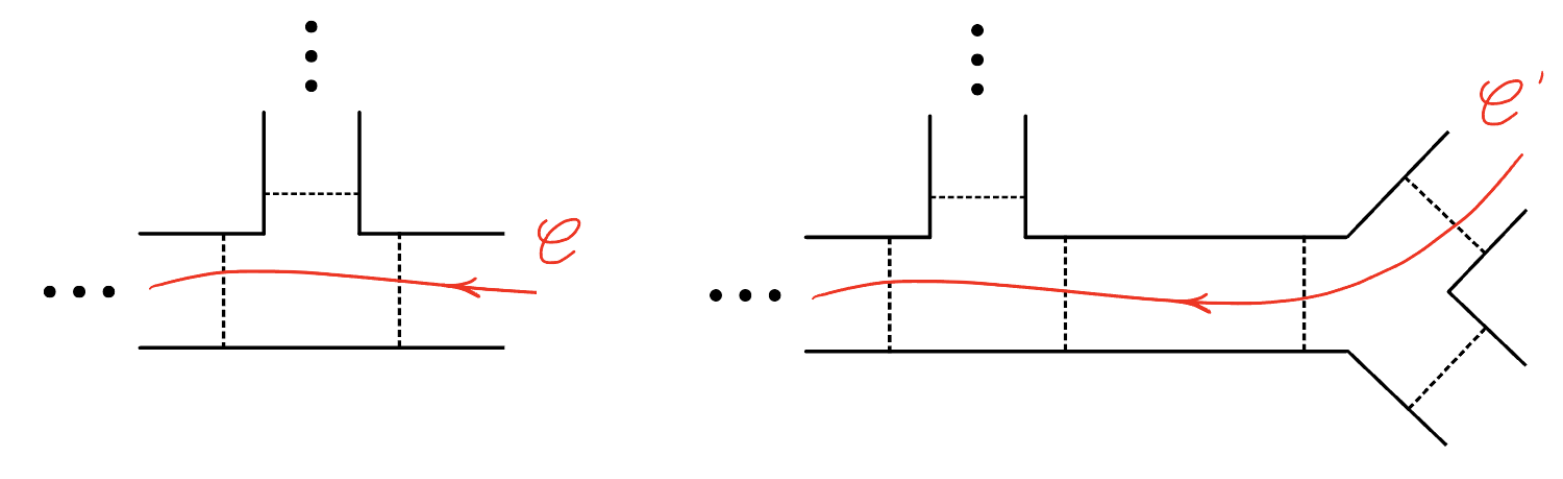}
\caption{Getting a new fatgraph.}
\label{fig:A21}
\end{center}
\end{figure}

\subsection{Headlight Functions}\label{sec:headlight}
It follows from the definition of $M_C$, as a product of the matrices in \eqref{eq:goncharov}, that
\begin{equation}
    \det M_C = \prod_{e\in C} y_e.
\end{equation}
Expanding the determinant, this gives
\begin{equation}\label{eq:ueq}
    1 = \frac{F_{-+}F_{+-}}{F_{--}F_{++}} + \frac{\prod y_e}{F_{--}F_{++}}.
\end{equation}
Motivated in part by this identity, define the \emph{$u$-variable} of a curve $C$ as the ratio
\begin{equation}
    u_C = \frac{F_{-+}F_{+-}}{F_{--}F_{++}}.
\end{equation}
These $u$-variables vastly generalise those studied in \cite{brown2009,Arkani_Hamed_2021}, and \eqref{eq:ueq} is a generalisation of the \emph{$u$-equations} studied there.

The \emph{headlight function} of a curve $C$ is the \emph{tropicalization} of the $u$-variable,
\begin{equation}
    \alpha_C = - \text{Trop\ }u_C.
\end{equation}
For a polynomial $F(y)$, its tropicalization captures the behaviour of $F$ at large values of $y_i$. Parametrise the $y_i$ as $y_i = \exp t_i$. Then, in the large $t$ limit,
\begin{align}
F(y) \rightarrow \exp \text{Trop\ } F (t).
\end{align}
For example, if $F(y) = 1+y_1+y_1y_2$, then $\text{Trop\ }F(t) = \max(0,t_1,t_1+t_2)$. In practice, $\text{Trop\ }F$ is obtained from $F$ by replacing multiplication with addition, and replacing sums with taking the maximum. 

In terms of the matrix $M_C$, the headlight function is
\begin{align}
\alpha_C =  \mathrm{Trop\ }M_C^{1,1} +  \mathrm{Trop\ }M_C^{2,2} - \mathrm{Trop\ }M_C^{1,2} -  \mathrm{Trop\ }M_C^{2,1}.
\end{align}

Headlight functions satisfy the following remarkable property:
\begin{align}\label{eq:alphaCgD}
\alpha_C ({\bf g}_D) = \left\{\begin{matrix} 1 & \qquad \text{if $C=D$} \\ 0 & \qquad \text{otherwise.} \end{matrix} \right.
\end{align}
This implies that headlight functions can be used to express any vector ${\bf g} \in V$ as a positive linear combination of the generators of a cone of the Feynman fan, by writing
\begin{align}\label{eq:decompC}
{\bf g} = \sum_C \alpha_C({\bf g}) \, {\bf g}_C.
\end{align}
This expansion has a geometrical interpretation. Any integer vector ${\bf g}\in V$ corresponds to some curve (or set of curves), $L$, possibly with self-intersections. Any intersections in $L$ can be uncrossed on $\Gamma$ using the \emph{skein relations}. Repeatedly applying skein relations, $L$ can be decomposed on the surface into a unique set of non-self-intersecting curves, and $\alpha_C(g)$ is the number of times the curve $C$ appears in this decomposition.

\subsection{Example: tree level at 5-points}\label{ex:A2:head}
\begin{figure}
\begin{center}
\includegraphics[width=0.6\textwidth]{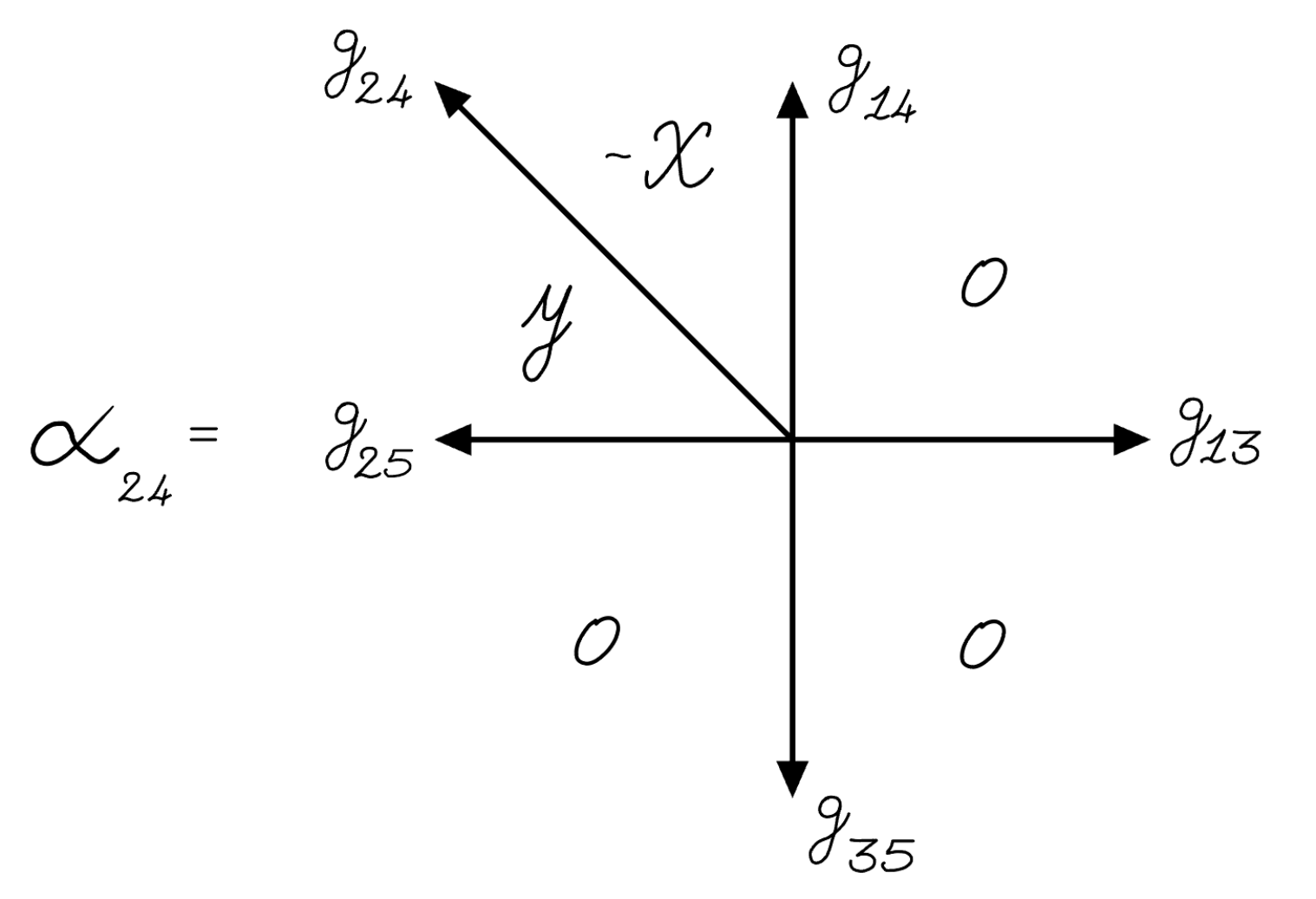}
\caption{The Schwinger parameter $\alpha_{24}$ on the Feynman fan.}
\label{fig:alpha24}
\end{center}
\end{figure}

The curves for the 5-points tree level amplitude were given in Section \ref{ex:A2:fan}. Their curve matrices, using the replacements \eqref{eq:goncharov}, are
\begin{align}
C_{13} &= Lx R & \longrightarrow \qquad \qquad & M_{13} = \begin{bmatrix} 1 & 1 \\ 1 & 1+x \end{bmatrix},\\
C_{14} &= LxLyR & \longrightarrow \qquad \qquad & M_{14} = \begin{bmatrix} 1 & 1 \\ 1+x & 1+x+xy \end{bmatrix},\\
C_{24} &= RxLyR & \longrightarrow \qquad \qquad & M_{24} = \begin{bmatrix} 1+x & 1+ x+ xy \\ x & x(1+y) \end{bmatrix},\\
C_{25} &= RxLyL & \longrightarrow \qquad \qquad & M_{25} =\begin{bmatrix} 1+x+xy & xy \\ x+xy & xy \end{bmatrix},\\
C_{35} &= RyL & \longrightarrow \qquad \qquad & M_{35} = \begin{bmatrix} 1+y & y \\ y & y \end{bmatrix}.
\end{align}
Given these matrices, the headlight functions are
\begin{align}
\alpha_{13}& = \max(0,x),\\
\alpha_{14}& = - \max(0,x) + \max(0,x,x+y),\\
\alpha_{24}& = - \max(0,x,x+y) + \max(0,x) + \max(0,y),\\
\alpha_{25}& = -x - \max(0,y) + \max(0,x,x+y),\\
\alpha_{35}& = -y + \max(0,y).
\end{align}
It can be verified that $\alpha_{ij}({\bf g}_C) = 1$ if $C = C_{ij}$, and that otherwise $\alpha_{ij}({\bf g}_C) = 0$. For example, the values taken by $\alpha_{24}$ are shown in Figure \ref{fig:alpha24}.

\subsection{Example: the non-planar 1-loop propagator}\label{ex:A11:head}
The mountainscapes for the non-planar 1-loop propagator are given in Section \ref{ex:A11:fan}. Using these, we can compute the headlight functions, and find:
\begin{align}
\alpha_n&=  f_n - 2f_{n-1} + f_{n-2},\qquad n\geq 0,\\
\alpha_n&=  g_n - 2g_{n+1} + g_{n+2},\qquad n<0.
\label{eq:alphaA11}
\end{align}
where the tropical functions $f_n$ and $g_n$ are given by
\begin{align}
f_n &= \max (0, (n+1) x, (n+1)x+ny), &\text{for}~& n\geq 0,\\
g_n &= \max (0, -(n+1)x,-(n+1)x - n y ),  &\text{for}~& n\leq -1,
\end{align}
with the following special cases:
\begin{align}
f_{-2}=0,~~f_{-1}=0,~~g_1=-2x-y,~~g_0=-x.
\end{align}
A full derivation of these functions using the matrix method is given in Appendix \ref{app:A11}.

It is easy to verify that these $\alpha_n$ satisfy the key property:
\begin{align}\label{eq:expectalpha11}
\alpha_n({\bf g}_m) = \left\{ \begin{matrix} 1 &~~\text{if $n=m$} \\ 0 & ~~\text{otherwise.} \end{matrix} \right.
\end{align}
For example, take $n,m\geq 0$. Then we find
\begin{align}
f_n({\bf g}_m) = \max (0, 1+n-m),
\end{align}
so that
\begin{align}
\alpha_n({\bf g}_m) & =  \max (0, 1+n-m) + \max (0, -1+n-m) - 2 \max (0, n-m).
\end{align}
This agrees with \eqref{eq:expectalpha11}.

\subsection{Spirals}\label{sec:spiralM}
Suppose $C$ is a curve that ends in a spiral around a loop boundary of $\Gamma$. If $1,2,...,m$ are the edges around that boundary, $C$ has the form
\begin{align}
C = W 1 L 2 L ... L m L 1 L 2 L ...,
\end{align}
for some subpath $W$. We can compute the transfer matrix for the infinite tail at the right end of $C$. The path for one loop around the boundary is
\begin{equation}
    \Delta := 1 L 2 L ... L m L,
\end{equation}
and the matrix for this path is
\begin{align}
M_\Delta = \begin{bmatrix} 1 & ~ 0 \\ F_* & ~ y_* \end{bmatrix},
\end{align}
where
\begin{equation}
y_* = \prod_{i=1}^m y_i, \qquad \text{and} \qquad F_* = y_1 + y_1y_2 + ... + y_1y_2...y_m.
\end{equation}
Now consider the powers, $M_\Delta^n$,
\begin{align}\label{eq:Minftyn}
M^n_\Delta = \begin{bmatrix} 1 & ~ 0 \\ F_*(1+y_*+\cdots+y_*^{n-1}) & ~ y_*^n \end{bmatrix}.
\end{align}
If the path $W$ has matrix
\begin{equation}
M_W = \begin{bmatrix} a & b \\ c & d \end{bmatrix},
\end{equation}
then the path $W \Delta^n$ has $u$-variable
\begin{equation}
u_{W\Delta^n} = \frac{b(c+d F_* (1+y_*+\cdots+y_*^{n-1})}{d(a+b F_* (1+y_*+\cdots+y_*^{n-1})}
\end{equation}
and in the $n\rightarrow \infty$ limit this tends to (assuming $y_* < 1$ for convergence)
\begin{equation}
u_{W\Delta^\infty} = \lim_{n\rightarrow \infty} u_{W\Delta^n} = \frac{b(c(1-y_*)+d F_*)}{d(a(1-y_*)+b F_*)},
\end{equation}
which is the $u$-variable for the full spiraling curve. Instead of computing this limit, it is convenient to instead compute this $u$-variable by multiplying the matrix $M_W$ with the matrix
\begin{equation}\label{eq:Minfty}
\widetilde{M}_\Delta = \begin{bmatrix} 1-y_* & 0 \\ F_* & 1 \end{bmatrix}.
\end{equation}
Note that $\widetilde{M}_\Delta$ is \emph{not} the $n\rightarrow\infty$ limit of $M_\Delta$. We can use this matrix, \eqref{eq:Minfty}, when computing the matrix for any curve that ends in a spiral: the spiraling part can be replaced by $\widetilde{M}_\Delta$ directly. Similarly, suppose a curve \emph{begins} with a spiral around the closed path $\Delta = RmR\cdots R2R1$. Then this infinite spiral contributes a factor of $\widetilde{M}_\Delta^T$ to the beginning of the matrix product, with $y_*$ and $F_*$ given as before.

\subsection{Example: the planar 1-loop propagator}\label{ex:D2:head}
We can put these formulas to work for the planar 1-loop propagator. The curves for this amplitude are given in Section \ref{ex:D2:fan}. Evaluating the curve matrices gives:
\begin{align}
M_{C_1} &= \begin{bmatrix} 1+x & ~1+x+xy\\ x & ~x+xy \end{bmatrix},~ & M_{C_2} &= \begin{bmatrix} 1+y& ~1+y+xy \\ y & ~y+xy \end{bmatrix},\\
M_{S_1} &=  \begin{bmatrix} 1+x & ~~1 \\ x(1+y) & ~~1\end{bmatrix},~ & M_{S_2} &= \begin{bmatrix} 1+y & ~~1 \\ y(1+x) & ~~1\end{bmatrix}.
\end{align}
The headlight functions are
\begin{align}
\alpha_{C_1} &=  \max(0,x) +\max(0,y)- \max(0,x,x+y),\\
\alpha_{C_2} &=  \max(0,x) +\max(0,y)- \max(0,y,x+y),\\
\alpha_{S_1} &=  - x - \max(0,y) + \max(0,x),\\
\alpha_{S_2} &=  - y - \max(0,x) + \max(0,y).
\end{align}
Once again, using the $g$-vectors from Section \ref{ex:D2:fan}, we verify that these functions satisfy
\begin{align}
\alpha_C({\bf g}_D) = \left\{ \begin{matrix} 1 &~~\text{if $C=D$} \\ 0 & ~~\text{otherwise.} \end{matrix} \right.
\end{align}

\subsection{Example: the genus one 2-loop vacuum}\label{ex:markov:head}
\begin{figure}
\begin{center}
\includegraphics[width=0.45\textwidth]{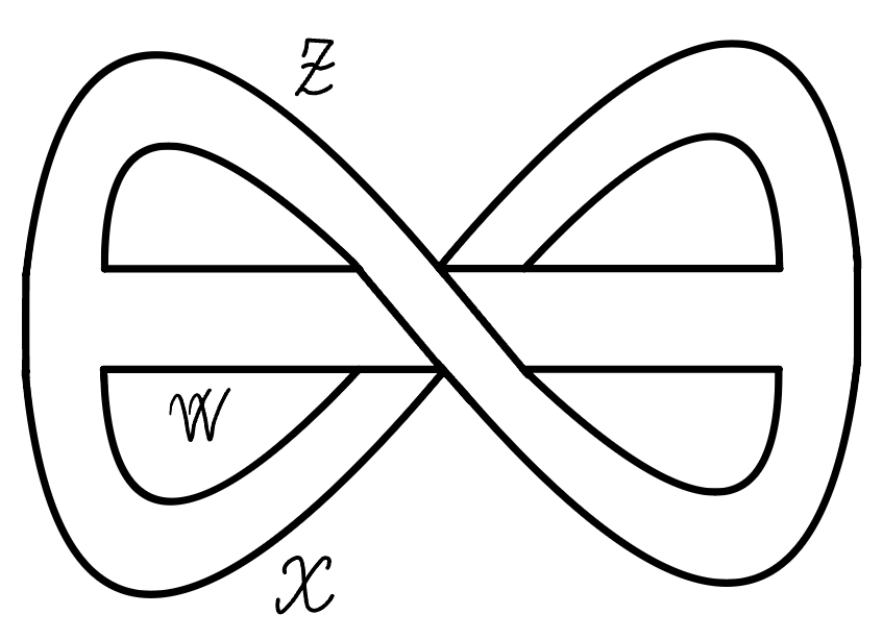}
\caption{The 2-loop vacuum graph with genus one.}
\label{fig:markovgraph}
\end{center}
\end{figure}
We now introduce a more complicated example: the 2-loop vacuum amplitude at genus one. A fatgraph for this amplitude, $\Gamma$, is given in Figure \ref{fig:markovgraph}. The colour factor of this graph has only one factor, $\text{tr}(1)$, because $\Gamma$ only has one boundary. In fact, the curves on $\Gamma$ must all begin and end in spirals around this one boundary. Using Figure \ref{fig:markovgraph} we can identify the curves which have precisely \emph{one valley} in their mountainscape: i.e. which only have one switch from turning right to turning left. These three curves are
\begin{align}
C_{1/0} &= (RwRzRx)^\infty R (wLxLzL)^\infty,\\
C_{0/1} &= (RxRwRz)^\infty R (xLzLwL)^\infty,\\
C_{1/1} &= (RzRxRw)^\infty R (zLwLxL)^\infty.
\end{align}
These curves are non-intersecting and form a triangulation. The surface associated to $\Gamma$ is the torus with one puncture, and the labels we assign to these curves are inspired by drawing the curves on the torus, pictured as a quotient of a $\mathbb{Z}^2$ lattice.

Besides $C_{1/1}$, we find that the only other curve compatible with both $C_{1/0}$ and $C_{0/1}$ is
\begin{align}
C_{-1/1} = (RxRwRz)^\infty RxLzR (xLzLwL)^\infty.
\end{align}
This curve has a peak at $z$, but no peaks at either $x$ or $w$ (which is what would result in an intersection with $C_{1/0}$ or $C_{0/1}$).

As we will see later, the four curves $C_{1/0}, C_{0/1}, C_{1/1}, C_{-1/1}$ are all we need to compute the 2-loop vacuum genus one amplitude. Evaluating these curves' matrices gives
\begin{align}
M_{1/0} &= \begin{bmatrix} w(1+x+xz)^2 +(1-xzw)^2 & ~~1+x+xz \\ w(1+x+xz) & ~~1 \end{bmatrix},\\
M_{0/1} &= \begin{bmatrix} x(1+z+zw)^2 +(1-xzw)^2 & ~~1+z+zw \\ x(1+z+zw) & ~~1 \end{bmatrix},\\
M_{1/1} &= \begin{bmatrix} z(1+w+wx)^2 +(1-xzw)^2 & ~~1+w+wx \\ z(1+w+wx) & ~~1 \end{bmatrix},\\
M_{-1/1} &= \begin{bmatrix} (1+x+xz)^2+z(x+xz+xzw)^2 & ~1+x+2xz+xz^2+xz^2w \\ x(1+x+2xz+xz^2+xz^2w) & ~x(1+z) \end{bmatrix}.
\end{align}
Because we restrict to anti-clockwise spiraling curves, all the $g$-vectors lie in the half-plane $x+z+w\leq 0$. Restricting to this half-plane, the headlight functions are given by
 \begin{align}
  \alpha_{1/0} &= \max(0,-w-2\max(0,x,x+z)),\\
   \alpha_{0/1} &= \max(0,-x-2\max(0,z,z+w)),\\
    \alpha_{1/1} &= \max(0,-z-2\max(0,w,w+x)),
\label{eq:alphaMarkov}    
\end{align}
and
\begin{gather}
\alpha_{-1/1} = \max(0,z) + \max\left(2\max(0,x,x+z),z+2x+2\max(0,z,z+w)\right)\\ - \max(0,x,x+z,x+2z,x+2z+w).
\end{gather}
We again verify that these $\alpha_C$ are $1$ on their corresponding $g$-vectors.

\section{Integrand Curve Integrals}\label{sec:integrands}
We want to compute the partial amplitudes of our theory. For some fatgraph $\Gamma$, let ${\cal A}$ be the amplitude that multiplies the colour factor $c_\Gamma$.

The momentum assignment rule in Section \ref{sec:mom} defines one set of loop momentum variables for all propagators contributing to the amplitude, even beyond planar diagrams. This means that ${\cal A}$ can be obtained as the integral of a single \emph{loop integrand} ${\cal I}$:
\begin{align}\label{eq:glsymA}
{\cal A} = \int\left( \prod_{i=1}^L d^D \ell_i \right) {\cal I}.
\end{align}
However, beyond planar diagrams, there is a price to pay for introducing our momentum assignment. For any triangulation by curves, $C_1,C_2,...,C_E$, we associate the product of propagators
\begin{align}\label{eq:term}
\frac{1}{X_{C_1}X_{C_2}\ldots X_{C_E}},
\end{align}
where $X_C$ is given by the momentum assignment rule. If we sum over every such term, \eqref{eq:term}, for all triangulations of $\Gamma$, we obtain some rational function ${\cal I}_\infty$. But the loop integral of ${\cal I}_\infty$ is not well defined if $\Gamma$ has a nontrivial mapping class group, $\MCG$. This is because two triangulations related by the $\MCG$ action integrate to the \emph{same} Feynman diagram. So the loop integral of ${\cal I}_\infty$ contains, in general, infinitely many copies of each Feynman integral.

Fortunately, we can compute integrands ${\cal I}$ for the amplitude by `dividing by the volume of $\MCG$'. As a function, ${\cal I}$ is not uniquely defined. But all choices for ${\cal I}$ integrate to the same amplitude.

We will compute integrands ${\cal I}$ using the headlight functions, $\alpha_C$. The formula takes the form of a \emph{curve integral},
\begin{align}\label{eq:glsymI}
{\cal I} = \int \frac{d^E{t}}{\MCG} \, e^{-S({\bf t})}.
\end{align}
Here, $E$ is the number of edges of the fatgraph $\Gamma$. We call it a \emph{curve integral} because the integral is over the $E$-dimensional vector space, $V$, whose integral points correspond to curves (or collections of curves) on $\Gamma$. As discussed in Section \ref{sec:fansyms}, the mapping class group $\MCG$ has a piecewise linear action on $V$, and we mod out by this action in the integral. We call $S(t)$ the \emph{curve action}. It is given by a sum
\begin{equation}
    S({\bf t}) = \sum_C \alpha_C(\mathbf{t}) X_C,
\end{equation}
where we sum over all curves, $C$, on the fatgraph.\footnote{We exclude \emph{closed curves} from this sum. Including the closed curves corresponds to coupling our colored field to an uncolored scalar particle. For simplicity, we delay the discussion of uncolored amplitudes} For a general derivation of this curve integral formula, see Appendix \ref{sec:app:ci}. In this section, we show how to practically use \eqref{eq:glsymI} to compute some simple amplitudes.

In fact, \eqref{eq:glsymI} also makes the loop integrals easy to do. This leads to a direct curve integral formula for the amplitude ${\cal A}$, which we study in Section \ref{sec:amplitudes}.

Later, in Section \ref{sec:recursion}, we also show that the integrands ${\cal I}$ can be computed recursively, starting from the curve integral formula, \eqref{eq:glsymI}. This result generalises the standard \emph{forward limit} method for 1-loop amplitudes to \emph{all} orders in the perturbation series.

\subsection{Example: the tree level 5-point amplitude}\label{ex:A2:int}
Curve integrals give new and simple amplitude formulas, even at tree level. Take the same fatgraph studied in Sections \ref{ex:A2:fan}, \ref{ex:A2:head} and \ref{ex:A2:int}. The kinematic variables for the curves on this graph are ($i<j-1$)
\begin{equation}
X_{ij} = (k_i+...+k_{j-1})^2+m^2.
\end{equation}
Then the amplitude, given by \eqref{eq:glsymA}, is
\begin{align}
  {\cal A}(12345) = \int dy_1dy_2\,Z,
\end{align}
where
\begin{align}
- \log Z = \alpha_{13} X_{13} + \alpha_{14} X_{14} + \alpha_{24} X_{24} + \alpha_{25} X_{25} + \alpha_{35} X_{35}.
\end{align}
Using the formulas for $\alpha_{ij}$ from Section \ref{ex:A2:head}, $Z$ can be further simplified to
\begin{align}
\log Z = X_{25}\,x + X_{35}\,y + s_{13}f_{13} + s_{14}f_{14} + s_{24}f_{24},
\end{align}
where $s_{ij} = 2k_i\cdot k_j$ and the $f_{ij}$ are the simple functions
\begin{equation}
    f_{13} = \max(0,x),\qquad f_{14} = \max(0,x,x+y),\qquad f_{24} = \max(0,y).
\end{equation}
The 5-point amplitude is then
\begin{align}
  {\cal A}(12345) = \int dy_1dy_2\,\exp \left( X_{25}\,x + X_{35}\,y + s_{13}f_{13} + s_{14}f_{14} + s_{24}f_{24} \right).
\end{align}
It is already interesting to note that the formula for the amplitude has been written in terms of the simple functions $f_{13},f_{14},f_{24},y_1,y_2$, and the Mandelstam invariants $s_{ij}$. These $s_{ij}$ are automatically summed together by the formula to form the appropriate poles of the tree level amplitude.

\subsection{Example: the planar 1-loop propagator}\label{ex:D2:int}
Consider again the 1-loop planar propagator (Sections \ref{ex:D2:fan} and \ref{ex:D2:head}). The amplitude is
\begin{align}
{\cal A} = \int d^D\ell \int\limits_{x+y\leq 0} dx dy {Z},
\end{align}
where
\begin{align}
- \log {Z} = \alpha_{C_1} X_{C_1} + \alpha_{C_2} X_{C_2} + \alpha_{S_1} X_{S_1}  + \alpha_{S_2} X_{S_2}.
\end{align}
We can assign the momenta of the curves to be
\begin{align}
P_{C_1} = 0, ~~P_{S_1} = \ell, ~~P_{S_2} = \ell+k, ~~P_{C_2} = 0.
\end{align}
Substituting these momenta (with $k^2+m^2=0$) into the integrand gives
\begin{align}
-\log {Z} = \ell^2(\alpha_{S_1}+\alpha_{S_2}) + 2\ell\cdot k \alpha_{S_2} + m^2(\alpha_{C_1} + \alpha_{C_2}  + \alpha_{S_1}).
\end{align}
At this point, we can either integrate over $x+y\leq 0$, or do the loop integral. Doing the loop integral first is a Gaussian integral, which gives
\begin{align}\label{eqn:exD2}
{\cal A} = \int\limits_{x+y\leq 0} dx dy \left(\frac{\pi}{\alpha_{S_1}+\alpha_{S_2}} \right)^{\frac{D}{2}} \exp \left(k^2 \frac{\alpha_{S_2}^2}{\alpha_{S_1}+\alpha_{S_2}} - m^2(\alpha_{C_1} + \alpha_{C_2}  + \alpha_{S_1}) \right).
\end{align}
This resembles the Symanzik formula for a single Feynman integral, but instead includes contributions from all three Feynman diagrams for this amplitude. Finally, substituting the headlight functions gives
\begin{align}
{\cal A} = \int\limits_{x+y\leq 0} dx dy \left(\frac{-\pi}{x+y} \right)^{\frac{D}{2}} \exp \left[m^2 \frac{(y+\max(0,x)-\max(0,y))^2}{x+y} - m^2|x| \right].
\end{align}

It is not immediately obvious that this reproduces the Feynman integrals for this amplitude. But note that, for example, restricting the domain of the integral to the negative orthant gives
\begin{align}
\int\limits_{x,y\leq 0} dx dy \left(\frac{-\pi}{x+y} \right)^{\frac{D}{2}} \exp \left(m^2 \left( \frac{y^2}{x+y} +x \right) \right).
\end{align}
After writing
\begin{equation}
    \frac{y^2}{x+y}+x = - \frac{xy}{x+y} +(x+y),
\end{equation}
this recovers the Feynman integral for the bubble graph. By extending the integral to the full region, $x+y\leq 0$, we recover not just this bubble integral, but the full amplitude!

\subsection{Example: the planar 1-loop 3-point amplitude}\label{ex:D3:int}
\begin{figure}
\begin{center}
\includegraphics[width=0.35\textwidth]{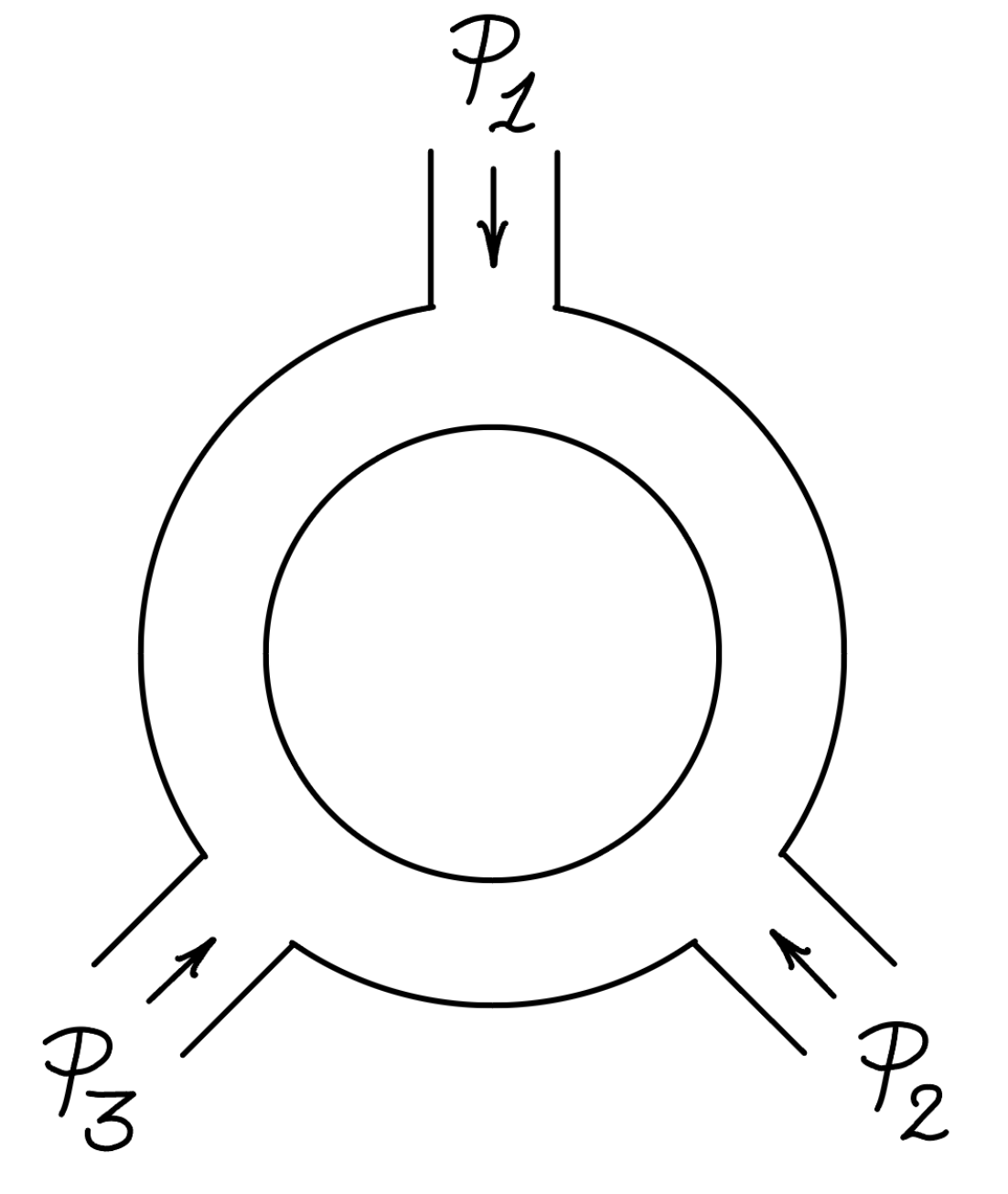}
\caption{A fatgraph for the 3-point 1-loop planar amplitude.}
\label{fig:D3graph}
\end{center}
\end{figure}

For a more complicated planar example, consider the 1-loop planar 3-point amplitude, with the fatgraph $\Gamma$, in Figure \ref{fig:D3graph}. There are nine curves on this graph: three curves $C_{i,i+2}$, connecting external lines $i,i+2$; three curves $C_{i,i}$, which loop around and come back to external line $i$; and three curves $C_{i,0}$ that start from the external line $i$ and end in a spiral around the closed loop.

In the planar sector, a convenient way to assign momenta is to use \emph{dual variables}. Let $z_i^\mu$ ($i=1,2,3$) be dual variables for the external lines, and $z_0$ be the dual variable for the closed loop. Then curves from external lines $i$ to $j$ have
\begin{equation}
    X_{i,j} = (z_j-z_i)^2+m^2,
\end{equation}
whereas a curve from $i$ that ends in a spiral around the loop has
\begin{equation}
    X_{i,0} = (z_i-z_0)^2 +m^2.
\end{equation}
If the external momenta are $p_1,p_2,p_3$, then we can take $z_1=0,z_2=p_1,z_3=p_1+p_2$. The closed loop variable, $z_0$, can be used as a loop momentum variable.

The 3-point one-loop planar amplitude is then
\begin{align}
    \mathcal{A} = \int d^D z_0 \int\limits_{\sum t_i \ge 0} d{\bf t} \, {Z},
\end{align}
where (taking cyclic indices mod $3$) 
\begin{align}
   -\log Z = \sum_{i=1}^3\alpha_{i,i+2}X_{i,i+2} + \sum_{i=1}^3 \alpha_{i,i} X_{i,i} + \sum_{i=1}^3 \alpha_{i,0} X_{i,0}.
\end{align}
The headlight functions for these curves are
\begin{align}\label{eq:planar1loop3param}
    &\alpha_{i,0} = t_i + g_{i+1} - g_{i}, \\
    &\alpha_{i,i+2} = g_i - f_i - f_{i+1}, \\
    &\alpha_{i,i} =  f_{i+1} + h_{i}-g_i - g_{i+1},
\end{align}
where
\begin{align}
f_i &= \max(0,t_i), \\
g_i &= \max(0,t_i,t_i+t_{i+1}), \\
h_i &= \max(0,t_i,t_{i}+t_{i+1},t_{i}+t_{i+1}+t_{i+2}).
\end{align}


\subsection{Note on factorization}\label{sec:factorization}
The integrands defined by curve integrals factorise in the correct way. Take again the curve integral
\begin{align}
{\cal I} = \int \frac{d^Et}{\MCG}\, Z.
\end{align}
In Appendix \ref{app:fact}, we show that the residue at $X_C=0$ is given by
\begin{align}\label{eqn:factfin}
\text{Res}_{X_C=0}\, {\cal I} = \int \frac{d^{E-1}t}{\MCG'} Z',
\end{align}
which is now the curve integral for the fatgraph $\Gamma_C$, obtained by cutting $\Gamma$ along $C$. In this formula, $\MCG'$ is the $\MCG$ of $\Gamma_C$, and the momentum $P_C^\mu$ of the curve $C$ is put on shell. In the fatgraph $\Gamma_C$, the curve $C$ gives two new boundaries, which are assigned momenta $\pm P_C^\mu$.

For example, before loop integration, the non-planar 1-loop fatgraph $\Gamma$ has loop integrand
\begin{align}
{\cal I} = \int dx dy\, \exp\left( - \sum_{n=-\infty}^\infty \alpha_n X_n \right).
\end{align}
Here, the momenta of the curves are $P_n^\mu = \ell^\mu + n k^\mu$. Consider the $X_0 = 0$ pole. The parameter $\alpha_0$ vanishes outside $x\geq 0$. In this region, the only non-vanishing parameters are $\alpha_1$ and $\alpha_{-1}$. The residue at $X_0=0$ is then
\begin{align}
\text{Res}_{X_{0}=0} {\cal I} = \int dy \, \exp\left( - \alpha_1' X_1 - \alpha_{-1}' X_{-1} \right),
\end{align}
where the restriction to $x=0$ gives $\alpha_1' = \max(0,y)$ and $\alpha_{-1}' = y - \max(0,y)$. This is the $n=4$ tree level amplitude, with external momenta are $k, \ell, -k, -\ell$, and $\ell^\mu$. The two propagators are $X_1 = (k+\ell)^2 +m^2$ and $X_{-1}=(k-\ell)^2+m^2$.

\section{Amplitude Curve Integrals}\label{sec:amplitudes}
Following the previous section, the curve integral formula for the full amplitude is
\begin{equation}
    {\cal A} = \int \frac{d^E\mathbf{t}}{\MCG} \int \left( \prod d^D \ell_a \right) \exp (-S({\bf t})). 
\end{equation}
The loop integration variables, $\ell_a$, appear quadratically in the curve action $S({\bf t})$. So, if we perform the loop integral \emph{before} performing the curve integral over the $t_i$, it is a Gaussian integral. The result is a curve integral
\begin{align}\label{eq:glsym20}
  {\cal A} = \int \frac{d^E\mathbf{t}}{\MCG} \,\left( \frac{\pi^L}{\mathcal{U}} \right)^{\frac{D}{2}} \exp\left(-\frac{\mathcal{F}_0}{\mathcal{U}} - \mathcal{Z} \right),
  \end{align}
where $\mathcal{U}, \mathcal{F}_0$ and $\mathcal{Z}$ are homogeneous polynomials in the $\alpha_C$'s that we call \emph{surface Symanzik polynomials}.

The curve integral \eqref{eq:glsym20} resembles the Schwinger form of a single Feynman integral, but it integrates to the full amplitude. Once again, it is important to mod out by the action of the mapping class group, to ensure that the integral does not overcount Feynman diagrams.

We now summarise how to compute the surface Symanzik polynomials, ${\cal U}, {\cal F}_0, {\cal Z}$. Suppose that a choice of loop momentum variables, $\ell_a^\mu$, has been fixed. The momentum assigned to a curve $C$ is of the form
\begin{equation}
    P_C^\mu = K_C^\mu + \sum h_C^a \ell_a^\mu,
\end{equation}
for some integers $h_C^a$. These $h_C^a$ geometrically can be understood in terms of intersections between $C$ and a basis of $L$ closed curves on the fatgraph. Using the $h_C^a$ intersection numbers, define an $L\times L$ matrix
\begin{equation}
    A^{ab} = \sum_C h_C^a h_C^b \alpha_C,
\end{equation}
and a $L$-dimensional vector (with momentum index $\mu$)
\begin{equation}
    B^{a,\mu} = \sum_C h_C^a \alpha_C K_C^\mu.
\end{equation}
The then surface Symanzik polynomials are
\begin{equation}\label{eq:surfsyms}
    \mathcal{U} = \det A,\qquad \frac{\mathcal{F}_0}{\cal U} = B^a_\mu \left(A^{-1}\right)_{ab} B^{b,\mu},\qquad \mathcal{Z} = \sum_C \alpha_C \left( K_C^2 + m^2 \right).
\end{equation}
These arise in the usual way by performing the Gaussian integral, as discussed in detail in Appendix \ref{sec:glsymder}.

In fact, the surface Symanzik polynomials have simple expressions when expanded as a sum of monomials. For a set of curves, $\mathcal{S} = \{C_1,...,C_L\}$, write $\alpha_\mathcal{S}$ for the corresponding monomial
\begin{align}
\alpha_\mathcal{S} = \prod_{i=1}^L \alpha_{C_i}.
\end{align}
The determinant, $\det A$, can be expanded to give
\begin{align}\label{eq:Uexpand}
 \mathcal{U} = \sum_{\substack{\mathcal{S}~\text{cuts}\,\Sigma\\\text{to disk}}} \alpha_\mathcal{S} \, ,
 \end{align}
 where we sum over all sets $\mathcal{S}$ whose curves cut $\Gamma$ down to a tree fatgraph. In other words, ${\cal U}$ is the sum over all \emph{maximal cuts} of the graph $\Gamma$. Moreover, using the Laplace expansion of the matrix inverse, $\mathcal{F}_0$ can be expanded to find
\begin{align}\label{eq:F0def}
 \mathcal{F}_0 = \sum_{\substack{\mathcal{S}'~\text{cuts}~\Sigma\\ \text{to 2 disks}}} \alpha_{\mathcal{S}'} \left( \sum_{C\in \mathcal{S}'}  K_C^\mu \right)^2,
 \end{align}
where the sum in this formula is now over sets $\mathcal{S}'$ of $L+1$ curves that factorise $\Gamma$ into two disjoint tree graphs. Each monomial in the sum is multiplied by the total momentum flowing through the factorisation channel.

A complete derivation of \eqref{eq:Uexpand} and \eqref{eq:F0def} is given in Appendix \ref{sec:glsymder}.

\subsection{Example: the planar 1-loop propagator}\label{ex:D2:amp}
We return to the planar 1-loop propagator (Sections \ref{ex:D2:fan}, \ref{ex:D2:head}, \ref{ex:D2:int}). Of the four curves $C_1,C_2,S_1,S_2$, only $S_1$ and $S_2$ carry loop momentum and cut $\Gamma$ open to a tree. The first surface Symanzik polynomial is therefore
\begin{align}
\mathcal{U} = \alpha_{S_1} + \alpha_{S_2} .
\end{align}
The $B$-vector is
\begin{equation}
    B^\mu = \alpha_{S_2} k^\mu,
\end{equation}
so that the second surface Symanzik polynomial is
\begin{align}
\mathcal{F}_0 = \alpha_{S_2}^2 k^2.
\end{align}
Finally,
\begin{equation}
    {\cal Z} = m^2 (\alpha_{S_1}+\alpha_{C_1}+\alpha_{C_2}).
\end{equation}
The amplitude is then given by the curve integral
\begin{align}
{\cal A} = \int\limits_{x+y\geq 0} dxdy \left( \frac{\pi}{\alpha_{S_1} + \alpha_{S_2}} \right)^{\frac{D}{2}} \exp\left(\frac{\alpha_{S_2} k^2}{\alpha_{S_1} + \alpha_{S_2}} - m^2 \left(\alpha_{S_1}+\alpha_{C_1}+\alpha_{C_2}\right) \right).
\end{align}
This again recovers the formula \eqref{eqn:exD2}, which we obtained by direct integration in the previous section.

\subsection{Example: the non-planar 1-loop propagator}\label{ex:A11:amp}
We return to the non-planar 1-loop propagator (Sections \ref{ex:A11:fan} and \ref{ex:A11:head}). The momentum of the curve $C_n$ is
\begin{align}
P_n^\mu = \ell^\mu + n p^\mu.
\end{align}
Every curve $C_n$ cuts $\Gamma$ to a tree graph with 4 external legs. So the first Symanzik polynomials is
\begin{align}
\mathcal{U} = \sum_{n=-\infty}^\infty \alpha_n,
\end{align}
where $\alpha_n$ is the headlight function for $C_n$. Every pair of distinct curves $C_n,C_m$ cuts $\Gamma$ into two trees, and so
\begin{align}
\mathcal{F}_0 = \sum_{n,m=-\infty}^\infty nm \alpha_n\alpha_m p^2.
\end{align}
Finally,
\begin{align}
\mathcal{Z} = \sum_{n=-\infty}^\infty \alpha_n (m^2+n^2p^2).
\end{align}
The amplitude is then
\begin{align}
  {\cal A} = \int \frac{dxdy}{\MCG} \left( \frac{\pi}{\mathcal{U}} \right)^{\frac{D}{2}} \exp\left(-\frac{\mathcal{F}_0}{\mathcal{U}} - \mathcal{Z} \right).
  \end{align} 
The MCG acts on the fan in this case as ${\bf g}_n \mapsto {\bf g}_{n+1}$. A fundamental domain for this action is clearly the positive orthant, spanned by ${\bf g}_0,{\bf g}_1$. In this orthant, the surface Symanzik polynomials are
\begin{align}
\mathcal{U} & = x+ y,\\
\mathcal{F}_0 &= y^2 p^2,\\
\mathcal{Z} & = x m^2.
\end{align}
So we find
\begin{align}
{\cal A} = \int\limits_{x,y\geq 0} dx dy \left(\frac{\pi}{x+y} \right)^{D/2} \exp\left( m^2 \left( - \frac{y^2}{x+y} - x \right) \right),
\end{align}
where we have put $p^\mu$ on shell, $p^2+m^2=0$. Or, equivalently,
\begin{align}
{\cal A} = \int\limits_{x,y\geq 0} dx dy \left(\frac{\pi}{x+y} \right)^{D/2} \exp\left( -p^2\frac{xy}{x+y} - m^2 (x+y) \right).
\end{align}

\subsection{Example: The non-planar 3-point amplitude}\label{ex:A21:amp}
Even at 1-loop, it is not always easy to identify the fundamental domain of the MCG. To see the problem, consider the non-planar one-loop 3-point amplitude. Let the first trace factor have external particle $p_1^\mu$, and the second trace factor have $p_2^\mu$ and $p_3^\mu$. The curves, $C_{ij}^n$, connecting a pair of distinct start and end points, $i,j$, are labelled by the number of times, $n$, they loop around the graph. The curves $C_{22}$ and $C_{33}$ begin and end at the same edge, and are invariant under the $\MCG$. Then, for a specific choice of loop momentum variable, we find the momentum assignments
\begin{equation}
    P_{12}^n = n p_1^\mu,\qquad P_{13}^n = np_1^\mu - p_2^\mu, \qquad P_{22} = 0,\qquad P_{33}=0.
\end{equation}

We can readily give the curve integral formula for the amplitude,
\begin{align}
{\cal A} = \int \frac{dxdydz}{\MCG} \left( \frac{\pi}{\mathcal{U}} \right)^{\frac{D}{2}} \exp\left(-\frac{\mathcal{F}_0}{\mathcal{U}} - \mathcal{Z} \right),
\end{align}
where the surface Symanzik polynomials are
\begin{equation}
\mathcal{U} = \sum_{n=-\infty}^\infty \alpha_{13}^n + \alpha_{12}^n,\qquad \mathcal{F}_0 = B^\mu B^\mu, \qquad {\cal Z} = m^2\left( \alpha_{22}+\alpha_{33} + \sum_{n=-\infty}^\infty \alpha_{12}^n \right).
\end{equation}
In the formula for ${\cal F}_0$, the $B$-vector is
\begin{equation}
B^\mu = \sum_{n=-\infty}^\infty n p_1^\mu \alpha_{12}^n + (np_1^\mu - p_2^\mu) \alpha_{13}^n.
\end{equation}

However, at this point we confront the problem of quotienting by $\MCG$. The MCG is generated by
\begin{align}
{\bf g}_{12}^n \mapsto {\bf g}_{12}^{n+1}, ~ {\bf g}_{13}^n \mapsto {\bf g}_{13}^{n+1},
\end{align}
and it leaves ${\bf g}_{22}$ and ${\bf g}_{33}$ invariant. Naively, we might want to quotient by the MCG by restricting the integral to the region spanned by: ${\bf g}_{12}^0,{\bf g}_{13}^0,{\bf g}_{22},{\bf g}_{33}$. However, this region is too small. It does not include any full cones of the Feynman fan. We could also try restricting the integral to the region spanned by: ${\bf g}_{12}^0,{\bf g}_{13}^0,{\bf g}_{12}^1,{\bf g}_{13}^1,{\bf g}_{22},{\bf g}_{33}$. But this region is too large! The amplitude has \emph{three} Feynman diagrams, but this region contains \emph{four} cones, so it counts one of the diagrams twice. 

As this example shows, it is already a delicate problem to explicitly specify a fundamental domain for the MCG action.

\subsection{Example: genus-one 2-loop amplitudes}\label{ex:markov:amp} 
The problem of modding by $\MCG$ becomes even more acute for non-planar amplitudes. The genus one 2-loop vacuum amplitude, considered in Section \ref{ex:markov:head}, is computed by a 3-dimensional curve integral. But the $\MCG$ action in this case is an action of ${\rm SL}_2\mathbb{Z}$. The action on $g$-vectors is of the form
\begin{equation}
    {\bf g}_{p/q} \mapsto {\bf g}_{(ap+bq)/(cp+dq)},\qquad \text{for}~\begin{bmatrix} a~&b\\c~&d\end{bmatrix} \in{\rm SL}_2\mathbb{Z}. 
\end{equation}
For the vacuum amplitude, a simple example of a fundamental region is the region spanned by ${\bf g}_{1/0},{\bf g}_{0/1},$ and ${\bf g}_{1/1}$. However, for the $n$-point genus one 2-loop amplitude, identifying a fundamental region of this ${\rm SL}_2\mathbb{Z}$-action becomes very difficult.

In the next section, we present a simple method to compute the integrals in our formulas, for any $\MCG$ action.

\section{Modding Out by the Mapping Class Group}\label{sec:mirz}
Our formulas for amplitudes and integrands take the form of integrals over $\mathbb{R}^E$ modulo the action of the Mapping Class Group, MCG,
\begin{equation}
    {\cal A} = \int \frac{d^E t}{\MCG}\, f(t),
\end{equation}
for some $\MCG$-invariant function, $f(t)$. One way to evaluate this integral is to find a fundamental domain for the MCG action. But it is tricky to identify such a region in general. Instead, it is convenient to mod out by the MCG action by defining a kernel, $\mathcal{K}$, such that
\begin{equation}
    {\cal A} = \int d^E t\, \mathcal{K}(t) f(t).
\end{equation}
In this section, we find kernels, $\mathcal{K}$, that can be used at all orders in perturbation theory, for all Mapping Class Groups.

\subsection{Warm up}
Consider the problem of evaluating an integral modulo a group action on its domain. For example, suppose $f(x)$ is invariant under the group of translations, $T$, generated by $x\mapsto x+a$, for some constant, $a$. We want to evaluate an integral
\begin{equation}\label{eq:quotex}
I = \int\limits_{\mathbb{R}/T} dx f(x).
\end{equation}
One way to do this is to restrict to a fundamental domain of $T$:
\begin{equation}
I = \int\limits_0^a dx f(x).
\end{equation}
But we can alternatively find a kernel $\mathcal{K}(x)$ such that
\begin{equation}
I = \int\limits_{-\infty}^\infty dx\, \mathcal{K}(x) f(x).
\end{equation}
One way to find such a kernel is to take a function $g(x)$ with finite support around $0$, say. Then we can write
\begin{equation}
    1 = \frac{\sum_{n=-\infty}^\infty g(x-na)}{\sum_{n=-\infty}^\infty g(x-na)},
\end{equation}
provided that $\sum_{n=-\infty}^\infty g(x-na)$ is nowhere vanishing. Inserting this into \eqref{eq:quotex},
\begin{equation}\label{eq:quotexfin}
I = \int\limits_{\mathbb{R}/T} dx\, \frac{\sum_{n=-\infty}^\infty g(x-na)}{\sum_{n=-\infty}^\infty g(x-na)} f(x) = \int\limits_{-\infty}^\infty dx\, \frac{g(x)}{\sum_{n=-\infty}^\infty g(x-na)} f(x).
\end{equation}
So that we can use
\begin{equation}
    \mathcal{K}(x) = \frac{g(x)}{\sum_{n=-\infty}^\infty g(x-na)}
\end{equation}
as a kernel to quotient out by the translation group. For example, suppose that we take $g(x) = \Theta(x+a)\Theta(-x+a)$, where $\Theta(x)$ is the Heaviside function. Inserting this into \eqref{eq:quotexfin} gives
\begin{equation}
    I = \int\limits_{-a}^a dx\, \frac{1}{2} f(x).
\end{equation}
The domain of this integral contains two copies of a fundamental domain for $T$, but this is compensated for by the $1/2$ coming from $\mathcal{K}(x)$ to give the correct answer.

\subsection{A Tropical Mirzakhani kernel}
The headlight functions, $\alpha_C$, give a very natural solution to the problem of defining an integration kernel, $\mathcal{K}$.

Consider the case when $\MCG$ has \emph{one generator}. Let $\mathcal{S}$ be the set of curves which are \emph{not} invariant under $\MCG$. The sum of their headlight functions,
\begin{align}
\rho = \sum_{C\in\mathcal{S}} \alpha_C,
\end{align}
is itself a $\MCG$-invariant function. Moreover, $\rho$ does not vanish on any top-dimensional cone (because no diagram can be formed without using at least one propagator from $\mathcal{S}$). So we can consider inserting the function
\begin{align}
1 = \frac{\rho}{\rho}
\end{align}
into our integrals.

The set $\mathcal{S}$ is the disjoint union of cosets under the MCG action, by the Orbit-Stabilizer theorem. When $\MCG$ has a single generator, these cosets are easy to describe. $\MCG$ does not alter the endpoints of curves. So if $C_{ij} \in \mathcal{S}$ is a curve connecting external lines $i$ and $j$, the orbit of $C_{ij}$ is a coset of $\mathcal{S}$. By the Orbit-Stabalizer theorem, these cosets are disjoint. So $\rho$ can be resumed as
\begin{align}
\rho = \sum_{i,j} \sum_{\gamma \in \MCG} \alpha_{\gamma C_{ij}}.
\end{align}
Given this, we can mod out by the $\MCG$ action by defining
\begin{align}\label{eq:mirzsing}
\mathcal{K} = \sum_{i,j} \frac{\alpha_{C_{ij}}}{\rho},
\end{align}
where we choose a distinguished representative, $C_{ij}$, for each coset. We call \eqref{eq:mirzsing} a \emph{tropical Mirzakhani kernel}, because it is a tropical version of the kernel introduced by Mirzakhani to compute Weil-Petersson volumes \cite{mirzakhani2007}. Each headlight function, $\alpha_{C_{ij}}$, is non-vanishing in a convex region $V_{C_{ij}}$ that is spanned by all the cones in the fan that contain ${\bf g}_{C_{ij}}$. These regions \emph{over-count} the diagrams, but this over-counting is corrected by the kernel, $\mathcal{K}$.

\subsection{Example: the non-planar 1-loop propagator}\label{ex:A11:mirz}
As a sanity check, let us repeat the calculation of the non-planar 1-loop propagator from Section \ref{ex:A11:amp}, but now using the tropical Mirzakhani kernel. The $\MCG$ has one generator, and no curves are $\MCG$-invariant. So take the set $\mathcal{S}$ to be the set of all curves, $C_n$, and write
\begin{align}
    \rho = \sum_{n =-\infty}^\infty \alpha_n.
\end{align}
Choose $C_0$, say, as the coset representative (all other curves are in the orbit of $C_0$). Then the tropical Mirzakhani kernel, \eqref{eq:mirzsing}, is
\begin{equation}
    \mathcal{K} = \frac{\alpha_0}{\rho}.
\end{equation}

Using this kernel, we find a pre-loop-integration integrand,
\begin{align}
\mathcal{I} = \int dxdy\, \mathcal{K}(x,y)\, \mathrm{exp}\left(-\sum_{i=-\infty}^\infty \alpha_i X_i\right).
\end{align}
The headlight functions for this example were given in \eqref{eq:alphaA11}. In particular, $\alpha_0 =  \max(0,x)$, which is vanishing outside of the region $x\geq 0$. In this region, the only other non-vanishing headlight functions are
\begin{align}
\alpha_{-1} = \mathrm{max}(0,y)\qquad\text{and}\qquad \alpha_{1} =- y+\mathrm{max}(0,y).
\end{align}
The formula is therefore
\begin{align}\label{eq:A11exmirz1}
\mathcal{I} = \int\limits_{x\geq 0}  dxdy\,  \frac{x}{x+|y|}  \mathrm{exp}\left(-\alpha_{-1}X_{-1} - \alpha_0X_0 - \alpha_1X_1\right).
\end{align}
We can now perform the loop integral. Recall that $X_n = (\ell + n k)^2+m^2$. Using this, the exponent, $Z$, in \eqref{eq:A11exmirz1} is
\begin{align}
     - \log Z = \rho\, \ell^2 + 2 \ell \cdot k (\alpha_1-\alpha_{-1}) + m^2 \alpha_0.
\end{align}
The Gaussian integral gives
\begin{align}\label{eq:A11exmirz2}
   {\cal A} = \int\limits_{x\geq 0} dxdy \frac{x}{x+|y|} \left(\frac{\pi}{x+|y|}\right)^{\frac{D}{2}} \mathrm{exp}\left( k^2 \frac{|y|^2}{x+|y|} - m^2x\right).
   \end{align}
This doesn't immediately look like the Feynman integral for the 1-loop bubble. However, writing
\begin{align}
\frac{2x}{x+y} = 1 + \frac{x-y}{x+y},
\end{align}
we find 
\begin{align}\label{eqn:A11exmir3}
{\cal A} =  \int\limits_{x,y\geq 0}dx dy \left( \frac{\pi} {x+y}\right)^{\frac{D}{2}} \exp\left(k^2\frac{y^2}{x+y}-m^2x\right).
\end{align}
since the integrand over $x,y\geq 0$ is even under $x\leftrightarrow y$, whereas $x-y$ is odd. This is still not exactly the same as the conventional integral. To recover the conventional form, note that the exponent can be rewritten as
\begin{equation}
    - \frac{y^2}{x+y} - x = \frac{xy}{x+y} - (x+y).
\end{equation}

\subsection{General Tropical Mirzakhani Kernels}\label{sec:mirz:U}
Tropical Mirzakhani kernels can be defined to \emph{any} mapping class group, with more than one generator. Fix some fatgraph $\Gamma$, with mapping class group $\MCG$.

A conceptually simple way to define a kernel is to consider the set of $L$-tuples of curves that cut $\Gamma$ to a tree graph. These define the \emph{first Symanzik polynomial},
\begin{equation}
\mathcal{U} = \sum_{\substack{S \\ \text{cuts to tree}}} \alpha_S,
\end{equation}
which can also be computed as a determinant of a matrix (Section \ref{sec:amplitudes}). This function does not vanish on top-dimensional cones of the Feynman fan, since every diagram contains a subset of propagators that cut $\Gamma$ to a tree. We can therefore insert
\begin{equation}
1 = \frac{\mathcal{U}}{\mathcal{U}}
\end{equation}
into our integrals. Under the $\MCG$ action, the set of $L$-tuples appearing in $\mathcal{U}$ is partitioned into cosets. Each coset represents an $\MCG$-inequivalent way of cutting $\Gamma$ down to a tree. By choosing a representative $L$-tuple for each such loop cut, we arrive at a kernel
\begin{equation}
\mathcal{K} = \sum_\text{distinct loop cuts} \frac{\alpha_S}{\mathcal{U}}.
\end{equation}
Our integrals can then be computed as a sum over maximal cuts:
\begin{equation}
    \mathcal{A} = \int \frac{d^Ey}{\MCG} {\cal I} = \sum_\text{distinct loop cuts} \int d^Ey \, \frac{\alpha_S}{\mathcal{U}}\, {\cal I}.
\end{equation}
The disadvantage of this formula is that it can be difficult to systematically identify a set of $\MCG$-inequivalent maximal cuts.

\subsection{The General Iterative Method}\label{sec:mirz:iter}
A more systematic way to quotient out by $\MCG$ is to break the $\MCG$-action one generator at a time. This iterative method has the advantage of being completely algorithmic. 

To apply the method, pick a trace-factor of $\Gamma$, $\beta$, which has some external particles, $1,...,m$. Let $\mathcal{S}_\beta$ be the set of curves that have at least one endpoint in $\beta$, excluding any curves that are $\MCG$-invariant, and write
\begin{align}\label{eq:rhodef}
\rho_\beta = \sum_{C \in \mathcal{S}_\beta} \alpha_C.
\end{align}
$\rho_\beta$ is $\MCG$-invariant. This is because the MCG action does not alter the endpoints of a curve. The set $\mathcal{S}_\beta$ therefore has a coset decomposition. For each MCG orbit in $\mathcal{S}_\beta$, pick a representative curve, so that
\begin{align}\label{eq:cosetexp}
\rho_\beta = \sum_{i=1}^{k}  \sum_{\gamma \in \mathrm{MCG}(\Sigma)} \alpha_{\gamma C_i},
\end{align}
for some $k=|\mathcal{S}_\beta / \text{MCG}(\Sigma)|$ coset representatives $C_1,...,C_k$. We give more details about how to pick a set of coset representatives below.

Every top-dimensional cone is generated by at least \emph{one} curve from the set $\mathcal{S}_\beta$, because otherwise that cone would not correspond to a complete triangulation of $\Gamma$. This means that $\rho_\beta$ is non-vanishing everywhere, except on some lower-dimensional cones. Away from this vanishing locus, we can write
\begin{align}\label{eq:fadpop}
1 = \frac{\rho_\beta}{\rho_\beta}.
\end{align}
Given this, we define a tropical Mirzakhani kernel
\begin{align}
\mathcal{K}_{\beta} = \sum_{i=1}^k \frac{\alpha_{C_i}}{\rho_\beta}.
\end{align}
This has the effect of breaking the MCG symmetry of the integrand, and reducing us to evaluating simpler integrals. In particular, we have
\begin{equation}
{\cal A} =  \int \frac{d^Et}{\MCG}\, \mathcal{I} = \sum_{i=1}^k\, \int \frac{d^E t}{\mathrm{Stab}(C_i)} \, \frac{\alpha_{C_i}}{\rho_\beta} \, \mathcal{I},
\end{equation}
where $\mathrm{Stab}(C_i)\leq \MCG$ is the \emph{stablizer subgroup} for $C_i$. The factor
\begin{equation}\label{eq:mirzCirho}
\frac{\alpha_{C_i}}{\rho_\beta}
\end{equation}
is \emph{itself} invariant under $\mathrm{Stab}(C_i)$. So the integrals,
\begin{equation}
    \int \frac{d^E t}{\mathrm{Stab}(C_i)} \, \frac{\alpha_{C_i}}{\rho} \, \mathcal{I},
\end{equation}
can themselves be evaluated by finding a Mirzkhani kernel for the new group, $\mathrm{Stab}(C_i)$. This iterative method ultimately yields an integral with no group action,
 \begin{equation}
 {\cal A} = \int \frac{d^Ey}{\MCG}\, \mathcal{I} =  \int d^n y \, \mathcal{K} \, \mathcal{I},
 \end{equation}
where $\mathcal{K}$ is a sum of products of kernels of the form \eqref{eq:mirzCirho}.

\begin{figure}
\begin{center}
\includegraphics[width=0.85\textwidth]{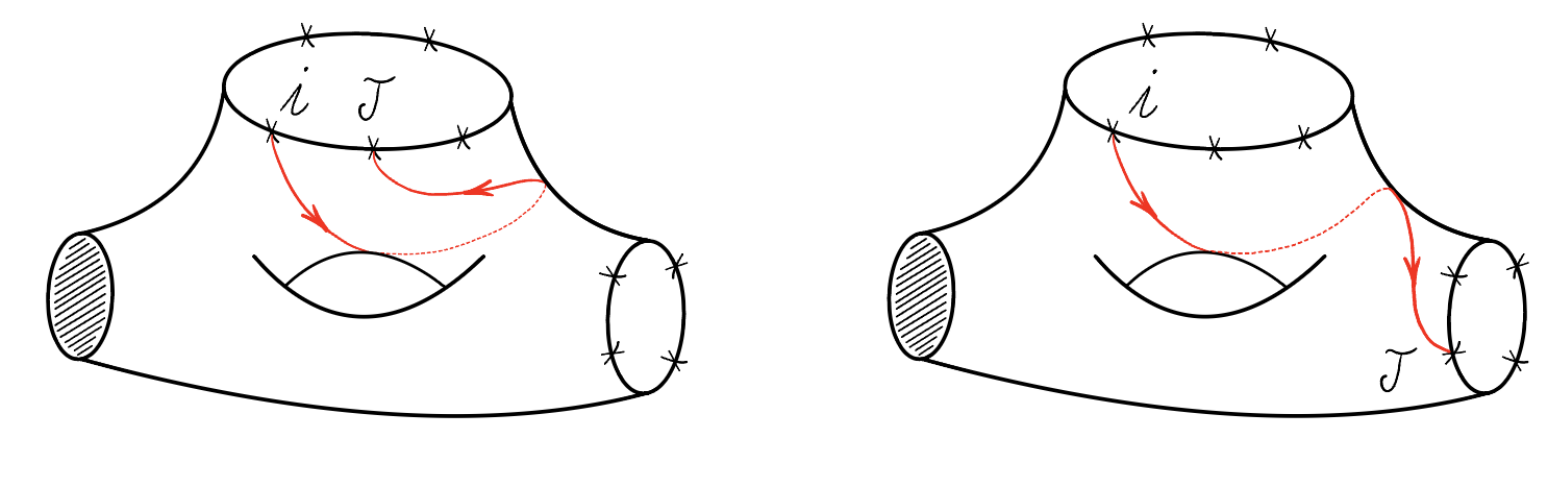}
\caption{The two types of curves that are not invariant under the MCG, drawn here on the surface $S(\Gamma)$ associated to a fatgraph: curves connecting distinct trace factors (right), and topologically nontrivial curves that begin and end on the same trace factor (left).}
\label{fig:mirzsets}
\end{center}
\end{figure}

To complete the description of the iterative method, we describe how to choose coset representatives from the set $\mathcal{S}_\beta$. The curves in this set break into two subsets, as in Figure \ref{fig:mirzsets}:
\begin{enumerate}
\item Curves $C$ whose endpoints lie in two distinct trace factors. These curves cut $\Gamma$ to a fatgraph $\Gamma_C$ which has one fewer trace factors.
\item Curves $C$ with both endpoints in the same trace factor. These curves cut $\Gamma$ to a fatgraph $\Gamma_C$ with one lower genus.
\end{enumerate}
Both of these subsets have decompositions into cosets specified by the endpoints of the curves. So, for every pair of particles, $i,j$ (with $i$ in trace factor $\beta$), pick \emph{any} curve $C_{ij}^0$ connecting them. These can be taken as coset representatives. The caveat is that, if $i,j$ are both in trace factor $\beta$, we must choose a curve $C_{ij}^0$ which is not $\MCG$-invariant. An $\MCG$-invariant curve generates a trivial coset. The first step to break the MCG is then to insert the kernel
\begin{equation}
    \sum_{i\in \beta}\sum_j \frac{\alpha_{ij}^0}{\sum_{{\cal S}_\beta}\alpha_C}.
\end{equation}

For amplitudes involving a large number of external particles, this iterative method naively requires a lot of work (growing like $n^L$ with the number of particles, $n$). However, this apparent complexity goes away completely if we choose an appropriate fatgraph, $\Gamma$, for our calculation. We use this to obtain simple formulas for amplitudes at all-$n$ in a separate paper, \cite{us_alln}. But for now we will focus on low-point amplitudes, to illustrate the method in its simplest form.

\subsection{Example: the genus one 2-loop vacuum amplitude}\label{ex:markov:mirz}
As an example, we briefly describe what happens for the genus one 2-loop vacuum amplitude (Sections \ref{ex:markov:head} and \ref{ex:markov:amp}). The $\MCG$ is now $\mathrm{SL}_2\mathbb{Z}$. In this case, there is only \emph{one} coset to consider, since every curve is related to every other by 
\begin{equation}
    {\bf g}_{p/q} \mapsto {\bf g}_{(ap+bq)/(cp+dq)},\qquad \text{for}~\begin{bmatrix} a~&b\\c~&d\end{bmatrix} \in{\rm SL}_2\mathbb{Z}. 
\end{equation}
For the first step of the iteration, we can take any curve, say $C_{1/0}$, as a coset representative. The kernel for the first step is
\begin{equation}
    {\cal K}_{1/0} = \frac{\alpha_{1/0}}{\sum_C \alpha_C}.
\end{equation}
The subgroup that leaves $C_{1/0}$ invariant is
\begin{equation}
   \text{Stab} \,C_{1/0} = \left\{ \begin{bmatrix} 1 & n \\ 0 & 1 \end{bmatrix} ~:~n\in\mathbb{Z} \right\} < {\rm SL}_2\mathbb{Z}.
\end{equation}
The curves compatible with $C_{1/0}$ form a single coset for the action of this subgroup. So, for the second step, we can choose just one of them, $C_{0/1}$, say, as a coset representative. The kernel for the second step is
\begin{equation}
    {\cal K}_{0/1} = \frac{\alpha_{0/1}}{\sum_{C'} \alpha_{C'}},
\end{equation}
where we sum only over curves, $C'$, that are non-intersecting with $C_{1/0}$. The final kernel is simply
\begin{equation}
    {\cal K} = \frac{\alpha_{1/0}}{\alpha_{1/0}+\alpha_{0/1}+\alpha_{1/1}+\alpha_{-1/1}} \,\frac{\alpha_{0/1}}{\alpha_{0/1}+\alpha_{1/1}+\alpha_{-1/1}},
\end{equation}
where the simplification arises because $C_{1/1}$ and $C_{-1/1}$ are the only curves compatible with both $C_{1/0}$ and $C_{0/1}$.

\section{Examplitudes}\label{sec:exam}
We now show how to use the tropical Mirzakhani kernels to evaluate curve integrals. We give detailed low-dimensional examples of amplitudes up to 3 loops.

\subsection{The non-planar 1-loop 3-point amplitude}\label{ex:A21:exam}
The formula for the 1-loop non-planar 3-point amplitude was given in Section \ref{ex:A21:amp}. However, we did not show how to quotient by the $\MCG$. Using the tropical Mirzakhani kernel, we now find the formula
\begin{align}\label{eq:exaA12-A}
    {\cal A} = \int d^3t\,  {\cal K}\, \left(\frac{\pi}{\cal U}\right)^{\frac{D}{2}}\,\exp \left( \frac{\mathcal{F}_0}{\mathcal{U}} - \mathcal{Z} \right),
\end{align}
where the Mirzakhani kernel is
\begin{equation}\label{eq:exaA12-K}
    {\cal K} = \frac{\alpha_{12}^0 + \alpha_{13}^0}{\rho},
\end{equation}
with $\rho$ the sum over all $\alpha_C$ (except for those curves which are invariant under the MCG, namely $C_{22}$, $C_{33}$). The surface Symanzik polynomials are, as before,
\begin{equation}
\mathcal{U} = \sum_{n=-\infty}^\infty \alpha_{13}^n + \alpha_{12}^n,\qquad \mathcal{F}_0 = B_\mu B^\mu, \qquad {\cal Z} = m^2\left( \alpha_{22}+\alpha_{33} + \sum_{n=-\infty}^\infty \alpha_{12}^n \right).
\end{equation}
In the formula for ${\cal F}_0$, the $B$-vector is
\begin{equation}
B^\mu = \sum_{n=-\infty}^\infty n p_1^\mu \alpha_{12}^n + (np_1^\mu - p_2^\mu) \alpha_{13}^n.
\end{equation}

Let us first see why \eqref{eq:exaA12-K} is a Mirzakhani kernel. The $\MCG$ has one generator. It leaves $C_{22}$ and $C_{33}$ invariant, but acts non-trivially on the set $\{ C_{12}^n, C_{13}^n \}$ of all curves that connect the first trace factor to the second trace factor. $\rho$ is the sum of $\alpha_C$ for all these curves,
\begin{align}
\rho = \sum_{n=-\infty}^\infty \left(\alpha_{12}^n + \alpha_{13}^n \right). 
\end{align}
This set has two $\MCG$ cosets, labelled by the start and end points of the curves. We can take $C_{12}^0$ and $C_{13}^0$ as the two coset representatives. $C_{12}^0$, for instance, represents the coset of all curves that begin at $1$ and end at $2$. (Recall Section \ref{sec:mirz}.)

Naively, it looks as if \eqref{eq:exaA12-A} involves infinitely many $\alpha_C$, which it would be laborious to compute. However, the Mirzakhani kernel ensures that only a few $\alpha_C$ are needed. To see how this works, consider, say, the first term in the kernel,
\begin{equation}\label{eq:exaA12-K12}
    {\cal K}_{12} = \frac{\alpha_{12}^0}{\rho}.
\end{equation}
In the region where $\alpha_{12}^0 \neq 0$, all other $\alpha_C$ are vanishing, except for:
\begin{align}
\alpha_{12}^{-1},~\alpha_{12}^1,~ \alpha_{13}^0,~\alpha_{13}^1,~\alpha_{22}.
\end{align}
So in this region, ${\cal U}$ and $B^\mu$ simplify to
\begin{align}
\mathcal{U} &= \alpha_{12}^0 +\alpha_{12}^1+ \alpha_{12}^{-1}+\alpha_{13}^0+\alpha_{13}^1,\\
B^\mu &=  - k_1^\mu \alpha_{12}^{-1} - k_2^\mu \alpha_{13}^0 + (k_1^\mu - k_2^\mu) \alpha_{13}^1.
\end{align}
When we compute these $\alpha$'s, using the matrix method, we find that they become simple functions in the region $x>0$, where $\alpha_{12}^0$ is non-zero. In this region, we have $\alpha_{12}^0 = x$. Moreover, the remaining 5 headlight functions become
\begin{align}
\alpha_{13}^1 &= - \max(0,y) + \max(0,y,y+z), & \alpha_{13}^0 &= \max(0,y),\\
\alpha_{12}^1 &= -y - \max(0,z) + \max(0,y,y+z),  & \alpha_{12}^{-1} &= -z + \max(0,z),\\
\alpha_{22} & = - \max(0,y,y+z) + \max(0,y) + \max(0,z). & 
\end{align}
These are precisely the headlight functions for the 5-point tree amplitude! We could have anticipated this, because cutting $\Gamma$ along $C_{12}^0$ yields a 5-point tree graph. Using these tree-like headlight functions, we can compute the contribution of ${\cal K}_{12}$ to the curve integral, \eqref{eq:exaA12-A}. The contribution from the second term in the Mirzakhani kernel is similar.

In this example, we find that we only need to know the headlight functions $\alpha_C$ for \emph{tree level} amplitudes, in order to compute the full 1-loop amplitude! In fact, we can prove that this happens \emph{in general}. Suppose a monomial, $\alpha_S$ (for some set of $L$ curves $S$), appears in the numerator of the kernel ${\cal K}$. In the region where $\alpha_S\neq 0$, all remaining $\alpha_C$'s simplify to become headlight functions for the tree-fatgraph obtained by cutting $\Gamma$ along all the curves in $S$. This general phenomenon is computationally very useful, and we study it in greater detail elsewhere.

\subsection{The genus one 2-loop vacuum amplitude}\label{ex:markov:exam}
We have already mentioned the 2-loop genus one vacuum computation in Sections \ref{ex:markov:head} and \ref{ex:markov:amp}. We now have all the tools to compute it properly. The result is the following simple integral
\begin{equation}
    {\cal A} = \int\limits_{R} dxdydz\,{\cal K}\,\left(\frac{\pi^2}{\cal U}\right)^{\frac{D}{2}} \exp\left(-{\cal Z}\right),
\end{equation}
where the kernel is (as given in Section \ref{ex:markov:mirz})
\begin{equation}
    {\cal K} = \frac{\alpha_{1/0}}{\alpha_{1/0}+\alpha_{0/1}+\alpha_{1/1}+\alpha_{-1/1}} \frac{\alpha_{0/1}}{\alpha_{0/1}+\alpha_{1/1}+\alpha_{-1/1}},
\end{equation}
and now with surface Symanzik polynomials
\begin{align}\label{ex:markov:UZ}
    {\cal U} &= \det A,\\
    {\cal Z} &= m^2 ( \alpha_{1/0}+\alpha_{0/1}+\alpha_{1/1}+\alpha_{-1/1}) .
    \label{eq:uzpolys}
\end{align}
Note that the region, $R$, where $\alpha_{1/0}\alpha_{0/1}\neq 0$ is, in the coordinates of Section \ref{ex:markov:head}, given by $w,x\leq 0$ and $x+2z\leq 0$. The curve integral is restricted by the Mirzakhani kernel to this region.

To see how this curve integral comes about, we need to understand how to assign momenta to the curves. The easiest way to assign momenta is to use the homology of curves on the torus, Section \ref{sec:homology}. Assign the A-cycle momentum $\ell_1$ and the B-cycle momentum $\ell_2$. The curve $C_{p/q}$ wraps the A-cycle $q$ times and the $B$-cycle $p$ times, and so it has momentum $p\ell_1+q\ell_2$ giving
\begin{equation}
    X_{p/q}  = (p \ell_1 + q \ell_2)^2 + m^2.
\end{equation}
With this momentum assignment, the matrix $A$, which records the dependence on chosen basis of loops, is
\begin{align}
    A^{ab} = \begin{bmatrix} \alpha_{1,0}+\alpha_{1,1}+\alpha_{-1,1} & \alpha_{1,1} - \alpha_{-1,1} \\ \alpha_{1,1} - \alpha_{-1,1} & \alpha_{0,1}+\alpha_{1,1}+\alpha_{-1,1} \end{bmatrix}.
\end{align}
Moreover, the momentum assigned to the curves has no non-loop part, so that
\begin{equation}
    {\cal Z} = m^2 \sum_C \alpha_C,
\end{equation}
which restricts to \eqref{eq:uzpolys} in the region $R$.

We now evaluate the amplitude. Once again, we will be aided by a striking simplification of the headlight parameters. The headlight parameters were given in Section \ref{ex:markov:head}. But in the region $R$, $\alpha_{1/1}$ and $\alpha_{-1/1}$ simplify to become tree-like headlight functions:
\begin{equation}
    \alpha_{1/1} = \max(0,-z) \qquad \text{and}\qquad \alpha_{-1/1} = \max(0,z).
\end{equation}
This corresponds to the fact that cutting $\Gamma$ along $C_{1/0}$ and $C_{0/1}$ gives a 4-point tree graph. Moreover, in this region,
\begin{equation}
\alpha_{1/0} = -w,\qquad \text{and}\qquad \alpha_{0/1} = - x - 2 \max(0,z),
\end{equation}
and so
\begin{align}
\alpha_{1/0}+\alpha_{0/1}+\alpha_{1/1}+\alpha_{-1/1} &= -w-x-z\\
\alpha_{0/1}+\alpha_{1/1}+\alpha_{-1/1} &= - x - z.
\end{align}
Moreover, ${\cal U}$ and ${\cal Z}$ become
\begin{equation}
    {\cal U} = \det A = wx+zw-(x+2\max(0,z))|z|,\qquad\text{and}\qquad {\cal Z} = - m^2 (w+x+z).
\end{equation}
So the vacuum amplitude is
\begin{align}
    {\cal A} = \int\limits_{R} dwdxdz \frac{w(x+2\max(0,z))}{(w+x+z)(x+z)} \left(\frac{\pi^2}{\cal U}\right)^{\frac{D}{2}}\,\mathrm{exp}\left(m^2 (w+x+z)\right).
\end{align}

It is not obvious that this is the correct answer. In the conventional calculation, the amplitude receives just a single contribution: the vacuum sunset Feynman diagram. Our formula resembles, but is not the same, as the Schwinger parameterisation for this diagram. To see that they are equivalent, note that $R$ can be divided into two regions: one where $z\geq 0$ and another where $z\leq 0$. By a simple change of variables, the integral over either one of these regions can be written as
\begin{equation}
I = \int\limits_{a,b,c\geq 0} dadbdc \frac{ab}{(a+b+c)(b+c)}\left(\frac{\pi^2}{ab+bc+ca}\right)^{\frac{D}{2}}\,\mathrm{exp}\left(-m^2 (a+b+c)\right).
\end{equation}
However, summing over the 6 possible permutations of $a,b,c$, note that
\begin{align}
\frac{ab}{b+c} + (\text{permutations of $a,b,c$}) = a+b+c.
\end{align}
It follows that $I$ is $1/6$ times the sunset integral with the usual parameterisation. Integrating over all of $R$, we then recover that
\begin{align}
    {\cal A} = \frac{1}{3} \int\limits_{x,y,z\geq 0} dxdydz \left( \frac{\pi^2}{xy+y|z|+|z|x} \right)^{\frac{D}{2}}\,\mathrm{exp}\left(-m^2 (x+y+|z|)\right).
\end{align}
This is $1/3$ times the vacuum sunset integral. The factor of $1/3$ corresponds to the fact that the fatgraph has $|\text{Aut}(\Gamma)| = 3$.

\subsection{The planar 2-loop tadpole}\label{exam:D1tilde}
\begin{figure}
\begin{center}
\includegraphics[width=0.45\textwidth]{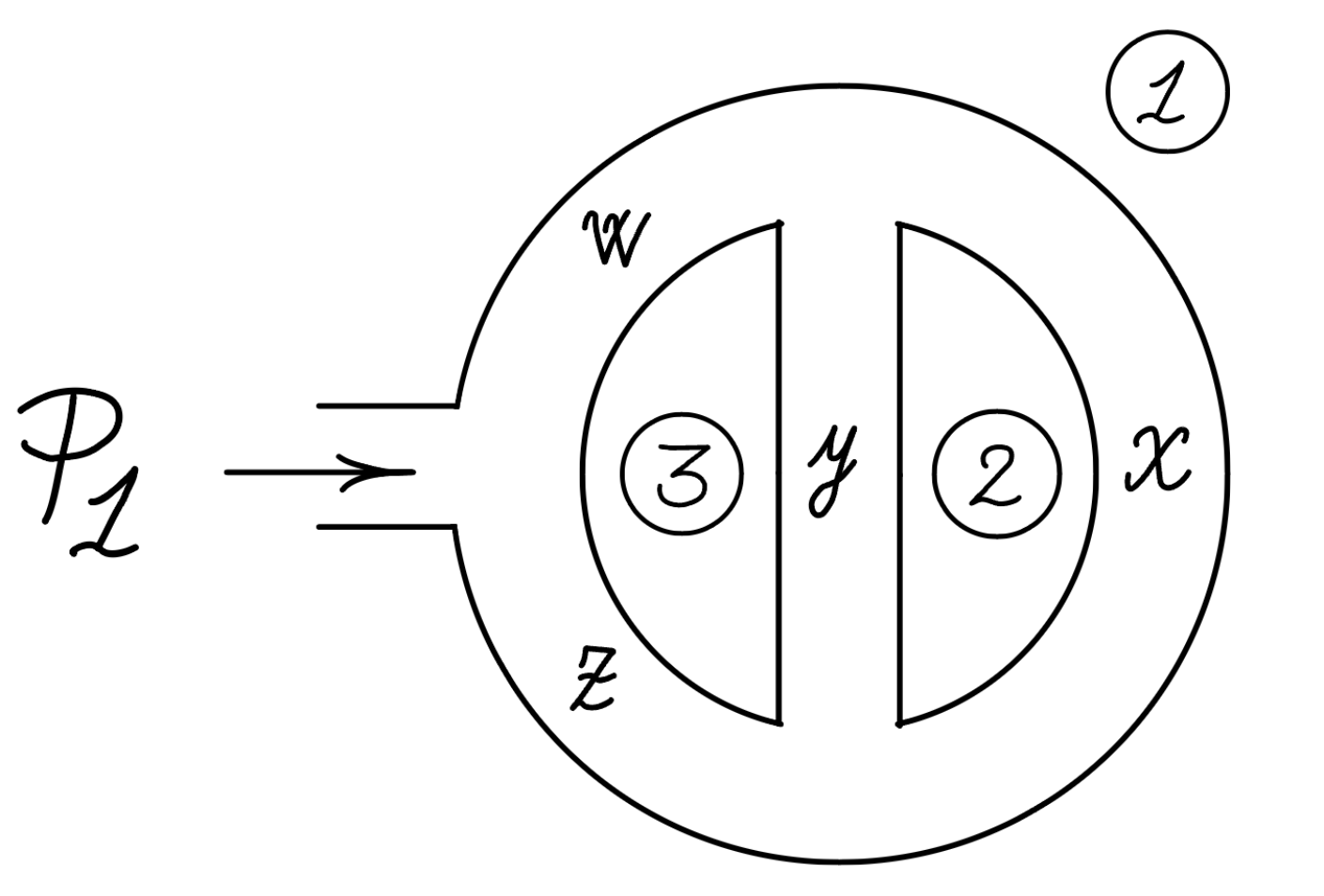}
\caption{A planar 2-loop tadpole graph.}
\label{fig:D1tilde}
\end{center}
\end{figure}

We can compute the planar 2-loop tadpole amplitude using the fatgraph $\Gamma$ in Figure \ref{fig:D1tilde}. The curves on this fatgraph can be labelled by their endings. We have two loop boundaries, labelled $2,3$ in the Figure. The curves are then $C_{23},C_{22},C_{33},C_{12}^n,C_{13}^n$, where $n$ indexes how many times the curves $C_{12}^n,C_{13}^n$ loop around before beginning their spiral. As usual, we will only need a small number of these curves to compute the amplitude.

Because $\Gamma$ is planar, we can introduce dual variables $z_1^\mu,z_2^\mu,z_3^\mu$ to parametrise the momenta of the curves. The propagator factors are then
\begin{equation}\label{eq:exaDtildeX}
    X_{12}^n = (z_2-z_1)^2+m^2,~~ X_{13}^n = (z_3-z_1)^2+m^2,~~X_{23} = (z_3-z_2)^2+m^2.
\end{equation}
It is convenient to take $z_3-z_1$ and $z_2-z_1$ as our loop momentum variables.

The curve integral for the amplitude is then
\begin{equation}
    {\cal A} = \int d^4 t\, {\cal K}\,\left(\frac{\pi^2}{\cal U}\right)^{\frac{D}{2}} \exp(-\mathcal{Z}),
\end{equation}
where
\begin{align}
    {\cal U} = \det A,\qquad\text{and}\qquad
    {\cal Z} = m^2 \left( \alpha_{23}+\alpha_{22}+\alpha_{33} + \sum_{n} (\alpha_{12}^n+\alpha_{13}^n) \right).
\end{align}
Moreover, using the momenta assignments from the dual variables, \eqref{eq:exaDtildeX}, $A$ is the $2\times 2$ matrix
\begin{equation}
    A = \begin{bmatrix} \alpha_{23}+\sum_{n=-1}^1 \alpha_{12}^n & \qquad \alpha_{23} \\ \alpha_{23} & \qquad \alpha_{23}+\sum_{n=-1}^1 \alpha_{13}^n \end{bmatrix}.
\end{equation}

\begin{figure}
\begin{center}
\includegraphics[width=0.85\textwidth]{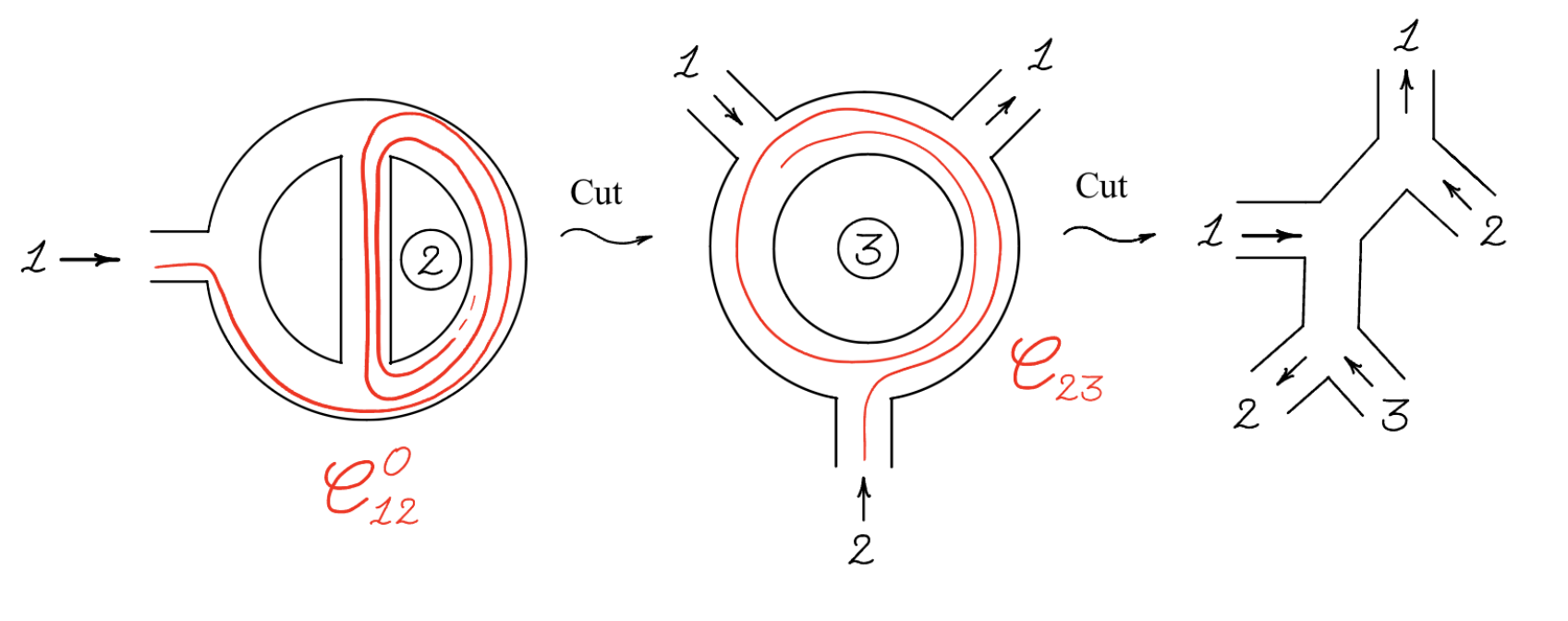}
\caption{A maximal cut of the planar 2-loop tadpole graph. The curve $C_{12}^0$ cuts $\Gamma$ to a 3-point 1-loop graph, and the curve $C_{23}$ cuts this further to a 5-point tree graph.}
\label{fig:D1tildecut}
\end{center}
\end{figure}

${\cal U}$ is the determinant of $A$, and each monomial in this determinant corresponds to a pair of curves that cut $\Gamma$ to a 5-point tree graph. Using the fact that $\alpha_C\alpha_D=0$ if $C,D$ intersect, we find
\begin{equation}
    {\cal U} = \sum_{n=-\infty}^\infty \left( \alpha_{23}\alpha_{12}^n + \alpha_{23}\alpha_{13}^n + \alpha_{12}^n\alpha_{13}^n + \alpha_{12}^n\alpha_{13}^{n+1} \right).
\end{equation}
Here, we have chosen a convention for the index $n$ such that $C_{12}^n,C_{13}^{n+1}$ are compatible, but $C_{12}^n,C_{13}^{n-1}$ intersect. The $\MCG$ has one generator, which acts on the index $n$. So it is clear that the monomials in ${\cal U}$ can be decomposed into four cosets (corresponding to the four terms in the sum). We therefore get a Mirzakhani kernel (of the type discussed in Section \ref{sec:mirz:U})
\begin{equation}
    {\cal K} = \frac{{\cal U}_0}{\cal U},
\end{equation}
with
\begin{equation}
    {\cal U}_0 = \alpha_{23}\alpha_{12}^0 + \alpha_{23}\alpha_{13}^0 + \alpha_{12}^0\alpha_{13}^0 + \alpha_{12}^0\alpha_{13}^{1}.
\end{equation}
In the region where ${\cal U}_0\neq 0$, only 12 $\alpha_C$'s are non-vanishing. In fact, each monomial in ${\cal U}_0$ defines a maximal cut of $\Gamma$, which cuts $\Gamma$ to a 5-point tree graph. See Figure \ref{fig:D1tildecut}. ${\cal A}$ is the sum of four terms,
\begin{equation}
    {\cal A} = {\cal A}_{C_{23},C_{12}^0}+{\cal A}_{C_{23},C_{13}^0}+{\cal A}_{C_{12}^0,C_{13}^0}+{\cal A}_{C_{12}^0,C_{13}^1},
\end{equation}
each corresponding to a different maximal cut of the fatgraph.

For instance, ${\cal A}_{C_{23},C_{12}^0}$ is given by the curve integral over the region $\alpha_{23}\alpha_{12}^0\neq 0$. In this region, only 5 other $\alpha_C$'s are non-vanishing. The curves correspond to the five curves on the 5-point tree graph obtained by cutting along $C_{23},C_{12}^0$. The 5 curves compatible with $C_{23},C_{12}^0$ are
\begin{equation}
C_{12}^1,~C_{12}^{-1},~C_{13}^0,~C_{13}^{1},~C_{22}.
\end{equation}
In this region, the headlight functions simplify, so that, similar to the previous examples, the curve integral only sees the headlight functions of the 5-point tree-level problem. Explicitly, in coordinates, we can take (in this region) $\alpha_{23}=w,~\alpha_{12}^0=x$, and the remaining headlight functions take the form of tree-level headlight functions (see Figure \ref{fig:D1tildecut}):
\begin{align}
\alpha_{13}^1 &= f_2-f_1-z, & \alpha_{13}^0 &= f_1-y,\\
\alpha_{22} &= f_2-f_3,  & \alpha_{12}^{1} &= f_3,\\
\alpha_{12}^{-1} & = f_1+f_3-f_2. & 
\end{align}
where
\begin{equation}
    f_1 = \max(0,y),~f_2 = \max(0,z,y+z),~f_3=\max(0,z).
\end{equation}
So, in this region, the $A$ matrix restricts to
\begin{equation}
    A' = \begin{bmatrix} w+ f_1-f_2+2f_3 & w\\ w & w + f_2-y-z\end{bmatrix},
\end{equation}
and ${\cal Z}$ restricts to
\begin{equation}
    {\cal Z}' = m^2(w+x-y-z + f_1+f_2+f_3).
\end{equation}
The contribution of this term to the amplitude is then
\begin{equation}
    {\cal A}_{C_{23},C_{12}^0} = \int\limits_{w,x\geq 0} dwdxdydz \, \frac{wx}{\det A'} \,\left(\frac{\pi^2}{\det A'}\right)^{\frac{D}{2}} \exp(-\mathcal{Z}').
\end{equation}
The other 3 cuts are similarly computed.

\subsection{The planar 3-loop vacuum amplitude}\label{exam:S03}
\begin{figure}
\begin{center}
\includegraphics[width=0.45\textwidth]{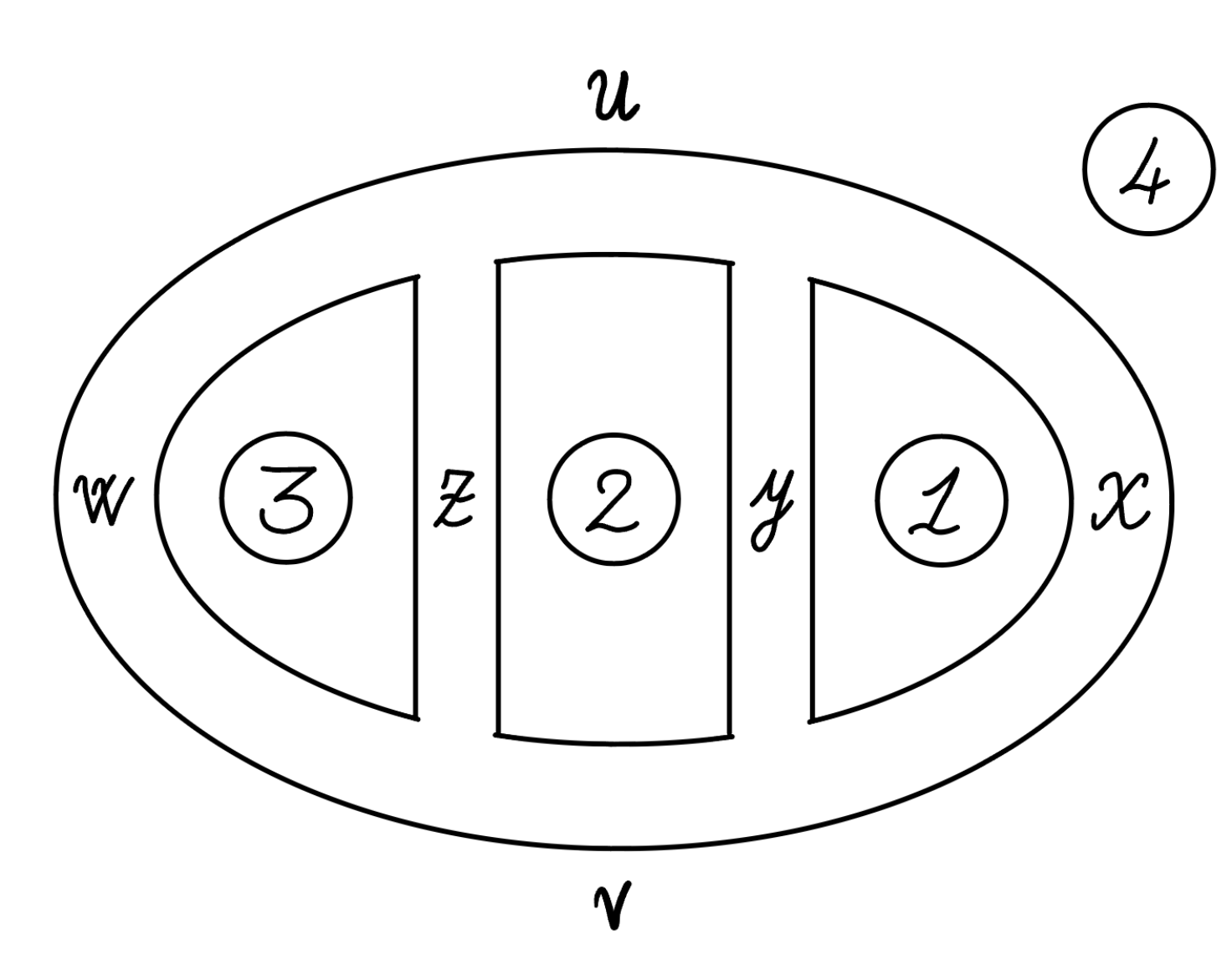}
\caption{Three loop.}
\label{fig:S4}
\end{center}
\end{figure}
We now consider a 3-loop example. The 3-loop vacuum amplitude can be computed using the 3-loop fatgraph, $\Gamma$, in Figure \ref{fig:S4}. The curves on $\Gamma$ all begin and end in a spiral. There are four loop boundaries, labelled $a=1,2,3,4$ in the Figure, that the curves can spiral around. Let $C_{ab}^\delta$ be the curves that begin spiralling around $a$, and end spiralling around $b$. There are infinitely many such curves, all related by the action of the $\MCG$. In fact, the $\MCG$ action in this case is quite complicated: it is an action of the braid group $B_3$. However, using a tropical Mirzakhani kernel, we can still compute the amplitude.

The momentum assignment to the curves is easy to describe, because $\Gamma$ is a planar graph. Introduce dual momentum variables, $z_a^\mu$, associated to the four boundaries, $a=1,2,3,4$. Then the propagator for $C_{ab}^\delta$ is just
\begin{equation}
    X_{ab} = (z_b^\mu - z_a^\mu)^2 + m^2.
\end{equation}
We can choose any three $z_a$ to be our loop momentum variables.

Our formula for the amplitude is then
\begin{equation}
    {\cal A} = \int d^6t\,{\cal K}\,\left(\frac{\pi^3}{\cal U}\right)^{\frac{D}{2}}\, \exp(-{\cal Z}),
\end{equation}
where the surface Symanzik polynomials are
\begin{align}
    {\cal U} = {\det}' \tilde{A},\qquad {\cal Z} = m^2 \sum \alpha_{ab}^\delta.
\end{align}
Here, we take a slightly different approach to presenting ${\cal U}$, adapted to the planar case, by using a reduced determinant, ${\det}'$, which excludes a row and column. The $4\times 4$ matrix $\tilde{A}$ is (for $a\neq b$)
\begin{align}
    \tilde{A}_{ab} = \sum_\delta \alpha_{ab}^\delta,\qquad     \tilde{A}_{aa} = - \sum_{c\neq a} \tilde{A}_{ac}.
\end{align}
By the matrix-tree theorem, the reduced determinant, ${\det}'\tilde{A}$, turns into a sum over all maximal cuts of the fatgraph $\Gamma$. In this case, a maximal cut is given by any three non-intersecting curves, $\{C_{ab}^{\delta},C_{cd}^{\delta'},C_{ef}^{\delta''}\}$, such that the pairs,---$ab$, $cd$, $ef$,---span a tree on the set $\{1,2,3,4\}$. So ${\det}'\tilde{A}$ indeed recovers the definition of ${\cal U}$ as the sum over maximal cuts of the fatgraph. Explicitly, it takes the form
\begin{equation}
    {\cal U} = \sum_{\delta,\delta',\delta''}\sum_{\text{trees}} \alpha_{ab}^{\delta} \alpha_{cd}^{\delta'}\alpha_{ef}^{\delta''}
\end{equation}

We can now use this formula for ${\cal U}$ to define a Mirzakhani kernel, ${\cal K}$. This set of triples appearing in ${\cal U}$ can be decomposed as a sum of cosets under the $\MCG$. The $\MCG$-action leaves the starts and ends of each curve unchanged. So we find that there are $16$ $\MCG$-inequivalent maximal cuts of $\Gamma$, corresponding to the $4^2$ distinct labelled trees in the set $\{1,2,3,4\}$. For each such labelled tree, we choose a coset representative.
\begin{equation}
\alpha_{ab}^{0}\alpha_{cd}^{0}\alpha_{ef}^{0},
\end{equation}
where the pairs $ab,cd,ef$ define the tree, and $C_{ab}^0,C_{cd}^0,C_{ef}^0$ is some choice of 3 non-intersecting curves. Let ${\cal U}_0$ be the sum of monomials for these $16$ coset representatives. It has the form
\begin{equation}
    {\cal U}^0 = \sum_{12~\text{perms}} \alpha_{12}\alpha_{23}\alpha_{34} + \sum_{4~\text{perms}} \alpha_{14}\alpha_{24}\alpha_{34}.
\end{equation}
Then
\begin{equation}
    {\cal K} = \frac{{\cal U}_0}{\cal U}
\end{equation}
is our Mirzakhani kernel.

An exercise in the intersection rules for mountainscapes shows that the following 6 curves are sufficient to build each of the 16 maximal cuts:
\begin{align}
    C_{14}^0 &= (xRyR)^\infty x (LvLwLuLx)^\infty,\\
    C_{24}^0 &= (uRyRvRzR)^\infty u (LxLvLwLu)^\infty,\\
    C_{34}^0 &= (wRzR)^\infty w (LuLxLvLw)^\infty,\\
    C_{12}^0 &= (yRxR)^\infty y (LuLzLvLy)^\infty,\\
    C_{23}^0 &= (zRuRyRvR)^\infty z (LwLz)^\infty,\\
    C_{13}^0 &= (RyRx)^\infty L v R (zLwL)^\infty.
\end{align}
This is because all of these curves are pairwise compatible. Using these curves, we can define a restricted matrix (for $a\neq b$)
\begin{align}
    \tilde{A}^0_{ab} = \alpha_{ab}^0,\qquad
 \tilde{A}^0_{aa} = - \sum_{c\neq a} \tilde{A}^0_{ac}
\end{align}
so that, by the matrix-tree theorem, ${\cal U}^0 = {\det}'\tilde{A}^0$. Our Mirzakhani kernel is then
\begin{equation}
    {\cal K} = \frac{{\det}'\tilde{A}^0}{{\det}'\tilde{A}}.
\end{equation}

For each of the 16 monomials in ${\cal U}^0$ we get a contribution to ${\cal A}$. For instance, take the monomial
\begin{equation}
    \alpha_{12}^0\alpha_{23}^0\alpha_{34}^0,
\end{equation}
corresponding to the tree $1-2-3-4$. The associated contribution to ${\cal A}$ only involves $\alpha_C$ for curves $C$ compatible with this maximal cut. This maximal cut gives a tree fatgraph, with colour ordering $(123432)$.\footnote{Cutting a curve that ends in a spiral around a loop boundary creates a new external line on that boundary.} So this contribution to the amplitude involves only the $9$ headlight functions for this 6-point tree fatgraph.

Finally, note that by permutation symmetry (with respect to the dual variables $z_a$), we only really need to evaluate two of the maximal cuts in our formula, say:
\begin{equation}
\alpha_{12}^0\alpha_{23}^0\alpha_{34}^0\qquad\text{and}\qquad \alpha_{14}^0\alpha_{24}^0\alpha_{34}^0.
\end{equation}
Then
\begin{equation}
    {\cal A} = 12 \,{\cal A}_{12,23,34} + 4 \,{\cal A}_{14,24,34},
\end{equation}
where each of ${\cal A}_{12,23,34}$ and ${\cal A}_{14,24,34}$ can be computed knowing only the headlight functions for a 6-point tree graph.

\section{A First Look at Recursion}\label{sec:recursion}
The tropical Mirzakhani kernels dramatically simplify the task of evaluating our amplitudes. Using these kernels, our formulas for amplitudes at $L$ loops end up expressed in terms of the headlight functions, $\alpha_C$, that we have already computed for lower loop level amplitudes. In this section, we show an alternative way to apply the Mirzakhani kernels to compute amplitudes, by using them to define a powerful recursion relation for the integrands, ${\cal I}$. 

Fix a fatgraph $\Gamma$. Its associated (pre-loop-integration) integrand is given by the curve integral
\begin{align}\label{eqn:intagain}
{\cal I} = \int \frac{d^n t}{\MCG} Z,\qquad Z = \exp\left(- \sum_C\alpha_C X_C\right).
\end{align}
To evaluate the curve integral, we introduce a tropical Mirzakhani kernel, as above. Take, for example, some trace factor $\beta$. The non-separating curves with endpoints on $\Gamma$ form a set $\mathcal{S}_\beta$, and which can be partitioned into $\MCG$ orbits with some coset representatives $C_1,\ldots, C_k$. Each of these curves, $C_i$, cuts $\Gamma$ to a fat graph $\Gamma_{C_i}$ with a smaller number of loops. The Mirzakhani kernel $\mathcal{K}_\beta$ then gives
\begin{align}\label{eqn:intagainmir}
{\cal I} = \sum_{i=1}^k \,\int \frac{d^n t}{\MCG}\, \frac{\alpha_{C_i}}{\rho} Z.
\end{align}
Introducing an auxiliary parameter, $\xi$, the $1/\rho$ can be incorporated into the exponential using
\begin{equation}
\frac{1}{\rho} = \int\limits_0^\infty d\xi \, e^{-\rho \xi}.
\end{equation}
Equation \eqref{eqn:intagainmir} then implies the following recursion formula:
\begin{align}
{\cal I} = \int\limits_0^\infty d\xi\, \sum_{i=1}^k \frac{-1}{(X_{C_i}+\xi)^2} {\cal I}_{\Gamma_{C_i}}(X'_C),
\label{eq:globfwd}
\end{align}
where the new dual variables $X'_C$ appearing in the integrand $I_{\Gamma_{C_i}}(X'_C)$ are given by
\begin{align}\label{eqn:globshift}
X'_C = \left\{ \begin{matrix} X_C + \xi & ~~~ \text{if}~C \in \mathcal{S}_\beta \\ X_C & \text{else.} \end{matrix} \right.
\end{align}
This formula, \eqref{eq:globfwd}, is a completely recursive way to obtain the rational functions ${\cal I}$ to all orders in the perturbation series. A detailed derivation of \eqref{eq:globfwd} is given in Appendix \ref{sec:app:recursion}.

For example, consider again the 1-loop non-planar propagator computed in Section \ref{ex:A11:amp}. The curves on $\Gamma$ are $\mathcal{S} = \{C_n\}$ as before, and their associated dual variables are
\begin{align}
X_n = (\ell + nk)^2.
\end{align}
The MCG has just one generator, and so we will only need to apply the global forward limit once. Taking $C_0$ as our coset representative, \eqref{eq:globfwd} gives 
\begin{align}
{\cal I}_\Gamma = \int\limits_0^\infty d\xi  \frac{-1}{(X_0+\xi)^2} {\cal I}_{\Gamma_{C_0}} (X_1+\xi,X_{-1}+\xi),
\end{align}
where $\Gamma_{C_0}$ is the 4-point tree graph obtained by cutting $\Gamma$ along $C_0$. The curves $C_1$ and $C_{-1}$ become the two possible propagators of $\Gamma_{C_0}$: on $\Gamma$, $C_1$ and $C_{-1}$ are the only two curves that do not intersect $C_0$. So we have,
\begin{align}
{\cal I}_\Gamma = - \int\limits_0^\infty d\xi  \left ( \frac{1}{(X_0+\xi)^2} \frac{1}{X_1+\xi} +  \frac{1}{(X_0+\xi)^2} \frac{1}{X_{-1}+\xi} \right).\end{align}
Evaluating the $\xi$ integral gives the following formula for the integrand:
\begin{align}\label{eq:1loopfwdex}
{\cal I}_\Gamma = \frac{1}{X_0(X_1-X_0)} + \frac{1}{X_0(X_{-1}-X_{0})}.
\end{align}
Here we see the appearance of \emph{linearised propagators}, of the form $1/(X_C - X_{C_i})$. Such linearised propagators have arisen in previous studies of forward limit \cite{He_2015, Agerskov_2020, Porkert_2023, geyer2018, 
Geyer_2018, geyer2019}. In the full sum, these linearised propagators sum to give back the ordinary loop integrand after identifications made using shifts of the loop momenta. In our current example, the loop momentum shift $\ell \mapsto \ell + k$ shifts the dual variables by $X_n \mapsto X_{n+1}$. Applying this shift to the second term in \eqref{eq:1loopfwdex} gives
\begin{align}
{\cal I}'_\Gamma = \frac{1}{X_0(X_1-X_0)} + \frac{1}{X_1(X_{0}-X_1)} = \frac{1}{X_0X_1}.
\end{align}
For higher loop integrands, we can use multiple iterations of \eqref{eq:globfwd} to write ${\cal I}$ as a sum over some tree amplitudes, with various shifts in the kinematic variables. 

Note that the recursion, \eqref{eq:globfwd}, continues to hold even when the $X_C$ variables are not all distinct. For example, if all $X_C$ are set equal to a constant, $X_C = X$, then ${\cal I}_\Gamma = C_\Gamma/X^E$, where $C_\Gamma$ is the number of Feynman diagrams contributing to the amplitude. In this case, \eqref{eq:globfwd} can be used to recursively compute the number of diagrams. Moreover, the recursion \eqref{eq:globfwd} also holds when there are higher poles in the integrand, arising from diagrams like bubbles. We give a more complete analysis of these recursions elsewhere.

\section{Outlook} 

The new representation of all-loop amplitudes we have studied in this paper has implications far beyond our understanding of scalar amplitudes, and has consequences for the understanding of particle and string scattering generally. We highlight a number of directions that are especially primed for immediate development.

The magic of the \emph{curve integral} formulas is that integrals over an $O(n)$ dimensional space, of an action built from $O(n^2)$ piecewise linear functions, automatically reproduces the full amplitudes, which are conventionally sums over $O(4^n)$ Feynman diagrams. The novelty of this formalism over conventional field theory must therefore become most manifest in the limit $n \to \infty$ of a large number of particles. In examples, we have found evidence that the external kinematical data can be chosen so that the large-$n$ limits the curve integrals are smooth, leading to formulas for amplitudes in the large-$n$ limit in terms of \emph{tropical path integrals}. Studying this limit might lead to a new understanding of the emergence of strings from colored particles at strong coupling. At strong coupling, the scattering for a small number of particles is exponentially small, and the amplitude is instead dominated by the emission of a huge number of particles, approximating field configurations that should more continuously connect to a string worldsheet picture. 

Even at finite $n$ the curve integral formalism offers radically new methods to compute amplitudes. For instance, it allows to evaluate amplitudes numerically by direct integration, thus avoiding the generation of Feynman diagrams altogether. The geometric properties of the fan suggest a new search for an optimal numerical integration strategy, uplifting recent breakthroughs in the numerical evaluation of Feynman integrals in parametric form to entire amplitudes \cite{Borinsky_2023,feyntrop}.

A second frontier ripe for immediate investigation is an understanding of gravity and gravity-like amplitudes. Just as the tr$\phi^3$ theory is a model for general colored amplitudes, a scalar model for gravity is given by an uncolored scalar $\sigma$ with cubic self-interaction $\sigma^3$. In special cases, it is now standard to think of uncolored and colored theories as related by double-copy or `gravity = gauge$^2$' formulas \cite{Bern:2010ue}. The stringy origin of these formulas, the KLT relations, is deeply connected to thinking about the string worldsheet in a fundamentally \emph{complex} fashion as a Riemann surface with a complex structure. But there are many reasons why our formulation of uncolored amplitudes will involve a very different sort of treatment. As we alluded to in the introduction, the existence of $\sigma$ is forced on us in the most elementary way by the structure of the Feynman fan, which has lower-dimensional `holes' that are beautifully completed by adding in new vectors corresponding to $\sigma$ particles. This does not remotely have the flavor of `gravity = gauge$^2$'. Moreover, as alluded to in the introduction, the $u$-variables central to our story are deeply connected to the string wordsheet (and Teichm\"uller space), but via \emph{hyperbolic geometry} and {\it not} through the conventional picture of Riemann surfaces with complex structure. All of this dovetails nicely with the many observations, in examples of gravity amplitudes, that there is vastly more structure to gravity amplitudes than is suggested by the `gravity=gauge$^2$' slogan. The striking way in which $\sigma$ is forced on us in our story is a new departure point for uncovering more of this hidden structure. 

Finally, our results here strongly suggest that there is way to describe fundamental particle physics in the real world from a more elementary starting point, with spacetime and quantum mechanics appearing as emergent principles. We believe that we have taken a major new step in this direction with the results we have begun to introduce in this paper. A number of major challenges remain before we can reach this goal. The first is to understand how fermions arise from this new perspective, which has so far only been applied to bosonic scattering. For Standard Model physics, describing chiral fermions will be especially interesting and important. Another challenge is that the key structures in our formulas stem from a fatgraph, which is most immediately connected to the adjoint representation of $U(N)$ gauge theories. But the quantum numbers of the Standard Model are more interesting. For instance, in the $SO(10)$ grand unified theory, the matter lives in ten fundamentals (higgses) together with three {\bf 16}'s for the fermions. How might the amplitudes for matter in these representations emerge from elementary combinatorial foundations?

\section*{Acknowledgments}
\small{We especially thank Song He and Thomas Lam for countless stimulating conversations on the topics of this paper over many years. We also thank Sebastian Mizera and Hofie Hannesdottir for many discussions, and Song He, Carolina Figueiredo, Daniel Longenecker, Qu Cao and Jin Dong for ongoing interactions related to the material of this paper over the past year. NAH is supported by the DOE under grant DE-SC0009988; further crucial contributions to his work were made possible by the Carl B. Feinberg cross-disciplinary program in innovation at the IAS. NAH also expresses sincere thanks to HF, PGP, GS and HT for restraining themselves from strangling him during the completion of this work. PGP is supported by ANR grant CHARMS (ANR-19-CE40-0017) and by the Institut Universitaire de France (IUF). PGP worked on this project while participating in {\it Representation Theory: Combinatorial Aspects and Applications} at the Centre for Advanced Study, Oslo. HF is supported by Merton College, Oxford. During this project HF received additional support from ERC grant GALOP (ID: 724638). During this project GS was supported by Brown University, Providence, the Perimeter Institute, Waterloo, and the Institute for Advanced Study, Princeton. GS was also funded by the European Union’s Horizon 2020 research and innovation programs {\it Novel structures in scattering amplitudes} (No. 725110) of Johannes Henn. GS thanks the groups of C. Anastasiou and N. Beisert at ETH Zurich for hospitality during the worst phase of the COVID-19 pandemic. HT was supported by NSERC Discovery Grant RGPIN-2022-03960 and the Canada Research Chairs program, grant number CRC-2021-00120.}

\appendix

\section{Deriving the Curve Integral Formula}\label{sec:app:ci}
To see why \eqref{eq:glsymI} is correct, let us write the amplitude explicitly. Write
\begin{align}
X_C = P_C^2+m^2
\end{align}
for the propagator factor associated to curve $C$ (with momentum $P_C^\mu$). Fix some fatgraph $\Gamma$ with some color factor $C_\Gamma$. The associated partial amplitude can be expressed with just one overall loop integration as
\begin{align}\label{eq:paragain}
{\cal A} =  \int\prod_{i=1}^L d^D \ell_i  \left( \sum_{\Gamma'} \prod_{C} \frac{1}{X_C} \right),
\end{align}
where sum over exactly one of every fatgraph $\Gamma'$ that has color factor $C_{\Gamma'} = C_\Gamma$. The integrand in this formula can be written as an integral over \emph{curve space}, $V$. To do this, recall that every top dimensional cone of the Feynman fan corresponds to some triangulation of $\Gamma$. Any vector ${\bf g}\in V_\Gamma$ can be expanded as a sum of the generators of the cone that it is in using
\begin{align}\label{eq:decomp}
{\bf g} = \sum_{C} \alpha_C({\bf g})\, {\bf g}_C,
\end{align}
where $\alpha_C$ are the headlight functions and ${\bf g}_C$ are the $g$-vectors of the curves, $C$. Consider the function on $V$ given by
\begin{align}
Z = \exp\left(-\sum_{C} \alpha_C(\mathbf{t}) X_C \right),
\end{align}
where the sum in the exponent is over all open curves $C$. Let $T$ be a triangulation corresponding to some top-dimensional cone, with curves $C_1,...,C_E$. Restricting $Z$ to this cone gives
\begin{align}
\left.Z\right|_{\text{cone}} =  \exp\left(-\sum_{i=1}^E \alpha_{C_i} (\mathbf{t}) X_{C_i} \right),
\end{align}
which follows from \eqref{eq:decomp}. Moreover, the generators of this top dimensional cone span a parallelopiped of unit volume, so there exist corresponding coordinates $y'_1,...,y'_E$ such that $d^E y = d^Ey'$ and so that any vector in this cone can be written as
\begin{align}
{\bf g} = \sum_{i=1}^E y'_i {\bf g}_{C_i}. 
\end{align}
The integral of $Z$ over this cone is then
\begin{align}
\int\limits_{\text{cone}} d^E{y} Z = \int\limits_{\geq 0} d^Ey' \, \exp\left( \sum_{i=1}^E - y_i' X_{C_i} \right) = 
\prod_{i=1}^E \frac{1}{X_C}.
\end{align}
It follows from this that the partial amplitude \eqref{eq:paragain} can be written as a curve integral over curve space:
\begin{align}\label{eq:glsym2}
  {\cal A} = \int \frac{d^E\mathbf{t}}{\MCG} \int \prod_{i=1}^L d^D \ell_i\, Z.
\end{align}
In this formula, we integrate over curve space modulo the action of the mapping class group. This ensures that we count each fatgraph $\Gamma$ only once. We explain how to compute these curve integrals, with non-trivial MCG actions, in Section \ref{sec:mirz}.

\section{Factorization in detail}\label{app:fact}
In the text, the factorization of the curve integral formula for integrands ${\cal I}$ is stated in \eqref{eqn:factfin}. This formula gives the residue of the pole $1/X_C$. To derive the formula, there are two possible cases to consider: either $C$ is MCG-invariant, or not.

\subsection{MCG invariant curve}
Suppose $C$ is MCG-invariant. The $X_C$ pole arises from the part of the integral over the region of curve space where $\alpha_C>0$. Since $\text{Stab}(C) = \text{MCG}(\Gamma)$, the MCG action has a well-defined restriction to this region and we have a well-defined curve integral
\begin{align}
{\cal I}' = \int\limits_{\alpha_C>0} \frac{d^Et}{\MCG} Z.
\end{align}
To compute ${\cal I}'$, take a triangulation containing $C$, with curves $C, D_1,...,D_{E-1}$. Take coordinates adapted to this cone:
\begin{align}
{\bf g} = t_C {\bf g}_C + \sum_{i=1}^{n-1} t_i' {\bf g}_{D_i}.
\end{align}
By the unit volume property, the integration measure is
\begin{align}
d^Et = dt_C d^{E-1}t'.
\end{align}
In these coordinates, the restriction of $Z$ to this region is
\begin{align}
\left.Z\right|_{t_C>0} = e^{-t_C X_C}\,\exp\left( - \sum_{D|C} \alpha_D X_D \right),
\end{align}
where the sum is over $D$ that do not intersect $C$. For these curves, $\alpha_D({\bf g}+{\bf g}_C) = \alpha_D({\bf g})$, so that the only $t_C$-dependence is in the $\exp(-t_C X_C)$ factor. Write $\alpha_D' = \alpha_D|_{t_C=0}$, for the headlight functions restricted to $t_C=0$. $\alpha_D'$ is the headlight function of $D$ considered as a curve on the cut fatgraph $\Gamma_C$. 

The $t_C$ integral gives
\begin{align}
{\cal I}' = \frac{1}{X_C} \int \frac{d^{E-1}t'}{\MCG} Z_C,
\end{align}
where
\begin{align}
Z_C = \exp\left( - \sum_{D|C} \alpha_D' X_D \right).
\end{align}
The full curve integral ${\cal I}$ is ${\cal I} = {\cal I}' + \ldots,$ where the $\ldots$ has no $X_C$ pole. So
\begin{align}
\text{Res}_{X_C=0} I =  \int \frac{d^{E-1}t'}{\MCG} Z_C,
\end{align}
where, on the RHS, $P_C^\mu$ is put on shell ($X_C \rightarrow 0$).

\subsection{MCG non-invariant curve}
If $\text{Stab}(C) < \text{MCG}$, we can use a Mirzakhani kernel to evaluate the $1/X_C$ pole. We choose $C$ as one of the coset representatives, so that the Mirzakhani kernel is
\begin{equation}
    {\cal K} = \frac{\alpha_C}{\rho} + \ldots.
\end{equation}
Then
\begin{align}
\int \frac{d^Et}{\MCG} Z = \int \frac{d^Et}{\text{Stab}C}\, \frac{\alpha_C}{\rho} Z + \ldots,
\end{align}
where the $\ldots$ are all terms without a $1/X_C$ pole. To guarantee that $X_C$ only appears in the first term, we can choose the other coset representatives $C_1,...,C_{L-1}$ so that all of these are curves that intersect $C$. We can put the $1/\rho$ in the numerator, by introducing an auxiliary integration variable $\xi$:
\begin{align}
\int \frac{d^Et}{\MCG} Z = \int\limits_0^\infty d \xi \int \frac{d^Et}{\text{Stab}(C)}\, \alpha_C e^{-\xi\rho} Z + \ldots.
\end{align}
Changing variables as before, and integrating over $t_C$ gives
\begin{align}
\int \frac{d^Et}{\MCG} Z = \int\limits_0^\infty d \xi \frac{-1}{(X_C+\xi)^2} \int \frac{d^{E-1}t'}{\text{Stab}(C)}\, Z' + \ldots,
\end{align}
where $Z'$ is obtained from $Z$ by shifting $X_D \mapsto X_D + \xi$ for all $D$ in the Mirzakhani set. Finally, integrating over $\xi$, and using
\begin{align}
\prod_{i=1}^m \frac{1}{X_i + \xi} = \sum_{i=1}^m \frac{1}{X_i+\xi} \prod_{j\neq i} \frac{1}{X_j - X_i},
\end{align}
we find
\begin{equation}
    \int \frac{d^Et}{\MCG} Z \rightarrow \frac{1}{X_C} \int \frac{d^{E-1}t'}{\text{Stab}(C)}\, Z_C + \ldots,
\end{equation}
where $-\log Z_C$ is the curve action given by summing over all curves, $D$, compatible with $C$:
\begin{equation}
    -\log Z_C = \sum_D \alpha_D X_D.
\end{equation}

Note that this calculation does not apply if the integrand has higher poles in $X_C$, such as if $X_C$ is a bubble propagator for a planar diagram.

\section{The Surface Symanzik polynomials}\label{sec:glsymder}
Fixing an assignment of momenta to the curves gives explicit formulas for the all the propagator factors
\begin{equation}
X_C = \left(K_C^\mu + \sum_{a=1}^L h_C^a\ell_a^\mu\right)^2+m^2,
\end{equation}
in terms of one set of loop momentum variables $\ell_a^\mu$. In terms of these loop variables, the curve action,
\begin{equation}
    -\log Z = \sum_C \alpha_C X_C,
\end{equation}
becomes
\begin{align}
-\log Z = - \ell_a^\mu A^{ab} \ell_b^\mu - 2B^a_\mu \ell_a^\mu - \mathcal{Z},
\end{align}
where $A,B,\mathcal{Z}$ are all linear functions in the generalised Schwinger parameters:
  \begin{align}
  A^{ab} & = \sum_C h_C^a h_C^b \alpha_C\\
  B^a_\mu &= \sum_C h_C^a \alpha_C K_{C\,\mu}\\
  \mathcal{Z} &= \sum_C \alpha_C (K_C^2+m^2)
  \end{align} 
Performing the Gaussian integral over the $\ell_a$ variables, in $D$ dimensions, gives
\begin{align}
  {\cal A} = \int \frac{d^E\mathbf{t}}{\MCG}\, \left( \frac{\pi^L}{\det A} \right)^{\frac{D}{2}} \exp\left(-B^TA^{-1}B - \mathcal{Z} \right).
  \end{align}
So we identify the surface Symanzik polynomials:
\begin{align}
\mathcal{U} = \det A,\qquad\text{and}\qquad \frac{\mathcal{F}_0}{\mathcal{U}} = B^T A^{-1} B.
\end{align}
These are the formulas used in the main text. In this appendix, we consider the explicit expansions of ${\cal U}$ and ${\cal F}_0$ in monomials.

\subsection{The first surface Symanzik}
Since $X^{ij}$ is linear in the parameters $\alpha_C$, the determinant $\det X$ is homogeneous of degree $L$. For a set of curves ${S} = \{C_1,...,C_L\}$, let us find the coefficient in $\det A$ of the monomial
\begin{align}
\alpha_{S} = \prod \alpha_{C_i}.
\end{align}
By the definition of the determinant, this coefficient is
\begin{align}\label{eq:Xdet}
\det A = \ldots + \alpha_{S} \, \left(\det \left. h\right|_{S}\right)^2 + \ldots\,,
\end{align}
where
\begin{align}
\det \left. h\right|_{S} = \epsilon_{i_1...i_L} h_{C_1}^{i_1}...h_{C_L}^{i_L}.
\end{align}
Note that the ordering of the curves $C_1,...,C_L$ does not matter, because this determinant only enters the formula for $\det A$ as a square.

We now make two observations. Firstly, $\det h|_S$ is only non-zero if the curves in $S$ cut $\Gamma$ to a tree graph. Secondly, for any conventional choice of loop variables (defined below), the determinants $\det h|_S$ are all either $0$ or $\pm 1$. So the result is that ${\cal U}$ is given by
\begin{align}
 \mathcal{U} = \sum_{\substack{{S}~\text{cuts}\,\Gamma\\\text{to tree}}} \alpha_{S}.
\end{align}

For the first statement, consider $L=1$. Then all curves have momenta of the form
\begin{align}
P_C = h_C^1\ell_1 + K_C^\mu.
\end{align}
If $h_C^1=0$, cutting $\Sigma$ along $C$ breaks it into two parts: one part with $L=1$, and a second part with $L=0$ (i.e. a disk). Whereas, if $h_C^1\neq 0$, cutting $\Gamma$ along $C$ cuts the loop open, giving a new surface with $L=0$ (i.e. a disk). So at 1-loop the first Symanzik polynomial is
\begin{align}
\mathcal{U} = \sum_{\substack{C~\text{cuts}\,\Gamma\\\text{to tree}}} \alpha_C \, \left(h_C^1\right)^2.
\end{align}
For $L>1$, the determinant $\det \left. h\right|_{S}$ is nonzero if and only if the linear transformation (in $H_1(\Gamma,\partial\Gamma)$ from $[L_1],...,[L_L]$ to $[C_1],...,[C_L]$ is invertible. By induction from the $L=1$ case, this means that the curves in $S$ cut $\Gamma$ to a disk. So
\begin{align}
 \mathcal{U} = \sum_{\substack{{S}~\text{cuts}\,\Gamma\\\text{to tree}}} \alpha_{S} \, \left(\det \left. h\right|_{S}\right)^2.
 \end{align}

Secondly, it turns out that $(\det h|_S)^2$ is either $0$ or $1$. We sketch how to prove this by fixing any genus $g$ fatgraph with $h$ trace-factor components. The loop order of such a fatgraph is
\begin{equation}
    L = 2g + h -1.
\end{equation}
A natural basis of loop-carrying curves can be given by picking some $2g$ curves $A_i,B_i$ wrapping the $A,B$-cycles of the graph, and $h-1$ curves $C_{i}$ connecting the $h$ trace factors. These give a set, $S$, of $L$ cures that cut $\Gamma$ to a tree, so $(\det h|_S)^2$=1. Moreover, we can choose our momentum assignment such that
\begin{equation}
    P_{A_i} = \ell_{2i-1},\qquad P_{B_i} = \ell_{2i},\qquad P_{C_i} = \ell_{2g+i}.
\end{equation}
Now consider the momenta of Dehn twists of these curves. For instance, taking one of the $C_i$, a Dehn twist $\gamma$ around one of its trace-factors gives a new curve
\begin{equation}
    P_{\gamma C_i} = P_{C_i} \pm k_\text{tf},
\end{equation}
where $k_\text{tf}$ is the total momentum of the trace factor. Moreover, any product of Dehn twists acting on a pair of A,B-cycles acts on their momenta as SL$_2\mathbb{Z}$:
\begin{equation}
    \begin{bmatrix} \ell_{2i-1}\\ \ell_{2i} \end{bmatrix} \mapsto X \begin{bmatrix} \ell_{2i-1}\\ \ell_{2i} \end{bmatrix}, 
\end{equation}
for some $X \in \text{SL}_2\mathbb{Z}$. In this way, we find that the momenta of any set, $S'$, that cuts $\Gamma$ to a tree, is obtained from the momenta of $S$ via translations by non-loop momenta, and SL$_2\mathbb{Z}$ transformations. Both of which leave the determinant unchanged:
\begin{equation}
    (\det h|_{S'})^2 = (\det h|_S)^2 = 1.
\end{equation}

 \subsection{The second surface Symanzik}
The second surface Symanzik polynomial is
\begin{align}
\frac{\mathcal{F}_0}{\mathcal{U}} = B^T A^{-1} B.
\end{align}
 The Laplace formula evaluates the inverse as
 \begin{align}
 \left(A^{-1}\right)^{ij} = \frac{(-1)^{i+j}}{\det A} |A|^{ij},
 \end{align}
 where $|A|^{ij}$ is the $i,j$ minor. Since $\mathcal{U}=\det A$,
 \begin{align}\label{app:F0step}
 \mathcal{F}_0 = 2 \sum_{C,D} \alpha_C\alpha_D K_C\cdot K_D \sum_{i,j} (-1)^{i+j} h_C^i h_D^j |A|_{ij}.
 \end{align}
As above, again write $S = \{C_1,...,C_L\}$ for a set of $L$ curves and $\alpha_{S}$ for the associated monomial. The minors of $A$ are
\begin{align}\label{app:minX}
|A|_{ij} = \sum_{S} \sum_{C\in {S}} \frac{\alpha_S}{\alpha_C} \, |h_{S}|^i_C|h_{S}|^j_C,
\end{align}
where $|h_{S}|^i_C$ is the $(i,C)$ minor of the matrix $h|_{S} = [h_{C_1}^i|...|h_{C_L}^i]$. By the definition of the determinant,
\begin{align}\label{app:detid}
\sum_{i=1}^L (-1)^i h_D^i |h_{S}|_C^i = \det h_{{S}_{C\rightarrow D}},
\end{align}
where ${S}_{C\rightarrow D}$ is the set obtained from ${S}$ by replacing $C$ with $D$. Substituting \eqref{app:minX} into \eqref{app:F0step}, and using the identity \eqref{app:detid}, gives (after reordering the summations)
 \begin{align}\label{app:F0step2}
 \mathcal{F}_0 = 2 \sum_{\substack{\mathcal{S}'\\ |\mathcal{S}'|=L+1}} \alpha_{\mathcal{S}'} \left( \sum_{C\in \mathcal{S}'} \left(\det h_{S'\backslash C} \right) K_C^\mu \right)^2,
 \end{align}
 where the sum is restricted to sets of $L+1$ curves $\mathcal{S}'$ such that \emph{any} $L$ subset of $\mathcal{S}'$ gives a nonvanishing determinant $\det h_{S'\backslash C}$.
 
 We make three observations to simplify this formula.
 
 First, by the previous section, any $L$-subset of ${S}'$ that has nonvanishing determinant cuts $\Gamma$ to a tree graph. It follows that the sum in this formula is over sets $\mathcal{S}'$ that \emph{factorize} $\Gamma$ into two trees!

Secondly, by the previous subsection, since each of the sets $S'\backslash C$ cuts $\Gamma$ to a tree, the determinants are all
\begin{equation}
    \det h_{S'\backslash C} = \pm 1.
\end{equation}

In fact, finally, note that both the vectors $h_C^i$ and the momenta $K_C^\mu$ are defined with respect to an orientation of $C$. For any subset $\mathcal{S}'$, these orientations can be chosen so that all the determinants $\det h_{S'\backslash C}$ are positive (say). For this choice,
 \begin{equation}
     \det h_{S'\backslash C} = 1.
 \end{equation}

Combining these three observations, the final formula for ${\cal F}_0$ is
 \begin{align}\label{app:F0}
 \mathcal{F}_0 =  \sum_{\substack{S'~\text{cuts}~\Gamma\\ \text{to 2 trees}}} \alpha_{{S}'} \left( \sum_{C\in {S}'}  K_C^\mu \right)^2,
 \end{align}
for an allowed choice of orientations of the momenta $K_C$. 

\section{The Recursion Formula}\label{sec:app:recursion}
For a fatgraph $\Gamma$, the curve integral for integrands is
\begin{align}
{\cal I} = \int \frac{d^E t}{\MCG} Z,
\end{align}
with
\begin{align}
-\log Z =  \sum_C\alpha_C X_C.
\end{align}
For some trace factor $\beta$ of $\Gamma$, we have the set of curves $\mathcal{S}$ that have one or two endpoints in $\beta$. Under the $\MCG$, this set has some, say $k$, coset representatives, $C_i$ ($i=1,\ldots,k$). Then 
\begin{align}
{\cal I} = \int \frac{d^E t}{\MCG} Z = \sum_{i=1}^k \int \frac{d^E t}{\text{Stab}(C_i)} \frac{\alpha_{C_i}}{\rho} Z,
\end{align}
where
\begin{align}
\rho = \sum_{C\in\mathcal{S}} \alpha_C.
\end{align}
Introducing an auxiliary parameter, $\xi$, we re-write this as
\begin{align}
{\cal I} = \sum_{i=1}^k  \int\limits_{0}^\infty d\xi \int \frac{d^E t}{\MCG} \, \alpha_{C_i} Z(\xi).
\end{align}
where the new integrand is
\begin{align}
-\log Z(\xi) =  \sum_{C\in\mathcal{S}} \alpha_C (X_C+\xi) + \sum_{D\not\in\mathcal{S}} \alpha_D X_D.
\end{align}
Integrating over the $\alpha_{C_i}$ direction in each term curve integral gives
\begin{align}
{\cal I} = \sum_{i=1}^k  \int\limits_{0}^\infty d\xi  \frac{-1}{(X_{C_i}+\xi)^2} \int \frac{d^{n-1} t'}{\text{Stab}(C_i)} \, Z'(\xi),
\end{align}
where
\begin{align}
-\log Z'(\xi) =  \sum_{C\in\mathcal{S}, C\neq C_i} \alpha'_C (X_C+\xi) + \sum_{D\not\in\mathcal{S}} \alpha'_D X_D,
\end{align}
and $\alpha'_C$ are the headlight functions obtained after integrating out the ${\bf g}_{C_i}$ direction. These are the headlight functions for the fatgraph $\Gamma_{C_i}$ obtained by cutting along $C_i$.

Note that we can evaluate the $\xi$ integral using identities such as
\begin{align}
\prod_{i=1}^m \frac{1}{X_i + t} = \sum_{i=1}^m \frac{1}{X_i+t} \prod_{j\neq i} \frac{1}{X_j - X_i}.
\end{align}
When all the $X_C$ propagator factors are distinct (i.e. there are no higher poles), we can perform the integral to find
\begin{align}
{\cal I} = \sum_{i=1}^k  \frac{1}{X_{C_i}} \int \frac{d^{n-1} t'}{\text{Stab}(C_i)} \, Z'(-X_{C_i}),
\end{align}

\section{Recursion Examples}

\subsection{The 3-point non-planar 1-loop amplitude}
Take $\Gamma$ to be the 3-point non-planar 1-loop diagram considered in Section \ref{ex:A21:exam}. The curves are $C_{12}^n, C_{13}^n, C_{22}, C_{33}$. For the Mirzakhani method, we have two cosets, with representatives $C_{12}^0, C_{13}^0$. Cutting $\Gamma$ along $C_{12}^0$ gives a 5-point tree fatgraph $\Gamma_{C_{12}^0}$. The curves compatible with $C_{12}^0$ are
\begin{align}
C_{12}^1, C_{13}^0, C_{12}^{-1},C_{13}^{-1}, C_{22}.
\end{align}
The global forward limit then computes $I_\Gamma$ as
\begin{align}
I_\Gamma = \frac{1}{X_{12}^0} I_{\Gamma_{C_{12}^0}}(X_{12}^1-X_{12}^0, X_{13}^0-X_{12}^0, X_{12}^{-1}-X_{12}^0,X_{13}^{-1}-X_{12}^0, X_{22}) + (2\leftrightarrow 3).
\end{align}
But the 5-point tree amplitude is
\begin{align}
I(X_1,X_2,X_3,X_4,X_5) = \sum_{i=1}^5 \frac{1}{X_i X_{i+1}}.
\end{align}
So the integrand is
\begin{multline}
I_\Gamma = \frac{1}{X_{12}^0(X_{12}^1-X_{12}^0)( X_{13}^0-X_{12}^0)} + \frac{1}{X_{12}^0(X_{13}^0-X_{12}^0)(X_{12}^{-1}-X_{12}^0)} + \frac{1}{X_{12}^0(X_{12}^{-1}-X_{12}^0)(X_{13}^{-1}-X_{12}^0)} \\ + \frac{1}{X_{12}^0(X_{13}^{-1}-X_{12}^0) X_{22}} + \frac{1}{X_{12}^0X_{22} (X_{12}^1-X_{12}^0)} + (2 \leftrightarrow 3).
\end{multline}
The momenta are explicitly
\begin{align}
P_{12}^n = \ell + n k_1,\qquad P_{13}^n = \ell + k_2 + n k_1,\qquad P_{22} = k_1,\qquad P_{33} = k_1+k_2.
\end{align}

\subsection{The 2-loop vacuum at genus one}
The 2-loop genus 1 vacuum amplitude has already been computed in Section \ref{ex:markov:exam}. Take again $\Gamma$ to be the 2-loop genus one vacuum diagram. The curves of $\Gamma$ are $C_{p/q}$, with momentum
\begin{align}
P_{p/q} = p \ell + q \ell'.
\end{align}
Every curve $C_{p/q}$ is in the same MCG-orbit. Pick, say, $C_{0/1}$ as the coset representative. The curves compatible with $C_{0/1}$ are $C_{1/n}$ for $n\in\mathbb{Z}$. Cutting $\Gamma$ along $C_{0/1}$ gives a 1-loop non-planar diagram $\Gamma_{C_{0/1}}$, and the curves $C_{1/n}$ can be identified with the curves we called `$C_n$' in the previous example. Applying the global forward limit once gives
\begin{align}\label{eqn:intmarkint}
I_\Gamma = \frac{1}{X_{0/1}}\, I_{\Gamma_{C_{0/1}}} (X_{1/n} - X_{0/1}).
\end{align}
However, we have already computed the 1-loop non-planar integrand, and found, up to loop-momentum shifts, that it is given by
\begin{align}
I_{\Gamma_{C_{0/1}}} (X_n) = \frac{1}{X_0X_1}.
\end{align}
Using this result in \eqref{eqn:intmarkint} gives
\begin{align}
I_\Gamma = \frac{1}{X_{0/1}(X_{1/0}-X_{0/1})(X_{1/1} - X_{0/1})}.
\end{align}
Loop re-definitions of $\ell$ and $\ell'$ can be used to cyclically permute the labels $0/1, 1/0, 1/1$. Summing over the possible three cyclic permutations (and dividing by $3$) gives
\begin{align}
I_{\Gamma} = \frac{1}{3} \frac{1}{X_{0/1}X_{1/0}X_{1/1}}.
\end{align}
The $1/3$ factor is expected because $|\text{Aut}(\Gamma)| = 3$. We therefore recover $1/3$ of the Feynman integral of the sunrise vacuum diagram.

\subsection{A comment on the 1-loop planar amplitudes}
Our formula for the 1-loop planar amplitudes can be computed directly, without topological recursion. The global Schwinger formula gives a well defined loop integrand for these amplitudes, without linearized propagators. However, we can arrive at a forward-limit-like formula for the 1-loop integrand by inserting the `trivial' Mirzakhani kernel
\begin{align}\label{eq:1loopmir}
1 = \sum_{i=1}^n \frac{\alpha_{0i}}{\sum_{i=j}^n \alpha_{0j}}
\end{align}
into the curve integral. Here, $\alpha_{0i}$ is the headlight function of $C_{0i}$, the curve from $i$ to the internal loop boundary, $0$. Equation \eqref{eq:1loopmir} then allows us to write the 1-loop planar n-point amplitude as a sum of $n$ disk amplitudes, with linearized propagators. Evaluating the integral, using the recursion \eqref{eq:globfwd}, the integrand is
\begin{equation}
I_\mathrm{1 loop}(1,...,n) = \left. \sum_{i=1}^n \frac{-1}{X_{0i}} A(12....i0i....n) \right|_{X_{0j}\mapsto X_{0j}-X_{0i}},
\end{equation}
where $A(12...n)$ are the tree-level partial amplitudes, but now with linearized propagators.

\section{Details for the non-planar 1-loop propagator}\label{app:A11}
The matrix for the curves $C_n$ with $n\geq 0$ is
\begin{align}
M_n = L D_x (L D_y R D_x)^n R.
\end{align}
Taking the transpose, we see that $M_n^T = M_n$. In particular,
\begin{align}
M_0 = \begin{bmatrix} 1 & 1 \\ 1 & 1+x \end{bmatrix}.
\end{align}
Given $M_0$, we can compute $M_n$ using
\begin{align}
M_{n+1} = M_n B_{+1},\qquad\text{where}~~B_{+1}=R^{-1} L D_y R D_x R = \begin{bmatrix} 0 & ~~- xy \\ 1 & ~~1+ x + xy \end{bmatrix}.
\end{align}
It follows that we can write
\begin{align}
M_n = \begin{bmatrix} F_{n-2} & F_{n-1} \\ F_{n-1} & F_{n} \end{bmatrix},
\end{align}
where
\begin{align}
F_{n+2} = (1+x+xy)F_{n+1} - xy F_n,
\end{align}
with initial conditions $F_{-2}=1, F_{-1}=1$. The first few examples are
\begin{align}
F_0 & = 1+x,\\
F_1 & = 1+2x+x^2+x^2y,\\
F_2 & = 1 + 3 x + 3 x^2 + x^3 + 2 x^2 y + 2 x^3 y + x^3 y^2.
\end{align}

Similarly, the matrix for the curves $C_n$ with $n<0$ is given by
\begin{align}
M_n = R D_y (R D_x L D_y)^{-n-1} L, \qquad  n <0.
\end{align}
These matrices are again symmetric, and
\begin{align}
M_{-1} = \begin{bmatrix} 1+y & ~y \\ y & ~y \end{bmatrix}.
\end{align}
We can evaluate $M_n$ using
\begin{align}
M_{n-1} =  M_n B_{-1},\qquad \text{where} ~B_{-1} = L^{-1} R D_x L D_y L = \begin{bmatrix} 1+ x+ xy & ~ xy \\ -1 & ~0 \end{bmatrix}.
\end{align}
This implies that $M_n$ ($n<0$) has the form,
\begin{align}
M_n = \begin{bmatrix} G_n & xy G_{n+1} \\ xy G_{n+1} & (xy)^2 G_{n+2} \end{bmatrix},
\end{align}
where the polynomials $G_n$ are determined by the recursion
\begin{align}
G_n = (1+x+xy) G_{n+1} - xy G_{n+2},
\end{align}
with initial condition $G_{1} = 1/(x^2y)$ and $G_0 = 1/x$. The first few polynomials are
\begin{align}
G_{-1} &= 1+y,\\
G_{-2} &= 1+x+2xy+xy^2,\\
G_{-3} &= (1+x+xy)^2 + x^2y(1+y)^2.
\end{align}

We now need to compute the tropicalizations of the polynomials $F_n$ ($n\geq -2$) and $G_n$ ($n\leq 1$). Write
\begin{align}
f_n = \text{Trop\ } F_n,\qquad\text{and}\qquad g_n = \text{Trop\ } G_n.
\end{align}
Then, for $n\geq 0$, we find
\begin{align}
f_n = \max (0, (n+1) x, (n+1)x+ny),
\end{align}
which follows by induction using that
\begin{align}
f_{n+2} = \max (\max(0,x,x+y) + f_{n+1}, \max(0,x+y) + f_n ).
\end{align}
Similarly, for $n\leq -1$,
\begin{align}
g_n = \max (0, -(n+1)x,-(n+1)x - n y ).
\end{align}
We also have that
\begin{align}
f_{-2}=0,~~f_{-1}=0,~~g_1=-2x-y,~~g_0=-x.
\end{align}
The headlight functions are
\begin{align}
\alpha_n&= - f_n + 2f_{n-1} - f_{n-2},\qquad n\geq 0,\\
\alpha_n&= - g_n + 2g_{n+1} - g_{n+2},\qquad n<0.
\end{align}

\bibliographystyle{unsrt}
\bibliography{refs}

\end{document}